\baselineskip = 18pt

\input epsf

\rightline{ROM2F/2007/18}
\rightline{KCL-MTH-08-05}

\vskip 2cm
\centerline
{\bf Charge multiplets and masses for $E_{11}$}
\vskip 1cm
\centerline{Paul P. Cook$^{a,b}$ and Peter West$^c$}
\vskip 0.5cm
\centerline{$^a$ Dipartimento di Fisica, Universita di Roma "Tor Vergata"}
\centerline{Via della Ricerca Scientifica, 1, Roma 00133, Italy}
\vskip 0.5cm
\centerline{$^b$ Scuola Normale Superiore,}
\centerline{Piazza dei Cavalieri, 7, Pisa, 56126, Italy}
\centerline{p.cook@sns.it}
\vskip 0.5cm
\centerline{$^c$ Department of Mathematics}
\centerline{King's College, London WC2R 2LS, UK}
\centerline{peter.west@kcl.ac.uk}

\vskip 2cm
\leftline{\sl Abstract}
\vskip .2cm
\noindent 
The particle, string and membrane charge multiplets are derived in detail from the decomposition of the $l_1$ (charge) representation of $E_{11}$ in three, four, five, six, seven and eight spacetime dimensions. A tension formula relating weights of the $l_1$ (charge) representation of $E_{11}$ to the fundamental objects of M-theory and string theory is presented. The reliability of the formula is tested by reproducing the tensions of the content of the charge multiplets. The formula reproduces the masses for the pp-wave, M2, M5 and the KK-monopole from the low level content of the $l_1$ representation of $E_{11}$. Furthermore the tensions of all the Dp-branes of IIA and IIB theories are found in the relevant decomposition of the $l_1$ representation, with the string coupling constant and $\alpha'$ appearing with the expected powers. The formula leads to a classification of all the exotic, KK-brane charges of M-theory.
\vskip 0.5cm
\bigskip
\bigskip
\eject
\bigskip
{\bf Introduction}
\medskip
The Kac-Moody algebra $E_{11}$ has been conjectured to be the algebra describing the symmetries of M-theory [1]. The arguments used to make this conjecture were based upon previously unnoticed properties of D=11 supergravity, leading to its formulation as a nonlinear realisation which included the Borel generators of $E_7$. These arguments lead to an $E_{11}$ algebra in eleven dimensions encoding the symmetries of M-theory. Generalised Kac-Moody algebras, such as $E_{11}$, are not well understood and analysis of their content is hampered by the lack of a simple way of applying the Serre relations to the putative generators of the algebra. However $E_{11}$ belongs to a class of algebras known as Lorentzian Kac-Moody algebras which have the property that the deletion of one node of the defining Dynkin diagram leaves behind a set of Dynkin diagrams of finite dimensional groups [2]. Progress has been made by decomposing the algebra into representations of finite dimensional sub-algebras which are graded by a level and analysing the content. Analysing the low-level content of the adjoint representation of $E_{11}$ in this way one can reproduce the dynamics of the bosonic sector of D=11 supergravity. In addition one finds all the bosonic fields of the D=11 supergravity theory at the lowest levels of the decomposition as well as the dual gravity field leading to a dualised gravity theory.  

However in the original non-linear realisation of the $E_{11}$ symmetry the translation generator had to be added in by hand. Space-time did not emerge from the adjoint representation of $E_{11}$. It was proposed that one could more naturally include spacetime by enlarging the algebra to include its first fundamental representation, or the $l_1$ representation, as well as the adjoint representation of $E_{11}$ [3]. The success of this approach is that the translation generator is associated to the highest weight of the $l_1$ representation and appears at the lowest level in the $l_1$ representation. Furthermore one finds that the $l_1$ representation contains generators having the correct index structure to be interpreted as the central charges of the supersymmetry algebra, even though the arguments used to conjecture an $E_{11}$ symmetry considered only the bosonic fields of supergravity. 

The associations between the $l_1$ representation and the adjoint representation of $E_{11}$ as well as other very-extended algebras have been studied in detail in [5]. It has been proposed that the full set of brane charges of M-theory are contained in the $l_1$ representation of $E_{11}$ [3] and, in fact, for every generator of the adjoint representation of $E_{11}$ one can associate a half-BPS brane solution [4] and in the $l_1$ representation one can find a corresponding generator associated with the conserved charge on the brane. Following the proposition that the $l_1$ representation contains the brane charges of M-theory, the brane charge multiplets in three dimensions which are predicted by U-duality [6,7,8,9,10] have been found inside the $l_1$ representation [11], and the particle multiplet has been identified in all dimensions from three to eight in [12]. To find the three dimensional charge multiplet the dimensional reduction on an 8-torus was carried out as a decomposition of the $l_1$ representation into representations of an $A_2\otimes E_8$ sub-algebra. In this paper we further the arguments in favour of the relevance of the $l_1$ representation by deriving all the possible charge multiplets from its algebra in three to eight spacetime dimensions.

If the conjecture is true that string theory, and M-theory, do carry a Kac-Moody algebra one may hope to better understand the algebras by making connection with areas of string theory that are well understood. One direction forward is by the introduction of group theoretical constructions that can be argued to have a clear interpretation in terms of the string theory. The most fundamental properties in a physical theory are those that give rise to the simplest dimensionful quantities. Space and time are already present inside the algebraic construction, via the local Lorentzian symmetry algebra, but a natural interpretation for mass is missing. In the case of string theory a non-perturbative concept amenable to study is the tension of BPS p-branes, or the mass per unit volume. Indeed previously [6,7,8,9,10] a group theoretical tension formula has been constructed as an empirical tool to discover the charge multiplets of M-theory and string theory by application of U-duality symmetries. It is our principle aim in this paper to derive this tension formula in the context of $E_{11}$ and to check our formula by duplicating the tensions found in the U-duality charge multiplets. 

A by-product of the tension formula that will be derived is the observation that most of the $l_1$ representation is associated to exotic charges carried by KK-branes, which are higher-dimensional generalisations of the KK-monopole and provide a higher dimensional origin to certain brane charges and KK-modes in lower dimensions (see, for example, [21,22]). Some of these KK-brane charges have been observed as a consequence of U-duality in [6,7,8,9,10]. The $l_1$ representation of $E_{11}$ includes all the exotic KK-brane charges expected by U-duality transformations, and these are organised into finite sets within the $l_1$ representation. The interpretation of KK-brane charges offers a new way to decompose the $E_{11}$ algebra, and conversely the $l_1$ representation also offers a classification of all the KK-brane charges expected in M-theory. We will present the full set of the simplest class of KK-brane charges expected in M-theory, as well as in the IIA and IIB string theories. 

The discussion in the paper will be split into two parts. First in sections 1 and 2 we will give explicit decompositions of the $l_1$ algebra. In the first section we will give the decomposition of the $l_1$ representation of $E_{11}$ relevant to the eleven dimensional theory as well as the decomposition to the two ten-dimensional theories. We present this original decomposition using an explicit basis for the root lattice vector space that has not been used in the $E_{11}$ literature previously that will greatly simplify our observations later in this paper. In section 2, we study the $l_1$ representation and identify how the charge multiplets are organised inside the representation, we then derive explicitly the charge multiplets of M-theory in three to eight dimensions. The abstract decomposition of the $l_1$ representation into representations of $A_{D-1}\otimes E_{11-D}$ for D spacetime dimensions is given in section 2.1. In section 2.2 we give the criteria for finding rank p charges in the $A_{D-1}$ algebra and find the corresponding representations in the $E_{11-D}$ algebra. In the remainder of section 3 we reproduce the particle, string and membrane charge multiplets in various dimensions and give the tensions of the exotic charges according to the formula that will be presented later in this paper. The tensions for the particle and string multiplets confirm the findings in the literature but the membrane multiplet charges have not been presented explicitly before. In the second part of the paper, commencing with section 3, we will present a tension formula and use it to derive the tensions associated to the fundamental objects of M-theory and string theory in section 3.3 from the roots of the $l_1$ representation, based on the computations in the section 1. Expressions for the tension associated to any root in the $l_1$ representation are given in section 3.3 and the simplest KK-brane charges are listed in their entirety. 
\medskip
{\bf 1 The $l_1$ representation of $E_{11}$}
\medskip
$E_{11}$ is described completely by its Dynkin diagram which is found by attaching three additional roots to the longest leg of the $E_8$ diagram, each extra simple root having the same length as any root of $E_8$.
\medskip
\centerline{\epsfbox{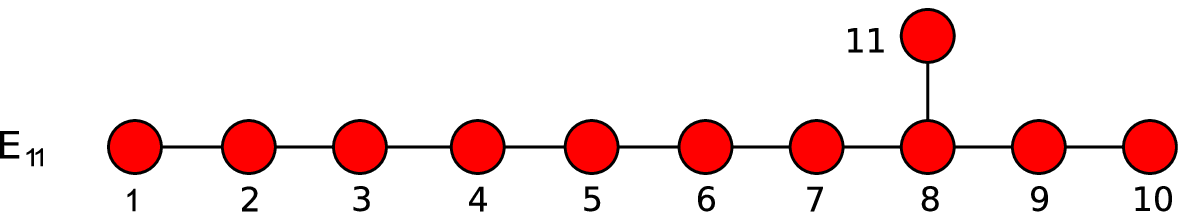}}
\medskip
By deleting the exceptional node, labelled eleven, one finds representations of the remaining $A_{10}$ sub-algebra, graded by the level, which is the number of times the deleted simple root must be added to the $A_{10}$ root to make it into an $E_{11}$ root. Decomposing the algebra in this way one finds the adjoint representation of $E_{11}$ and at the first few levels one finds the gravitational field, a three-form field, a six-form field and the dual to gravity field. With the exception of the dual to gravity these fields are well-known bosonic fields from eleven dimensional supergravity. Dimensionally reducing the algebra to D dimensions corresponds to deleting different nodes of the Dynkin diagram leaving representations of $A_{D-1}$ sub-algebras. For example in the reduction to ten dimensions there is choice of which nodes to delete and this gives rise to one of the most beautiful aspects of the construction, namely that the two choices correspond to the choice of IIA or IIB theories in ten dimensions. More explicitly one deletes nodes of the $E_{11}$ Dynkin diagram so as to leave behind an $A_9$ Dynkin diagram (a line of nine connected nodes). One can do this in two ways, by deleting nodes $11$ and $10$ or by deleting node $9$, which yields the bosonic fields of the IIA and IIB theories respectively in an elegant way.

The representations of $E_{11}$ other than the adjoint are also interesting and of direct relevance to theoretical physics. The $l_1$, or charge, representation of $E_{11}$ is believed to contain all the brane charges of M-theory in the $E_{11}$ weight lattice [3]. The decomposition of this algebra to different spacetime dimensions also corresponds to the deletion of different nodes of its associated Dynkin diagram. Of interest to physicists are the decompositions giving representations of $SL(11)$ and $SL(10)$ sub-groups, which give generators in the algebra conjectured to be the brane charges of M-theory and the two ten-dimensional supergravity theories respectively. In section 1.1 we will give the decomposition relevant to M-theory and in sections 1.2 and 1.3 we make the decompositions connected to the IIA and IIB supergravity theories. The presentation in this section will differ cosmetically from much of the literature in that it will make use of a vector space basis denoted herein by $\{e_i\}$ vectors instead of the more usual simple root basis, $\alpha_i$.

We recall that the $l_1$ representation of $E_{11}$ takes the first fundamental weight of $E_{11}$, $l_1$, associated to the translation generator, and treats it as the highest weight of a representation in the $E_{11}$ weight lattice [3]. Alternatively one can obtain the $l_1$ representation of $E_{11}$ by extending the $E_{11}$ Dynkin diagram with a node attached by a single line to the longest leg of the $E_{11}$ diagram, giving the Dynkin diagram of $E_{12}$ shown below, and one then restricts to just those roots with the coefficient of the extra root, $\alpha_*$, set to one. In other words one decomposes $E_{12}$ by the deletion of the node $\alpha_*$ and the $l_1$ representation of $E_{11}$ is found at level one with highest weight $l_1$, the first fundamental weight of $E_{11}$.
\medskip
\centerline{\epsfbox{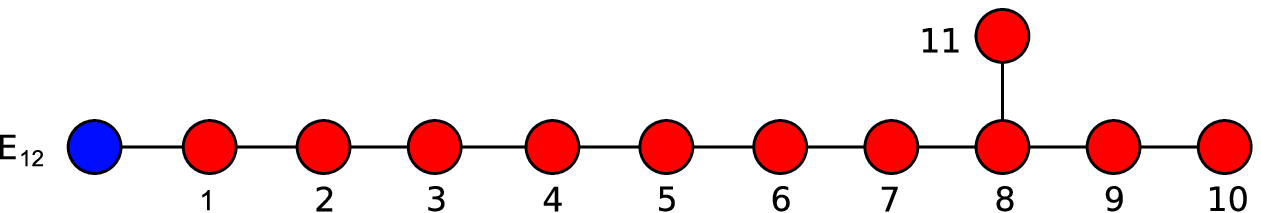}}
\medskip
A general root appearing in the $E_{12}$ root lattice with $m_*=1$ has the form:
$$\beta=\alpha_*+\sum_{i=1}^{11}m_i\alpha_i$$
Before beginning the decompositions of our algebra a few comments about the root lattices of generalised Kac-Moody algebras are in order. Whether or not the generic root $\beta$ appears in the root lattice depends upon the application of the Serre relations which, in terms of the Chevalley generators, are,
$$[E_i \ldots [E_i,E_j]\ldots]=0 \qquad [F_i \ldots [F_i,F_j]\ldots]=0$$
Here there are $(1-A_{ij})$ $E_i$ generators ($F_i$ generators in the second relation) where $A_{ij}$ is the Cartan matrix associated to the Dynkin diagram of the algebra. One consequence of the Serre relations is that $\beta^2\leq 2$ for roots in the $l_1$ lattice. Another consequence is that any root appearing in the lattice must have connected support on the Dynkin diagram. These and another readily defined condition, on the index structure of generators appearing in the algebra to be discussed later, are sufficient to capture much of the the Serre relations, and are far easier to apply computationally. The application of the Serre relations in a simple manner is an open problem in mathematics and later on when we give explicit roots in the lattice at high levels these are the criteria that will have been applied. 
\medskip
{\bf 1.1 The eleven-dimensional theory}
\medskip
Our aim here is to decompose the $l_1$ representation into representations of a preferred $A_{10}$ sub-algebra, giving representations of $SL(11)$. This decomposition has been made previously [5] but here we will present the results using a vector space basis $\{e_*,e_1,\ldots e_{11}\}$ for the root lattice that will simplify later calculations in the paper but has not been much used in the literature previously. The preferred $A_{10}$ algebra is given by the Dynkin diagram of $E_{12}$ shown above when the nodes indicated by a $*$ and 11 are deleted and is called the gravity line in this decomposition. The generators will be $SL(11)$ tensors graded by a level and are believed to be the brane charges of M-theory.

We first decompose the roots into components in the $E_{11}$ lattice and a vector orthogonal to it, which we shall call $y$, by writing
$$\alpha_*=y-l_1$$
Where $l_1$ is the first fundamental weight of $E_{11}$, we recall that fundamental weights are dual to the corresponding simple roots of an algebra:
$$<\lambda_i,\alpha_j>=\delta_{ij}$$
Where $\lambda_i$ denotes the i'th fundamental weight and $\alpha_i$ are the simple roots indicated by the Dynkin diagram of the algebra. In this case, $\alpha_*^2=2$ and $l_1^2={1\over2}$, so that $y^2={3\over 2}$. It will be useful later to consider an explicit vector space basis for our root system. We introduce the basis $\{e_*,e_1,\ldots e_{11}\}$ endowed with the Lorentzian inner product:
$$<a,b>=\sum_{i=*}^{11} a_ib_i-{1\over 9}\sum_{i=*}^{11} a_i \sum_{j=*}^{11} b_j$$
Where $a=\sum_{i=*}^{11}a_ie_i$ and similarly for $b$. We represent the simple roots of $E_{12}$ in this basis with,
$$\alpha_i=e_i-e_{i+1}\qquad i=*,1,2\ldots 10$$
$$\alpha_{11}=e_9+e_{10}+e_{11}$$
So that the inner products between the simple roots encoded in the Cartan matrix are reproduced, and all roots have length-squared normalised to two. In this notation the vector $y$ is,
$$y=e_*-{1\over 2}(e_1+\ldots +e_{11})$$
A generic root in the $l_1$ representation therefore takes the form:
$$\beta=y-l_1+\sum_{i=1}^{11}m_i\alpha_i$$
And defines the weight vectors in the $E_{11}$ lattice descended from the highest weight $l_1$:
$$\Lambda=l_1-\sum_{i=1}^{11}m_i\alpha_i$$
We may further decompose\footnote{$^1$}{We note that in the $E_{11}$ literature $m_{11}$, the coefficient of $\alpha_{11}$, is often referred to as the level of the decomposition and denoted $l$, but here, since we will consider decompositions characterised by more than one level we will continue using the notation $m_{11}$.} the root into components with roots in the $A_{10}$ lattice and a vector orthogonal to it, $z$. We write,
$$\alpha_{11}=z-\lambda_8$$
Where $\lambda_i$ is the i'th fundamental weight of $A_{10}$, defined in relation to the simple roots of $A_{10}$ by $<\lambda_i,\alpha_j>=\delta_{ij}$. Explicitly,
$$z={3\over 11}(e_1+\ldots +e_{11})$$
Where $z^2=-{2\over 11}$. The fact that the orthogonal component has an imaginary length is the sign of an indefinite algebra - from the Serre relations which define the Kac-Moody algebra we deduce that all roots in the algebra have length-squared less than or equal to two. It is the existence of a negative contribution to the root length that leads to an infinite algebra. It also allows us to sensibly decompose the infinite algebra into infinite copies of a finite algebra graded by the orthogonal imaginary component. 

We note that $\alpha_*=y+{3\over 2}z-\lambda_1$ in this decomposition. So that,
$$\beta=y+({3\over 2}+m_{11})z-\hat{\Lambda}$$
Where,
$$\hat{\Lambda}=\lambda_1+m_{11}\lambda_8-\sum_{i=1}^{10}m_i\alpha_i\equiv\sum_{i=1}^{10}p_i\lambda_i$$
$\hat{\Lambda}$ is a highest weight representation in the weight space of $A_{10}$, $SL(11)$. By taking the inner product with the fundamental weight $\lambda_j$ we find the coefficients of the simple roots in $\beta$,
$$\eqalignno{m_j&=\cases{{j\over 11}(3m_{11}+A-1)-B_j+1,  \qquad & $j\leq 8$  \cr
{j\over 11}(-8m_{11}+A-1)-B_j+8m_{11}+1,& $j>8$}}$$
Where we have defined the useful, integer expressions $A\equiv \sum_{i=1}^{10}ip_i$ and $B_j\equiv \sum_{i=1}^{j}ip_i+j\sum_{i>j}^{10}p_i$.The simple root coefficients must be positive integers. So a general solution is given by 
$$m_{11}={1\over 3}(-A+1+11k)\eqno{(1.1)}$$
Where $k$ is an integer which is bounded from below by the condition that $m_1\geq 1$, implying that $k\geq \sum p_i$.
Let us explicitly realise the lower bound on $k$ by rewriting $k$ as 
$$k\equiv \sum_{i=1}^{10}p_i + C$$ 
Where $C$ is a constant greater than or equal to zero. Substituting this expression for $k$ into the expression for the level, $m_{11}$, we have,
$$m_{11}={1\over 3}(\sum_{i=1}^{10}(11-i)p_i+1+11C)$$
At each level $m_{11}$ we will find representations of $SL(11)$ described by the $p_i$ and the new parameter $C$. We can now give a direct interpretation of $C$ in terms of the blocks of antisymmetrised indices that appear on the $SL(11)$ generators. We recall that the $l_1$ representation as a representation of $SL(11)$ is formed from its highest weight, the translation generator, $P_a$, by its commutator action with the three-form generator, $R^{a_1a_2a_3}$. At level $m_{11}$ the three form generator acts $m_{11}$ times on the translation generator. Consequently we can express the number of indices, $\#$, of any generator appearing at level $m_{11}$ as
$$\#=3m_{11}-1$$
By the same analysis one can see that number of indices on the generators of the adjoint representation of $E_{11}$ is three times the level [20]. Substituting the expression for $m_{11}$ we have,
$$\#=\sum_{i=1}^{10}(11-i)p_i + 11C$$
The first term gives the index structure of an $SL(11)$ tensor, that is the $p_i$ control the blocks of indices in our generator of length less than eleven, while $C$, controls the number of blocks of eleven antisymmetrised indices. Blocks of eleven indices are proportional to the completely antisymmetric tensor, or volume form $\epsilon$, in eleven dimensions, and correspond to trivial representations of $SL(11)$. We therefore note that an interesting choice is to set $C=0$ and so $k=\sum p_i$, if we do this we exclude from our algebra any generators containing the volume form, $\epsilon$. 

Substituting our expression for $m_{11}$ into the simple root coefficient expressions above we have:
$$\eqalignno{m_j&=\cases{jk-B_j+1,  \qquad & $j\leq 8$  \cr
m_{11}(8-j)+jk-B_j+1,& $j>8$}}$$
To reiterate we have solved the problem of finding which roots occur in the $E_{12}$ lattice at a particular level $m_{11}$ (with $m_*=1$) in terms of an integer $k$ which is bounded above and below. We observe that $m_1\geq1$ since we are considering irreducible representations meaning that roots in the algebra must have connected support on the Dynkin diagram\footnote{$^2$}{That is, the non-zero coefficients of a root in $E_{12}$ are all connected by the links of the Dynkin diagram when laid out on top of the diagram.} (and do not form a sub-algebra) and, by construction $m_*=1$, therefore any root in the $l_1$ representation must have $m_1\geq1$. In our notation above, the variable $k$ is bounded from below in terms of the weight coefficients of the $A_{10}$ representation at each level:
$$k\geq \sum_{i=1}^{10}p_i$$
Consequently we find the generic root $\beta$ corresponding to the $l_1$ representation of $E_{11}$:
$$\beta=e_*+\sum_{n=1}^{10}(k-\sum_{i=n}^{10}p_i)e_n+ke_{11}\eqno{(1.2)}$$
Such that,
$$\beta^2={1\over 9}(8-A^2+2A-22k^2-22k+4Ak)+\sum_{i=1}^{10}p_iB_i$$
The fact that $\beta^2=2,0,-2\ldots$ gives an upper bound on $k$. An interesting class of roots occurs when we consider the lower bound for $k$, i.e. $k=\sum_i{p_i}$, which corresponds to generators having no blocks of eleven antisymmetrised indices, i.e. not including volume forms, $\epsilon$. In this case,
$$\beta^2={1\over 9}[8+\sum_{i=1}^{10}p_i^2(11-i)(i-2)-2\sum_{i=1}^{10}p_i(11-i)+2\sum_{j>i}p_ip_j(11-j)(i-2)]\eqno{(1.3)}$$
We recall that a putative root exists when $\beta^2= 2,0,-2\ldots$ is satisfied by the weight labels $p_i$ for a given level $m_{11}$.  
At low levels one finds the translation generator, $P_a$, a two-form, $Z^{ab}$, and a five-form, $Z^{abcde}$. These are tensors which have the correct index structure to be interpreted as the central charges of the supersymmetry algebra in eleven dimensions. This is quite surprising not least because we have been working solely with bosonic fields but also because the charge algebra is infinite and continues beyond the well-known supersymmetry charges. We list the low-level content derived in this section in table A.1 in the appendix. The results of this computational problem were originally given in [5] and in the notation of this section, and to much higher levels, in [16].
\medskip
{\bf 1.2 The ten-dimensional IIA theory}
\medskip
We now carry out the reduction from the eleven-dimensional theory to the ten dimensional one, which has previously been carried out in [5], but as noted earlier we will use a different notation useful to our later computations. The IIA theory may be obtained from the eleven dimensional supergravity theory by dimensional reduction on a circle [23]. In terms of the algebra we delete a node of the $A_{10}$ lattice that gives an $A_9$ lattice. For the $l_1$ representation this means the deletion of $\alpha_{10}$ in addition to the deletions of $\alpha_*$ and $\alpha_{11}$ given in section 1.1. The procedure is an extension of that carried out in the previous section. We denote,
$$\alpha_{10}=w-\nu_9$$
Where $\nu_i$ are the fundamental weights of the $A_9$ lattice containing the roots $\alpha_1,\alpha_2,\ldots \alpha_9$. So that $w^2={11\over10}$ and 
$$w={1\over 10}(e_1+\ldots e_{10})-e_{11}$$
In this decomposition we have,
$$\eqalign{\alpha_*&=y+{3\over 2}z-{1\over11}w-\nu_1 \cr
\alpha_{11}&=z-\nu_8-{8\over 11}w}$$
Therefore,
$$\beta=y+({3\over2}+m_{11})z+(-{1\over 11}-{8\over11}m_{11}+m_{10})w-\Lambda$$
Where $\Lambda$ is a highest weight in the $A_9$ algebra, specifically we have,
$$\Lambda=\nu_1+m_{11}\nu_8+m_{10}\nu_9-\sum_{i=1}^{9}m_i\alpha_i\equiv \sum_{i=1}^{9}p_i\nu_i$$
Taking the inner product with $\nu_j$ we find,
$$\eqalignno{m_j&=\cases{{1\over 10}(-A+8m_{11}+9m_{10}+1),\qquad & $j=9$ \cr
{j\over 10}(A+2m_{11}+m_{10}-1)-B_j+1, & $j\leq 8$}}$$
Where $A=\sum_{i=1}^9ip_i$ and $B_j\equiv \sum_{i=1}^{j}ip_i+j\sum_{i>j}^{9}p_i$. To find integer coefficients we find solutions parameterised by two variables, $q$ and $k$, and related to the two deleted nodes of the $E_{11}$ part of the Dynkin diagram:
$$\eqalign{m_{11}&={1\over 2}(-A-q+9k+1),  \cr
m_{10}&=q+k, \cr
m_9&=m_{11}+q, \cr
m_j&=kj-B_j+1, \qquad j\leq 8}$$
We note that $k\geq\sum_{i=1}^9p_i$ and $q\geq -k$. The solution corresponds to $SL(10)$ tensors with $2m_{11}+m_{10}-1$ indices, associated to $m_{11}$ adjoint actions of $R^{a_1a_2}$, $m_{10}$ actions of $R^b$ and one action of the translation generator, $P_c$ in the algebra. Denoting the number of indices on a generator appearing in the algebra by $\#$, we have,
$$\eqalignno{\#&\equiv 2m_{11}+m_{10}-1 \cr
&= -A+10k \cr
&= \sum_{i=1}^{9}(10-i)p_i+10C}$$
Where in the last equality we have expressed the lower bound on the variable $k$ by writing $k=\sum_{i=1}^{10}{p_i}+C$ where $C\geq0$ is a constant. We see that $C$ controls the blocks of ten antisymmetrised indices appearing in the algebra, which are related to trivial representations of $SL(10)$. By setting $C=0$ and so $k=\sum_{i=1}^{10}p_i$ we neglect these trivial representations whose generators carry blocks of ten antisymmetrised indices.
However for general $k$ we have,
$$\beta=e_*+\sum_{n=1}^{9}(k-\sum_{i=n}^{10}p_i)e_n+ke_{10}+(m_{11}-m_{10})e_{11}\eqno{(1.4)}$$
And,
$$\beta^2=1+2q^2+2k^2+A(q-k)-6qk-q-k+\sum_{i=1}^9p_iB_i$$
As mentioned, the interesting set of solutions is given by considering $k=\sum_{i=1}^9p_i$ and let us rewrite $q=k+a$ where we have shifted $q$ to simplify the notation and $a\geq -2k$ is an integer. In this case we have,
$$\beta^2=1+a(2a-1)+\sum_{i=1}^{9}p_i^2(i-2)+\sum_{i=1}^{10}p_i(a(i-2)-2)+2\sum_{j>i}p_ip_j(i-2)$$ 
The results of this section are readily used to compute the charge algebra corresponding to the $IIA$ theory and the weights at low levels are shown in table A.2 in the appendix. Amongst the charges we find a scalar, $Z$, the charge of the fundamental string $Z^a$, the NS5 brane charge, $Z^{a_1\ldots a_5}$ and even forms $Z^{a_1a_2}$, $Z^{a_1\ldots a_4}$, $Z^{a_1\ldots a_6}$, $Z^{a_1\ldots a_8}$ corresponding to the Ramond-Ramond brane charges of the D0, D2, D4, D6 and D8 branes of type IIA string theory.

\medskip
{\bf 1.3 The ten-dimensional IIB theory}
\medskip
The charge algebra corresponding to the IIB theory, which has not been presented in the literature before, is obtained by deleting $\alpha_*$, $\alpha_9$ and $\alpha_{10}$ from the Dynkin diagram of $E_{12}$ and finding representations of the $A_9$ algebra whose positive simple roots are $\alpha_1, \alpha_2, \ldots \alpha_8$ and $\alpha_{11}$. In this decomposition we delete,
$$\eqalign{\alpha_*&=y+x-{1\over2}v-\mu_1  \cr
\alpha_{10} &=x-{5\over 2}v \cr
\alpha_9 &=v-\mu_8}$$
Where we denote by $\mu_i$ the fundamental weights of the $A_9$ Dynkin diagram dual  to the roots $\{\alpha_1, \ldots \alpha_8$, $\alpha_{11}\}$. We have introduced the vectors,
$$\eqalign{x&={1\over 2}(e_1+\ldots e_{10})+e_{11}, \qquad x^2=-{1\over2} \cr
v&= {1\over 5}(e_1+\ldots e_9)-{1\over 5}(e_{10})+{4\over 5}e_{11}, \qquad v^2={2\over5}}$$
Now,
$$\beta=y+(1+m_{10})x+(-{1\over2}+m_9-{5\over2}m_{10})z-\Lambda$$
Where,
$$\Lambda=\mu_1+m_9\mu_8-\sum_{i=1}^{8,11}m_i\alpha_i\equiv \sum_{i=1}^{8,11}p_i\mu_i$$
By taking inner products with the fundamental weights of the $A_9$ we find,
$$\eqalignno{m_j&=\cases{{1\over 10}(-A+8m_9+1),\qquad & $j=11$ \cr
{j\over 10}(A-1+2m_9)-B_j+1, & $j\leq 8$}}$$
We parameterise the roots using $m_9={1\over2}(-A+1+10k)$ and find,
$$\eqalignno{m_j&=\cases{m_9-k,\qquad & $j=11$ \cr
jk-B_j+1, & $j\leq 8$}}$$
In this case we have $A=\sum_{i=1}^8ip_i+9p_{11}$ and $B_j=\sum_{i=1}^jip_i+j\sum_{i>j}^{8,11}p_i$. 
The solution gives tensors of $A_{10}$ which have $2m_9-1$ indices, counting the number of actions ($m_9$) of the two form generator $Z^{a_1a_2}$ contracted with the translation generator $P_a$. Let us use $\#$ to denote the number of indices on an arbitrary generator appearing in the algebra, then,
$$\eqalignno{\#&\equiv 2m_9-1\cr
&=-A+10k\cr
&=\sum_{i=1}^{8}(10-i)p_i +p_{11} +10 C}$$
The last term counts blocks of ten antisymmetrised indices, corresponding to trivial representations of $SL(10)$, and by setting $C=0$, and thus $k=\sum_{i=1}^{8,11}p_i$, we disregard these representations.
Note that in this case we have an SL(2) symmetry corresponding to the $\alpha_{10}$ root that was deleted and not directly attached to the $A_9$ gravity line of roots. This SL(2), given by a Dynkin diagram with a single node, has its own weight lattice. The fundamental weight in this $A_1$ diagram is $\nu\equiv {1\over 2}\alpha_{10}$. One can find the components of the vectors orthogonal to the $A_9$ lattice, $x, v$ which are in the $\nu$ direction:
$${1\over x^2}<x,\nu>=-{1\over 2}\qquad {1\over v^2}<v,\nu>=-1$$
Therefore by writing the $A_1$ highest weight as $q\nu$ we find, taking the inner product with $\nu$,
$$m_{10}={1\over 2}(m_9-q)$$
If we parameterise $q=m_9-2l$, we have,
$$m_{10}=l$$
Consequently,
$$\beta=e_*+\sum_{n=1}^{9}(k-\sum_{i=n}^{8,11}p_i)e_n+(l-k)e_{10}+(m_9-k-l)e_{11}\eqno{(1.5)}$$
And,
$$\beta^2=1+2l(l-m_9)+10k^2-2kA+\sum_{i=1}^{8,11}p_iB_i \eqno{(1.6)}$$
We write the root length squared in this form to illustrate the point that if $m_{10}=l$ gives a solution (i.e. root length squared less than or equal to two) then so does $m_{10}=m_9-l$ which is due to the Weyl reflection perpendicular to $\alpha_{10}$.
For the special case $k=\sum_{i=1}^{8,11}p_i$ then,
$$\beta^2=1+l(2l-1)+\sum_{i=1}^{8}p_i^2(10-i)+p_{11}^2-\sum_{i=1}^{8}p_i(10-i)-p_{11}+2\sum_{j>i}^{j=8}p_ip_j(10-j)+2\sum_{i<11}p_ip_{11}$$
The results of this section are used to compute the charge algebra corresponding to the $IIB$ theory to low-levels in table A.3 of the appendix. Amongst the charges we find odd forms $Z^{a\alpha}$, $Z^{a_1\ldots a_3}$, $Z^{a_1\ldots a_5\alpha}$, $Z^{a_1\ldots a_7(\alpha\beta)}$ and $Z^{a_1\ldots a_9(\alpha\beta\gamma)}$ corresponding to the brane charges of the fundamental string and the D1 brane; the D3 brane; the D5 brane and the NS5 brane; D7, D9 branes as well as their S-dual charges of IIB string theory. The Greek indices $\alpha, \beta \ldots$ transform under the $SL(2)$ symmetry. 
\medskip
{\bf 2 Exotic charges}
\medskip
The U-duality transformations of toroidally compactified M-theory are the Weyl reflections of an $E_n$ root lattice with different brane states being represented by weights of $E_n$ [6,7,10,17,18]. A group-theoretic approach to uncovering U-duality $E_n$ multiplets was given in [8,9] and we will recover aspects of this analysis. The weight vector corresponding to the brane solution encodes the tension of the BPS brane states and application of the Weyl reflections of $E_n$ fills out the U-duality brane charge multiplets. Foe example the particle multiplet was discovered by encoding the well-known particle brane solution as a weight vector and then applying the U-duality transformations to it. The new solutions found under the action of all the combinations of the U-duality transformations completed the particle multiplet. In this construction the tension weight vector was introduced as an empirical tool. The brane multiplets containing the particle and string charges were recognised as fundamental representations of the $E_n$ algebra. For example upon dimensional reduction to three dimensions (n=8) the particle multiplet has highest weight $\lambda_1$ ({\bf 248}) and the string multiplet $\lambda_7$ ({\bf 3875}), where $\lambda_i$ are the fundamental weights of $E_n$. This work is discussed in detail in the original papers [6,7,10,17,18], but especially in the review [9]. 
 
In addition to the expected brane charges, coming from the dimensional reduction of charges associated to brane solutions in eleven dimensional supergravity, many exotic charges were also present in the U-duality charge multiplets whose higher dimensional origin was unknown. Exotic charges, in other words, are not derived from the dimensional reduction of the central charges appearing in the decomposition of the antiicommutator of the eleven dimensional supercharges [25],
$$\{Q_\alpha,\bar{Q}_\beta\}=\Gamma_{\alpha\beta}P_M+{1\over 2}\Gamma^{MN}_{\alpha\beta}Z_{MN}+{1\over 2}\Gamma^{MNPQR}_{\alpha\beta}Z_{MNPQR}$$
For example by dimensionally reducing $P_M,Z_{MN},Z_{MNPQR}$ to five dimensions one recovers the ${\bf 6\oplus 15\oplus 6}$ ${\bf=27}$ of SU(6) by assigning all the central charge indices to be internal indices. This agrees with the U-duality particle multiplet in five dimensions found in [6,7,10,17,18]. However the particle multiplet in four dimensions derived from the central charges of the eleven dimensional supersymmetry algebra reproduces only the ${\bf 7\oplus 21\oplus 21=49}$ of the degrees of freedom of the ${\bf 56}$ of SU(6), which is the representation of the particle multiplet in four dimensions. The extra seven charges found in the particle multiplet in four dimensions are derived from the dimensional reduction of charges other than the usual central charges of the eleven dimensional superalgebra and are exotic charges. These extra states in $D=4$ were associated to the compactification of the KK6 monopole, having charge $Z_{MNPQRST,R}$ which does not appear in the supersymmetry algebra, wrapping the $T^7$ in the compactification. In $D=3$ many more exotic states were uncovered in the charge multiplets whose eleven dimensional origin was unclear.

The conjecture that the $l_1$ representation of $E_{11}$ contains the full set of brane charges of M-theory gave an eleven-dimensional origin to these observations [5]. The Weyl group of the $E_n$ sub-algebras, appearing upon dimensional reduction, being implicitly included from the outset. Furthermore the inclusion of the charge associated to the dual to gravity field in the eleven-dimensional algebra, the field giving rise to the KK6 monopole, from the outset explains the appearance of the exotic states appearing upon reduction to $D=4$. The $l_1$ representation of $E_{11}$ contains an infinite set of charges in addition to the central charges  of the supergravity algebra and the charge of the KK6 monopole. It is these extra charges which give an eleven dimensional origin to the exotic charges appearing in the U-duality brane charge multiplets. These exotic charges are all associated to KK-branes, which we will consider further in section 3. 

In this paper we will derive essentially the same tension formula that played a crucial role in uncovering the U-duality multiplets in [8,9] but the expression given here will have a simple origin in the $E_{11}$ conjecture. Compared to the original work uncovering the U-duality multiplets we will work in a reverse sense. First we will derive the brane charge multiplets from $E_{11}$ and then second we will apply a tension formula to the results. Our aim will be to demonstrate the validity of the tension formula, but to do so we will first decompose the $l_1$ representation of $E_{11}$ to various dimensions ($3\leq D\leq 8$) and identify the brane charge multiplets. We note that the reduction to three dimensions of the $l_1$ representation of $E_{11}$ has previously been carried out in [11] and the particle multiplet has been derived from $l_1$ representation in dimensions, D, where $3\leq D\leq 8$ in [12] in all cases the results were in perfect agreement with the U-duality multiplet of charges.

In this section we extend the work of [11] and outline the decomposition of the $l_1$ algebra to arbitrary dimensions, $D<11$, which is the algebraic equivalent of compactification on a (11-D)-torus and then apply the tension formula, which will be derived in section 3, to the charges that are found. 
\medskip
{\bf 2.1 General decomposition}
\medskip
Let us give the decomposition of the $l_1$ representation of $E_{11}$ in terms of its $A_{D-1}$ and $E_{11-D}$ sub-algebras. The details of this decomposition can be found in [11].

To find the $l_1$ representation of $A_{D-1}\otimes E_{11-D}$ we delete the node labelled D, in figure 2.1, to obtain,
\medskip
\halign{ \centerline{ #} \cr
\epsfbox{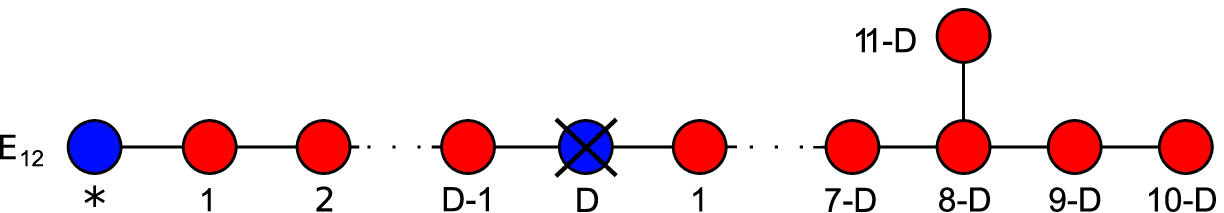} \cr
Figure 2.1 The decomposition of $E_{12}$ to $A_{D-1}\otimes E_{11-D}$ \cr}
\medskip
For example, to find representations of $A_4\otimes E_6$ we delete the fifth node. The two deleted roots may be expressed in terms of vectors in the root space of $E_{11}$ with components orthogonal to the vectors in the remaining root space after deletion of nodes $*$ and $D$. We carry out the decomposition to $A_{D-1}\otimes E_{11-D}$ by expressing the D'th root, $\alpha_D$, in the form,
$$\alpha_D = -\nu + x, \qquad \nu = -\sum_{i=1, i\neq D}^{i=11} \lambda_{i}(A_{(E_{11})})_{iD}$$
Where $(A_{(E_{11})})_{iD}$ is the $i,D$'th component in the Cartan matrix of $E_{11}$, $x$ is a vector in the root lattice orthogonal to all the remaining roots after the deletion of $\alpha_{D}$, and now $\lambda_i$ for $i=1\ldots D-1$ are the fundamental weights of $A_{D-1}$, and when $i=(D+1) \ldots 11$ they are fundamental weights of $E_{11-D}$. Denoting the fundamental weights of the decomposed $E_{11}$ by $l_i$ we have,
$$l_D={1\over x^2}x$$
$$l_i=\lambda_i^{(r)}-{<\nu,\lambda_i^{(r)}> \over x^2}x$$
Where $r$ is either 1, referring to the $A_{D-1}$ sub-algebra, or 2, referring to the $E_{11-D}$ sub-algebra. In this notation, 
$$\alpha_*=-\lambda_1^{(1)}-{<\lambda^{(1)}_1,\lambda^{(1)}_{D-1}>\over x^2}x+y$$
We now return to our consideration of $E_{12}$, a general root of which is given by, 
$$\beta = m_*\alpha_* + m_D\alpha_D + \sum_{i=1,i\neq D}^{11} m_i\alpha_i$$
For a positive root, $m_*$, $m_D$ and $m_i$ are positive integers. Substituting our expressions for $\alpha_*$ and $\alpha_D$, we obtain,
$$\beta = m_*y+(m_D-m_*{<\lambda_1^{(1)},\lambda_{D-1}^{(1)}> \over x^2})x-\sum_{(r=1,2)} \Lambda^{(r)}$$
Where,
$$\Lambda^{(r)}=-\sum_{i=1,i\neq D}^{11} m_i^{(r)}\alpha_i^{(r)}+m_D\nu^{(r)}+m_*\lambda_1^{(r)}\delta_{(r,1)}$$
Adopting the notation $m_i^{(1)}=m_i$, $m_i^{(2)}=n_i$, $\lambda_i^{(1)}=\mu_i$ and $\lambda_i^{(2)}=\lambda_i$, we have,
$$\Lambda^{(1)}=-\sum_{i=1}^{D-1} m_i\alpha_i^{(1)}+m_D\mu_{D-1}+m_*\mu_1\equiv \sum_i^{D-1}q_i\mu_i$$
$$\Lambda^{(2)}=-\sum_{i=1}^{11-D} n_i\alpha_i^{(2)}+m_D\lambda_1\equiv \sum_i^{11-D}p_i\lambda_i$$
Taking the inner product with $\mu_j$ and $\lambda_j$ respectively we obtain expressions for the root coefficients, $n_i$ and $m_i$,
$$\eqalign{A_{D-1}:\quad &-m_j=\sum_i q_i<\mu_i,\mu_j>-m_D<\mu_{D-1},\mu_j>-m_*<\mu_1,\mu_j> \cr
E_{11-D}: &-n_j=\sum_i p_i<\lambda_i,\lambda_j>-m_D<\lambda_1,\lambda_j> } \eqno{(2.1)}$$
We make use of formulae for the inner products of the fundamental weights of $E_{11-D}$ derived in appendix A of reference [11],
$$\eqalign{\lambda_i=\cases{\hat{\mu_i}+{3i \over D-2}z, \qquad \qquad & $i=1,\ldots , 8-D$ \cr
\hat{\mu}_i+{(8-D)(11-D-i)\over D-2}z, &$i=9-D,10-D$ \cr
{(11-D)\over (D-2)}z, &$i=11-D$}}\eqno{(2.2)}$$
These weights are derived by deleting the n'th node with respect to an $E_n$ diagram, $z$ is the vector in the root space corresponding to the linear independence of the n'th node and $\hat{\mu}_i$ are the weights of the $A_{n-1}$ subalgebra, where $n=11-D$. We note that $z^2={D-2\over 11-D}$ and,
$$\lambda_1^2=<\hat{\mu}_1,\hat{\mu}_1>+{9\over (D-2)(11-D)}={D-1\over D-2}$$
Consequently,
$$\alpha_D^2=x^2+{\mu_{D-1}^2}+{\lambda_1^2}=x^2+{D-1\over D}+{D-1\over D-2}$$
Normalising $\alpha_D^2=2$ gives,
$$x^2={-2\over D(D-2)}$$
We note that the vector $x$ is given explicitly in our basis by:
$$x={1\over D}(e_1+\ldots e_D)+{1\over D-2}(e_{D+1}+\ldots e_{11})$$
\medskip
{\bf 2.2 Rank $p$ charges}
\medskip
We commence by looking for p-brane charges, that is rank p charges with all indices in the spacetime algebra, $Z^{a_1\ldots a_p}$.  This corresponds to a representation of the $A_{D-1}$ sub-algebra with highest weight $\mu_{D-p}$, where we recall that $\mu_i$ are the fundamental weights of $A_{D-1}$.  The condition we must satisfy is that $$\sum_i q_i\mu_i =\mu_{D-p}$$ A $\mu_{D-p}$ weight in the $A_{D-1}$ sub-algebra of our decomposition leads to constraints upon the values to be taken by the root coefficient $m_D$. 

We are interested in the decomposition of the $l_1$ representation of $E_{11}$ and as such we take $m_*=1$ and from the $A_{D-1}$ equation in (2.1) we obtain,
$$\eqalign{-m_j&=<\mu_{(D-p)},\mu_j>-m_D<\mu_{(D-1)},\mu_j>-<\mu_1,\mu_j>\cr
&=\cases{(D-p-1)-{j\over D}(D-p-1+m_D),\qquad  &$j \geq (D-p)$\cr
-1+{j\over D}(p+1-m_D), &$j \leq (D-p)$}}\eqno{(2.3)}$$
Since $-m_j$ must be integer valued and negative we find a simple set of solutions for $m_D$ having the form,
$$m_D=p+1+kD\eqno{(2.4)}$$
Where $k$ is a constant bounded from below because $m_1\geq 1$, implying that $k\geq 0$. The solution corresponds in the algebra to $m_D$ adjoint actions of a generator $R^a$ with the translation generator $P_c$. The resulting generator in the algebra has $m_D-1$ contravariant indices. Let us denote the indices by $\#$, then,
$$\eqalignno{\#&\equiv m_D-1 \cr
&=p+kD}$$
The variable $k$ corresponds to blocks of $D$ antisymmetrised indices in the $SL(D)$ algebra, and indicates the occurrence of trivial representations of $SL(D)$. By setting $k=0$ we may neglect these representations and simplify the algebra.

Having found a criterion for $m_D$ corresponding to a p-brane charge in the $A_{D-1}$ sub-algebra, we now turn our attention to restrictions on specific weights of the $E_{11-D}$ sub-algebra consistent with the values of $m_D$ corresponding to a rank p charge. We commence by finding conditions for representations of single fundamental weights of $E_{11-D}$, $\lambda_i$, for which we set $\sum_i p_i\lambda_i =\lambda_i$ in (2.1) and making use of equation (2.2) we find,
$$\eqalign{-n_j&=<\lambda_i,\lambda_j>-m_D<\lambda_1,\lambda_j> \cr
&=\cases{i\leq 8-D
\cases{
i-m_D+{j \over D-2}(i-m_D), \qquad &$j\leq 8-D, i\leq j$ \cr
j-m_D+{j\over D-2}(i-m_D), &$j\leq 8-D, i\geq j$ \cr
{2(11-D-j)\over D-2}(i-m_D), &$j=9-D, 10-D$ \cr
{3\over D-2}(i-m_D), &$j=11-D$} 
\cr
i=9-D,10-D
\cases{
-m_D+ {j\over D-2}(2(11-D-i)-m_D), \qquad &$j\leq 8-D$ \cr
{11-D-j\over D-2}((8-D)^2-i(6-D)-2m_D), &$j=9-D,10-D, i\leq j$ \cr
{1 \over D-2}(4(11-D-i)-2m_D), &$j=9-D,10-D, i\geq j$ \cr
{1\over D-2}((8-D)(11-D-i)-3m_D), &$j=11-D$ }
\cr
i=11-D 
\cases{
-m_D+{j\over D-2}(3-m_D), \qquad &$j\leq 8-D$ \cr
{11-D-j\over D-2}(8-D-2m_D), &$j=9-D,10-D$ \cr
{1\over D-2}(11-D-3m_D), &$j=11-D$}
}}\eqno{(2.5)}$$
The simplest case with a solution is dependent upon the choice of fundamental weight $\lambda_i$ and is
$$\eqalign{m_D=\cases{
i+l(D-2), \qquad  &$i\leq 8-D$ \cr
2(11-D-i)+l(D-2), \qquad &$i=9-D,10-D$ \cr
3+l(D-2), \qquad &$i=11-D$ 
}}\eqno{(2.6)}$$
Where $l$ is a positive integer or zero, which indicates trivial representations of $E_{11-D}$.
Substituting the solution for $m_D$ into the expression for the simple root coefficients, $n_j$, given above we have,
$$\eqalignno{-n_j&=<\lambda_i,\lambda_j>-m_D<\lambda_1,\lambda_j> \cr
&=\cases{i\leq 8-D
\cases{
-l(D-2)-lj, \qquad &$j\leq 8-D, i\leq j$ \cr
j-i-l(D-2)-jl &$j\leq 8-D, i\geq j$ \cr
-2l(11-D-j), &$j=9-D, 10-D$ \cr
-3l, &$j=11-D$} 
\cr
i=9-D,10-D
\cases{
-2(11-D-i)-l(D-2)-lj, \qquad &$j\leq 8-D$ \cr
(11-D-j)((D-10)+(2l+i)), &$j=9-D,10-D, i\leq j$ \cr
-2l, &$j=9-D,10-D, i\geq j$ \cr
(-11+D+i-3l), &$j=11-D$ } 
\cr
i=11-D 
\cases{
-m_D-jl, \qquad &$j\leq 8-D$ \cr
(11-D-j)(-1-2l), &$j=9-D,10-D$ \cr
-1-3l, &$j=11-D$}
}}$$
The dependence on the parameter $l$ is made manifest and we see that $l\geq0$. Notice that when $E_{11-D}$ is decomposed into representations of $SL(11-D)$ by deleting node $(11-D)$, the parameter $l$ indicates blocks of $11-D$ antisymmetrised indices. Explicitly, deletion of node $(11-D)$ together with the deletion of node $D$ leads to tensors with $3n_{11-D}-m_D$ which we denote $\#$ so that,
$$\#=\cases{-i+l(11-D), \qquad & $i\leq 8-D$ \cr
(11-D-i)+l(11-D)\qquad & $i=9-D,10-D$ \cr
l(11-D)\qquad & $i=11-D$ }$$
This allows us to see that $l$ controls the appearance of blocks of $11-D$ antisymmetrised indices. By setting $l=0$ we disregard the trivial representations of $E_{11-D}$. We note the subtlety that a trivial representation in the $E_{11-D}$ algebra, after the action of the local sub-group, may give rise to non-trivial representations in the spacetime $SL(D)$ algebra and a non-trivial representation of $E_{11-D}$, we will discuss this possibility in section 2.6.

We choose values of $m_D$ which give a rank p charge in the $A_{D-1}$ sub-algebra and then read off the value of the corresponding fundamental weight in the $E_{11-D}$ sub-algebra. This amounts to equating our two conditions for $m_D$, equations (2.4) and (2.6), 
$$\eqalign{p=
\cases{
i+l(D-2)-kD-1, \qquad & $i\leq 8-D$ \cr
2(11-D-i)+l(D-2)-kD-1, \qquad & $i=9-D,10-D$\cr
2+l(D-2)-kD, & $i=11-D$}}$$
In particular for $l=k=0$ we find the charge content indicated in table 2.1, where the charges are indicated by $SL(D)\otimes SL(11-D)$ tensors. We use the index $a_i$ to indicate an index in the spacetime associated to the weight of $A_{D-1}$ being considered, and the index $j_i$ to indicate internal coordinates coming from the representation of $E_{11-D}$ that the charge transforms under. The fundamental weights of $E_{11-D}$ belong to a given representation of $SL(11-D)$, so we may label them by the highest weight of their $SL(11-D)$ representation. For example, if we consider the representation of $E_{11-D}$ whose highest weight is $\lambda_{11-D}$ it belongs to the representation of $SL_{11-D}$ with highest weight $Z^{(9-D)(10-D)(11-D)}$. 
In table 2.1 it is the highest weight of $SL(11-D)$ that we have used to indicate the $E_{11-D}$ representation.

For representations of $E_{11-D}$ the internal indices in the decomposition to $SL(D)\otimes SL(11-D)$ tensors will vary throughout the multiplet, but the spacetime indices will remain unaltered. Let us consider the example of the particle charge multiplet in $D=3$ which is the first fundamental representation of $E_8$ with highest weight, $\lambda_1$. The $\bf 248$ has $SL(D)\otimes SL(11-D)$ tensors:
$$P_j (8), \quad Z^{j_1j_2} (28), \quad Z^{j_1\ldots j_5} (56), \quad Z^{j_1\ldots j_7,k} (63),\quad Z^{j_1\ldots j_8} (1), \quad Z^{j_1\ldots j_3} (56), \quad Z^{j_1\ldots j_6} (28), \quad Z^{j} (1)$$  
It is convenient to indicate the tensor charge associated to just the highest weight in the representation, it is this that is indicated by the charges in table 2.1. 
From table 2.1 we read that such a charge $Z^{a_1j_1}$ is associated to the (10-D)'th fundamental representation (with highest weight $\lambda_{10-D}$) of $E_{11-D}$, and so on for the other charge multiplets.
$$\halign{\centerline{#} \cr
\vbox{\offinterlineskip
\halign{\strut \vrule \quad \hfil # \hfil\quad &\vrule \quad \hfil # \hfil\quad 
&\vrule \quad \hfil # \hfil\quad  &\vrule #
\cr
\noalign{\hrule}
Index of $\lambda_i$& Weight of $E_{11-D}\otimes A_{D-1}$ & Highest Weight Charge &\cr
\noalign{\hrule}
$i\leq 8-D$&$\lambda_i\otimes\mu_{D-i}$&$Z^{a_1\ldots a_{i-1}j_1\ldots j_{11-D-i}}$ &\cr
\noalign{\hrule}
$i=9-D$&$\lambda_{9-D} \otimes\mu_{D-3}$&$Z^{a_1a_2a_3j_1j_2}$&\cr
\noalign{\hrule}
$i=10-D$&$\lambda_{10-D}\otimes\mu_{D-1}$&$Z^{a_1j_1}$&\cr
\noalign{\hrule}
$i=11-D$&$\lambda_{11-D}\otimes\mu_{D-2}$&$Z^{a_1a_2j_1j_2j_3}$&\cr
\noalign{\hrule}
}
}\cr
Table 2.1 Charges associated to fundamental weights of $E_{11-D}$ upon dimensional reduction \cr indicated by $SL(D)\otimes SL(11-D)$ tensors\cr}$$
It may be useful to associate the various charge multiplets associated to a fundamental weight of $E_{11=D}$ with their nodes on the $E_{11-D}$ Dynkin diagram,
\medskip
\centerline{\epsfbox{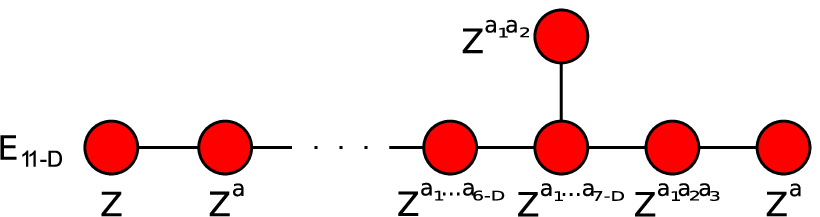}}
\medskip
For the case of $D=3$ [11], the charges associated to the fundamental weights of nodes $11-D, 10-D$ and $1$ of $E_{11-D}$ are the highest weights of the membrane, string and particle multiplets [9]. 
The set of charges up to rank $D-1$ where $D$ ranges from three to eight derived from the $l_1$ multiplet is shown in table 2.2. Many of the charge multiplets indicated therein are associated to representations of $E_{11-D}$ whose highest weight is a sum of fundamental weights. We deal with these cases in section 2.6, where we extract the full content of table 2.2 from the $l_1$ algebra.
$$\halign{\centerline{#} \cr
\vbox{\offinterlineskip
\halign{\strut \vrule \quad \hfil # \hfil\quad &\vrule  \quad \hfil # \hfil\quad &\vrule \hfil # \hfil 
&\vrule \hfil # \hfil  &\vrule \hfil # \hfil &\vrule \hfil # \hfil &\vrule \hfil # \hfil &\vrule \hfil # \hfil &\vrule \hfil # \hfil &\vrule \hfil # \hfil &\vrule#
\cr
\noalign{\hrule}
D&G&$Z$&$Z^{a}$&$Z^{a_1a_2}$&$Z^{a_1\ldots a_{3}}$&$Z^{a_1\ldots a_{4}}$&$Z^{a_1\ldots a_{5}}$&$Z^{a_1\ldots a_6}$&$Z^{a_1\ldots a_7}$&\cr 
\noalign{\hrule}
8&$SU(3)\otimes SU(2)$&$\bf (3,2)$&$\bf (3,1)$&$\bf (1,2)$&$\bf (3,1)$&$\bf (3,2)$&$\bf (1,3)$&$\bf (3,2)$&$\bf (6,1)$&\cr 
&&&&&&&$\bf (8,1)$&$\bf (6,2)$&$\bf (18,1)$&\cr 
&&&&&&&$\bf (1,1)$&&$\bf (3,1)$&\cr 
&&&&&&&&&$\bf (6,1)$&\cr 
&&&&&&&&&$\bf (3,3)$&\cr 
\noalign{\hrule}
7&$SU(5)$&$\bf 10$&$\bf 5$&$\bf 5$&$\bf 10$&$\bf 24$&$\bf 40$&$\bf 70$&-&\cr 
&&&&&&$\bf 1$&$\bf 15$&$\bf 50$&-&\cr 
&&&&&&&$\bf 10$&$\bf 45$&-&\cr 
&&&&&&&&$\bf 5$&-&\cr 
\noalign{\hrule}
6&$SO(5,5)$&$\bf 16$&$\bf 10$&$\bf 16$&$\bf 45$&$\bf 144$&$\bf 320$&-&-&\cr
&&&&&$\bf 1$&$\bf 16$&$\bf 126$&-&-&\cr 
&&&&&&&$\bf 120$&-&-&\cr 
\noalign{\hrule}
5&$E_6$&$\bf 27$&$\bf 27$&$\bf 78$&$\bf 351$&$\bf 1728$&-&-&-&\cr 
&&&&$\bf 1$&$\bf 27$&$\bf 351$&-&-&-&\cr 
&&&&&&$\bf 27$&-&-&-&\cr 
\noalign{\hrule}
4&$E_7$&$\bf 56$&$\bf 133$&$\bf 912$&$\bf 8645$&-&-&-&-&\cr 
&&&$\bf 1$&$\bf 56$&$\bf 1539$&-&-&-&-&\cr 
&&&&&$\bf 133$&-&-&-&-&\cr 
&&&&&$\bf 1$&-&-&-&-&\cr 
\noalign{\hrule}
3&$E_8$&$\bf 248$&$\bf 3875$&$\bf 147250$&-&-&-&-&-&\cr
&&$\bf1$&$\bf248$&$\bf 30380$&-&-&-&-&-&\cr
&&&$\bf 1$&$\bf 3875$&-&-&-&-&-&\cr
&&&&$\bf 248$&-&-&-&-&-&\cr
&&&&$\bf 1$&-&-&-&-&-&\cr
\noalign{\hrule}
}}\cr
Table 2.2 Charge multiplet representations of the group, G, from the $l_1$ representation of $E_{11}$ in $3\leq D \leq 8$\cr}$$
\medskip
{\bf 2.3 The Particle Multiplet}
\medskip
We identified the highest weight of this multiplet above,  and we now discuss the identification of the weights of the $l_1$ representation associated to particle, or zero brane, charges in detail. The particle charge multiplet will consist of objects having no spacetime indices $(a_i)$, and a set of internal indices $(j_i)$. From table 2.1 we see that this charge corresponds to the representation with highest weight $(\lambda_1\otimes \mu_{D-1})$, having a highest weight charge indicated by an $SL(11-D)$ tensor, $Z^{j_1\ldots j_{10-D}}$, i.e. it has only internal indices and it transforms in the first fundamental representation of the $E_{11-D}$ sub-group. 

To find the roots explicitly we recall that this is the representation whose highest weight is the first fundamental weight of $E_{11-D}$, so that in $D=3$ it is the ${\bf 248}$ of $E_8$, in $D=4$ it is the ${\bf 56}$ of $E_7$ and so on. In section 2.2 the condition on the highest weight representation of $SL(D)$ has been written in terms of $p$, where the charge multiplet's highest weight is the charge associated to a p-brane. From equation (2.3), with $p=0$ for the particle charge multiplet we find that,
$$m_j=1 \qquad j=1,\ldots D-1$$
From equation 2.4 we have $m_D=1$. Since we are considering the $l_1$ representation $m_*=1$ then the particle multiplet contains roots with root coefficients $(1^{D+1},m_{D+1},\ldots m_{11})$. We use the superscript notation as a shorthand to indicate that a number of sequential roots share the same coefficient, e.g. 
$$(1^{D+1},m_{D+1},\ldots m_{11})\equiv \alpha_*+\alpha_1+\ldots+\alpha_D+\sum_{i=D+1}^{11}m_i\alpha_i$$
The simple root coefficients $\{m_{i>D}\}$ correspond to the weights of the first fundamental representation of $E_{11-D}$ which we will identify. For the reduction to $D$ dimensions we should find the particle multiplet being made up of roots whose first $D+1$ root coefficients are $1$'s. The most complicated case is that of the reduction to $D=3$ which has been studied already in [11], and the $l_1$ representations have been shown to form the $\bf{248}$ of $E_8$. In addition to this condition we must identify the root vectors of the fundamental representation of $E_8$ amongst these roots.  By finding all roots of the form $(1^{D+1},m_{D+1},\ldots m_{11})$ we identify the $\bf{248\oplus 1}$ indicated in table 2.2. To distinguish the roots of $\bf{248}$ we must identify the root vectors of the fundamental $\bf{248}$ amongst the candidate roots. In more complicated cases to follow this process will be less trivial than in the present example. From the $l_1$ representation of $E_{11}$ we can read off the charges in the particle multiplet and these are listed in table 2.3. In section 3 of this paper we will give a formula, equation (3.3), relating a root in the $E_{12}$ lattice with a mass. In table 2.3 we give the result of applying this formula to the roots in the table, the resulting masses exactly match those found by U-duality [8,9].
$$\halign{\centerline{#} \cr
\vbox{\offinterlineskip
\halign{\strut \vrule \quad \hfil # \hfil\quad &\vrule \quad \hfil # \hfil\quad 
&\vrule \quad \hfil # \hfil\quad  &\vrule \quad \hfil # \hfil\quad &\vrule #
\cr
\noalign{\hrule}
$E_{12}$ root&Charge&Dimension of $SL(8)$ tensor&Mass (highest weight)&\cr 
\noalign{\hrule}
$(1^5,0^7)$&$P_j$&$8$&${1\over R_4}$&\cr
\noalign{\hrule}
$(1^9,0^2,1)$&$Z^{j_1j_2}$&$28$&${R_{10}R_{11}\over l_p^3}$&\cr
\noalign{\hrule}
$(1^7,2,3,2,1,2)$&$Z^{j_1\ldots j_5}$&$56$&${R_7\ldots R_{11}\over l_p^6}$&\cr
\noalign{\hrule}
$(1^5,2,3,4,5,6,4,2,3)$&$Z^{j_1\ldots j_8}$&$1$&${V\over l_p^9}$&\cr
$(1^5,2,3,4,5,3,1,3)$&$Z^{j_1\ldots j_7,k}$&$63$&${VR_{11}\over R_4l_p^9}$&\cr
\noalign{\hrule}
$(1^4,2,3,4,5,6,4,2,4)$&$Z^{j_1\ldots j_8,k_1\ldots k_3}$&$56$&${VR_9\ldots R_{11}\over l_p^{12}}$&\cr
\noalign{\hrule}
$(1^4,2,3,5,7,9,6,3,5)$&$Z^{j_1\ldots j_8,k_1\ldots k_6}$&$28$&${VR_6\ldots R_{11}\over l_p^{15}}$&\cr
\noalign{\hrule}
$(1^4,3,5,7,9,11,7,3,6)$&$Z^{j_1\ldots j_8,k}$&$8$&${V^2R_{11}\over l_p^{18}}$&\cr
\noalign{\hrule}
}}\cr
Table 2.3 The particle charge multiplet in $D=3$/The {\bf 248} of $E_8$ \cr}$$
In table 2.3 we have used $V$ to indicate $R_{D+1}R_{D+2}\ldots R_{11}$ indicating the volume of the internal space. From table 2.3 we can identify generators which commute  with each other to form generators proportional to multiple copies of the volume form($\epsilon^{\mu_1\ldots \mu_{11-D}}$). We call such pairs of generators dual. The Young tableaux of a generator and its dual may be combined to form a rectangular tableau whose height is the dimension of the internal space. Such a Young tableau is proportional to multiple volume tensors. For example, in table 2.3 the generators at levels $0,1,2$ are dual to those at $4,5,6$, and those at level $3$ are self-dual, that is
$$[P_j,Z^{l_1\ldots l_8,k_1\ldots k_8,j}]\propto \epsilon^2,\qquad [Z^{l_1l_2},Z^{j_1\ldots j_8,k_1\ldots k_6}]\propto \epsilon^2,\qquad [Z^{j_1\ldots j_5}, Z^{l_1\ldots l_8,k_1\ldots k_3}]\propto \epsilon^2$$
$$[Z^{j_1\ldots j_8},Z^{k_1\ldots k_8}]\propto \epsilon^2,\qquad [Z^{j_1\ldots j_7,k},Z^{l_1\ldots l_7,m}]\propto \epsilon^2$$
We are identifying a bilinear Casimir for the representation. By using the mass formula (to be derived in section 3) one can see that the mass associated to a generator, ${\cal M}$, and the mass associated to its dual generator, ${\cal M'}$, multiply to give an invariant squared mass, ${\cal MM'=M}^2$, for the representation. For the $\bf 248$ shown in table 2.3 the invariant mass squared is:
$${\cal M}^2_{\bf 248}\sim {V^2\over l_p^{18}}$$
Let us look at the reduction to $D=4$ where the particle multiplet should belong to representations of $E_7$. From tables of $E_{12}$ roots [16], we find the $l_1$ content listed in table 2.4.
$$\halign{\centerline{#} \cr
\vbox{\offinterlineskip
\halign{\strut \vrule \quad \hfil # \hfil\quad &\vrule \quad \hfil # \hfil\quad 
&\vrule \quad \hfil # \hfil\quad  &\vrule \quad \hfil # \hfil\quad &\vrule #
\cr
\noalign{\hrule}
$E_{12}$ root&Charge&Dimension of $SL(7)$ tensor&Mass&\cr 
\noalign{\hrule}
$(1^5,0^7)$&$P_j$&$7$&${1\over R_5}$&\cr
\noalign{\hrule}
$(1^9,0,0,1)$&$Z^{j_1j_2}$&$21$ & ${R_{10}R_{11}\over l_p^3}$&\cr
\noalign{\hrule}
$(1^7,2,3,2,1,2)$&$Z^{j_1\ldots j_5}$&$21$ &${R_7\ldots R_{11}\over l_p^6}$&\cr
\noalign{\hrule}
$(1^5,2,3,4,5,3,1,3)$&$Z^{j_1\ldots j_7,k}$&$7$ &${VR_{11}\over l_p^9}$&\cr
\noalign{\hrule}
}}\cr
Table 2.4 The particle charge multiplet in $D=4$/The {\bf 56} of $E_7$ \cr}$$
By identifying the dual generators in table 2.4 one can associate an invariant mass squared to the multiplet,
${\cal M}^2_{\bf 56}\sim {V\over l_p^9}$.
In the reduction to $D=5$ we look for the $\bf{27}$ of $E_6$, and the corresponding roots are in table 2.5.
$$\halign{\centerline{#} \cr
\vbox{\offinterlineskip
\halign{\strut \vrule \quad \hfil # \hfil\quad &\vrule \quad \hfil # \hfil\quad 
&\vrule \quad \hfil # \hfil\quad  &\vrule \quad \hfil # \hfil\quad &\vrule #
\cr
\noalign{\hrule}
$E_{12}$ root&Charge&Dimension of $SL(6)$ tensor&Mass&\cr 
\noalign{\hrule}
$(1^6,0^6)$&$P_j$&$6$&${1\over R_6}$ &\cr
\noalign{\hrule}
$(1^9,0,0,1)$&$Z^{j_1j_2}$&$15$&${R_{10}R_{11}\over l_p^3}$&\cr
\noalign{\hrule}
$(1^7,2,3,2,1,2)$&$Z^{j_1\ldots j_5}$&$6$&${R_7\ldots R_{11}\over l_p^6}$ &\cr
\noalign{\hrule}
}}\cr
Table 2.5 The particle charge multiplet in $D=5$/The {\bf 27} of $E_6$\cr}$$
For the reduction to $D=6$ we look for the $\bf{16}$ of $SO(5,5)$. From the $l_1$ we find the states of table 2.6.
$$\halign{\centerline{#} \cr
\vbox{\offinterlineskip
\halign{\strut \vrule \quad \hfil # \hfil\quad &\vrule \quad \hfil # \hfil\quad 
&\vrule \quad \hfil # \hfil\quad  &\vrule \quad \hfil # \hfil\quad &\vrule #
\cr
\noalign{\hrule}
$E_{12}$ root&Charge&Dimension of $SL(5)$ tensor&Mass&\cr 
\noalign{\hrule}
$(1^7,0^5)$&$P_j$&$5$&${1\over R_7}$&\cr
\noalign{\hrule}
$(1^9,0,0,1)$&$Z^{j_1j_2}$&$10$&${R_{10}R_{11}\over l_p^3}$&\cr
\noalign{\hrule}
$(1^7,2,3,2,1,2)$&$Z$&$1$&${R_7\ldots R_{11}\over l_p^6}$ &\cr
\noalign{\hrule}
}}\cr
Table 2.6 The particle charge multiplet in $D=6$/The {\bf 16} of $SO(5,5)$ \cr}$$
We repeat the process for $D=7$ where we look for the $\bf{10}$ of $SU(5)$, the appropriate roots are listed in table 2.7.
$$\halign{\centerline{#} \cr
\vbox{\offinterlineskip
\halign{\strut \vrule \quad \hfil # \hfil\quad &\vrule \quad \hfil # \hfil\quad 
&\vrule \quad \hfil # \hfil\quad  &\vrule \quad \hfil # \hfil\quad &\vrule #
\cr
\noalign{\hrule}
$E_{12}$ root&Charge&Dimension of $SL(4)$ tensor&Mass&\cr 
\noalign{\hrule}
$(1^8,0^3)$&$P_j$&$4$&${1\over R_8}$&\cr
\noalign{\hrule}
$(1^9,0,0,1)$&$Z^{j_1j_2}$&$6$&${R_{10}R_{11}\over l_p^3}$&\cr
\noalign{\hrule}
}
}\cr
Table 2.7 The particle charge multiplet in $D=7$/The {\bf 10} of $SL(5)$ \cr}
$$
The formulae used in this section are even robust up to $D=8$, where we identify the ${\bf (3,2)}$ of $SU(3)\otimes SU(2)$ contained in the $l_1$ representation from the roots $(1^9,0,0,1)$ and $(1^9,0^3)$, associated to generators $Z^{j_1j_2}$ and $P_j$, both having dimension three. $$\halign{\centerline{#} \cr
\vbox{\offinterlineskip
\halign{\strut \vrule \quad \hfil # \hfil\quad &\vrule \quad \hfil # \hfil\quad 
&\vrule \quad \hfil # \hfil\quad  &\vrule \quad \hfil # \hfil\quad &\vrule \quad \hfil # \hfil\quad &\vrule\quad \hfil # \hfil\quad &\vrule\quad \hfil # \hfil\quad &\vrule\quad \hfil # \hfil\quad &\vrule\quad \hfil # \hfil\quad &\vrule#
\cr
\noalign{\hrule}
D&$Z$&$Z^j$&$Z^{j_1j_2}$&$Z^{j_1\ldots j_3}$&$Z^{j_1\ldots j_4}$&$Z^{j_1\ldots j_5}$&$Z^{j_1\ldots j_6}$&$Z^{j_1\ldots j_7}$&\cr 
\noalign{\hrule}
3&&&&&&&&$8^{(0)}$&\cr 
&&&$28^{(1)}$&&&$56^{(2)}$&&&\cr 
&$64^{(3)}$&&&$56^{(4)}$&&&$28^{(5)}$&&\cr 
&&$8^{(6)}$&&&&&&&\cr 
\noalign{\hrule}
4&&&&&&&$7^{(0)}$&-&\cr 
&&&$21^{(1)}$&&&$21^{(2)}$&&-&\cr 
&&$7^{(3)}$&&&&&&-&\cr 
\noalign{\hrule}
5&&&&&&$6^{(0)}$&-&-&\cr 
&&&$15^{(1)}$&&&$6^{(2)}$&-&-&\cr 
\noalign{\hrule}
6&&&&&$5^{(0)}$&-&-&-&\cr 
&&&$10^{(1)}$&&&-&-&-&\cr 
&$1^{(2)}$&&&&&-&-&-&\cr 
\noalign{\hrule}
7&&&&$4^{(0)}$&-&-&-&-&\cr 
&&&$6^{(1)}$&&-&-&-&-&\cr 
\noalign{\hrule}
8&&&$3^{(0)}$&-&-&-&-&-&\cr 
&&&$3^{(1)}$&-&-&-&-&-&\cr 
\noalign{\hrule}
}}\cr
Table 2.8 The particle charge multiplets in $3\leq D \leq 8$\cr}$$
We summarise the appearance of the particle multiplet from the $l_1$ representation in table 2.8.
The table shows the dimension of $SL(11-D)$ tensors together with the level the generator appears at in the decomposition, so that $X^{(l)}$ indicates a tensor carrying $X$ degrees of freedom appearing in the decomposition at level $l$. The columns indicate the number of indices in the charge modulo blocks of $11-D$ indices.
\medskip
{\bf 2.4 The String Multiplet}
\medskip
The string multiplet contains charges with one spacetime index. From table 2.1 we see that such charges are associated to the representation of $E_{11-D}$ with highest weight $\lambda_{10-D}$. In $D=3$ this is the ${\bf 3875}$ of $E_8$, in $D=4$ it is the ${\bf 133}$ of $E_7$ and so on. From section 2.2. we recall that the string multiplet corresponds to 1-brane charges, by putting $p=1$ and $k=0$ in equation (2.4) we find the simple root coefficient $m_D=2$ for string charges. From equation (2.3), with $p=1$ for the string, we find that,
$$m_j=1 \qquad j=1,\ldots D-1$$
The candidate roots in the $E_{12}$ lattice corresponding to the string multiplet have simple root coefficients $(1^{D},2,m_{D+1},\ldots m_{11})$. The $m_{i>D}$ take values which vary according to the weight vector of the representation of $E_{11-D}$ with highest weight $\lambda_{10-D}$. 

Using equation (2.5) we find the root in the $E_{12}$ lattice corresponding to the weight $\mu_{10-D}$ of $E_{11-D}$ ($p_{10-D}=1$ and all other $p_i=0$), that is the highest weight of the representation we are seeking to identify, is $(1^{D},2^{9-D},0,0,1)$. This root does not appear in the $l_1$ representation however, as discussed in [11], it is related to the root $(1^9,0,0,1)$ in the $l_1$ representation, with charge $Z^{8.11}$ by a series commutators with the generators ${K^D}_{D+1}$, ${K^{D+1}}_{D+2}$, $\ldots$ ${K^7}_8$, giving the charge $Z^{D.11}$. This charge has one compact index (j=11) and one non-compact index (a=D), and so is a component of a string charge $Z^{aj}$. We can also apply the generators ${K^a}_{j}$ to the roots of the particle multiplet and find string charges in a similar manner. 
$$\halign{\centerline{#} \cr
\vbox{\offinterlineskip
\halign{\strut \vrule \quad \hfil # \hfil\quad &\vrule \quad \hfil # \hfil\quad 
&\vrule \quad \hfil # \hfil\quad  &\vrule \quad \hfil # \hfil\quad &\vrule #
\cr
\noalign{\hrule}
$E_{12}$ Root&Charge&Dimension of $SL(8)$ tensor&Mass&\cr 
\noalign{\hrule}
$(1^3,2^6,0,0,1)$&$Z^{a,j_1}$&$8$&${R_3R_{11}\over l_p^3}$&\cr
\noalign{\hrule}
$(1^3,2^5,3,2,1,2)$&$Z^{a,j_1\ldots j_4}$&$70$&${R_3R_8\ldots R_{11}\over l_p^6}$&\cr
\noalign{\hrule}
$(1^3,2^3,3,4,5,3,1,3)$&$Z^{a,j_1\ldots j_6,k}$&$216$&${R_3R_6\ldots R_{10}R^2_{11}\over l_p^9}$&\cr
$(1^3,2^2,3,4,5,6,4,2,3)$&$Z^{a,j_1\ldots j_7}$&$8$&${R_3R_5\ldots R_{11}\over l_p^9}$ &\cr
\noalign{\hrule}
$(1^3,2^2,3,4,5,6,4,2,4)$&$Z^{a,j_1\ldots j_7,k_1\ldots k_3}$&$420$&${R_3VR_9R_{10}R_{11}\over R_4l_p^{12}}$ &\cr
$(1^3,2,3,4,5,6,7,4,1,4)$&$Z^{aj_1\ldots j_8,k_1k_2}$&$28$&${R_3V(R_{10}R_{11})^2\over l_p^{12}}$ &\cr
$(1^3,2,3,4,5,6,7,4,2,4)$&$Z^{aj_1\ldots j_8,(k_1,l_1)}$&$36$&${R_3VR_{11}^2\over l_p^{12}}$ &\cr
\noalign{\hrule}
$(1^3,2^2,3,5,7,9,6,3,5)$&$Z^{aj_1\ldots j_7,k_1\ldots k_6}$&$168$&${R_3VR_6\ldots R_{11}\over R_4l_p^{15}}$ &\cr
$(1^3,2,3,4,5,7,9,6,3,5)$&$Z^{aj_1\ldots j_8,k_1\ldots k_5}$&$56$&${R_3VR_7\ldots R_{11}\over l_p^{15}}$ &\cr
$(1^3,2,3,4,5,6,8,5,2,5)$&$Z^{aj_1\ldots j_8,k_1\ldots k_4,l_1}$&$504$&${R_3VR_8\ldots R_{10}R_{11}^2\over l_p^{15}}$ &\cr
\noalign{\hrule}
$(1^3,2,4,6,8,10,12,8,4,6)$&$Z^{aj_1\ldots j_8,k_1\ldots k_8}$&$1$&${R_3V^2\over l_p^{18}}$ &\cr
$(1^3,2,3,5,7,9,11,7,3,6)$&$Z^{aj_1\ldots j_8,k_1\ldots k_7,l_1}$&$63$&${R_3V^2R_{11}\over R_4l_p^{18}}$ &\cr
$(1^3,2,3,4,6,8,10,6,3,6)$&$Z^{aj_1\ldots j_8,k_1\ldots k_6,l_1l_2}$&$720$&${R_3V^2R_{10}R_{11}\over R_4R_5l_p^{18}}$ &\cr
$(1^3,2,3,5,7,9,11,7,3,6)$&$Z^{aj_1\ldots j_8,k_1\ldots k_7,l_1}$&$63$&${R_3V^2R_{11}\over R_4l_p^{18}}$ &\cr
\noalign{\hrule}
$(1^3,2,3,5,7,9,12,8,4,7)$&$Z^{aj_1\ldots j_8,k_1\ldots k_7,l_1\ldots l_4}$&$504$&${R_3V^2R_8\ldots R_{11}\over R_4 l_p^{21}}$ &\cr
$(1^3,2,4,6,8,10,12,8,4,7)$&$Z^{aj_1\ldots j_8,k_1\ldots k_8,l_1\ldots l_3}$&$56$&${R_3V^2R_9\ldots R_{11}\over l_p^{21}}$ &\cr
$(1^3,2,4,6,8,10,12,7,3,7)$&$Z^{aj_1\ldots j_8,k_1\ldots k_8,l_1l_2,m_1}$&$168$&${R_3V^2R_{10} R_{11}^2\over l_p^{21}}$ &\cr
\noalign{\hrule}
$(1^3,2,3,6,9,12,15,10,5,8)$&$Z^{aj_1\ldots j_8,k_1\ldots k_7,l_1\ldots l_7}$&$36$&${R_3V^3\over R_4^2 l_p^{24}}$ &\cr
$(1^3,2,4,6,9,12,15,10,5,8)$&$Z^{aj_1\ldots j_8,k_1\ldots k_8,l_1\ldots l_6}$&$28$&${R_3V^3\over R_4R_5l_p^{24}}$ &\cr
$(1^3,2,4,6,8,11,14,9,4,8)$&$Z^{aj_1\ldots j_8,k_1\ldots k_8,l_1\ldots l_5,m_1}$&$420$&${R_3V^3R_{11}\over R_4R_5R_6 l_p^{24}}$ &\cr
\noalign{\hrule}
$(1^3,2,5,8,11,14,17,11,5,9)$&$Z^{aj_1\ldots j_8,k_1\ldots k_8,l_1\ldots l_8,m_1}$&$8$&${R_3V^3R_{11}\over l_p^{27}}$ &\cr
$(1^3,2,4,7,10,13,16,10,5,9)$&$Z^{aj_1\ldots j_8,k_1\ldots k_8,l_1\ldots l_7,m_1m_2}$&$216$&${R_3V^3R_{10}R_{11}\over R_4 l_p^{27}}$ &\cr
\noalign{\hrule}
$(1^3,2,5,8,11,14,18,12,6,10)$&$Z^{aj_1\ldots j_8,k_1\ldots k_8,l_1\ldots l_8,m_1\ldots m_4}$&$70$&${R_3V^3R_8\ldots R_{11}\over l_p^{30}}$ &\cr
\noalign{\hrule}
$(1^3,2,5,9,13,17,21,14,7,11)$&$Z^{aj_1\ldots j_8,k_1\ldots k_8,l_1\ldots l_8,m_1\ldots m_7}$&$8$&${R_3V^4\over R_4l_p^{33}}$ &\cr
\noalign{\hrule}
}
}\cr
Table 2.9 The string charge multiplet in $D=3$/The {\bf 3875} of $E_8$ \cr}$$
To commence we rediscover the $E_{12}$ roots of the string multiplet in $D=3$. These roots were originally given in [11] and we list them in table 2.9 together the associated masses given by equation (3.3) which gives expressions in agreement with the string multiplet listed explicitly in appendix B of [9]. The procedure of searching the low levels of the $l_1$ representation for roots with root coefficients of the form $(1^3,2,m_4, \ldots m_{11})$ reproduces the ${\bf 3875\oplus 248\oplus 1}$ of $E_8$, from which the ${\bf 3875}$ can be identified. It is a simple matter to pair the dual generators in the string multiplet, to find an invariant mass squared for the multiplet:
$${\cal M}^2_{\bf3875} \sim {R_a^2V^4\over l_p^{36}}$$
We find the string multiplet in $D=4$ has $E_{12}$ roots with root coefficients $(1^4,2,m_5\ldots m_{11})$. In $D=4$ we look for the ${\bf 133}$ of $E_7$ and the corresponding roots from the $l_1$ are listed in table 2.10.
$$\halign{\centerline{#} \cr
\vbox{\offinterlineskip
\halign{\strut \vrule \quad \hfil # \hfil\quad &\vrule \quad \hfil # \hfil\quad 
&\vrule \quad \hfil # \hfil\quad  &\vrule \quad \hfil # \hfil\quad &\vrule #
\cr
\noalign{\hrule}
$E_{12}$ root&Charge&Dimension of $SL(7)$ tensor&Mass&\cr 
\noalign{\hrule}
$(1^4,2^5,0,0,1)$&$Z^{aj_1}$&$7$&${R_4R_{11}\over l_p^3}$&\cr
\noalign{\hrule}
$(1^4,2^4,3,2,1,2)$&$Z^{aj_1\ldots j_4}$&$35$&${R_4R_8\ldots R_{11}\over l_p^6}$&\cr
\noalign{\hrule}
$(1^4,2^2,3,4,5,3,1,3)$&$Z^{aj_1\ldots j_6,k}$&$49$&${R_4VR_{11}\over R_5l_p^9}$&\cr
\noalign{\hrule}
$(1^4,2,3,4,5,6,4,2,4)$&$Z^{aj_1\ldots j_7,k_1\ldots k_3}$&$35$&${R_4VR_9\ldots R_{11}\over l_p^{12}}$&\cr
\noalign{\hrule}
$(1^4,2,3,5,7,9,6,3,5)$&$Z^{aj_1\ldots j_7,k_1\ldots k_6}$&$7$&${R_4V^2\over R_5l_p^{15}}$ &\cr
\noalign{\hrule}
}
}\cr
Table 2.10 The string charge multiplet in $D=4$/The {\bf 133} of $E_7$\cr}$$
The invariant mass squared for the multiplet is,
$${\cal M}^2_{\bf133} \sim {R_a^2V^2\over l_p^{18}}$$
In $D=5$ we look for the ${\bf 27}$ of $E_6$. This corresponds to roots of the form $(1^5,2, m_6 \ldots m_{11})$, which we list in table 2.11.
$$\halign{\centerline{#} \cr
\vbox{\offinterlineskip
\halign{\strut \vrule \quad \hfil # \hfil\quad &\vrule \quad \hfil # \hfil\quad 
&\vrule \quad \hfil # \hfil\quad  &\vrule \quad \hfil # \hfil\quad &\vrule #
\cr
\noalign{\hrule}
$E_{12}$ root&Charge&Dimension of $SL(6)$ tensor&Mass&\cr 
\noalign{\hrule}
$(1^5,2^4,0,0,1)$&$Z^{aj_1}$&$6$&${R_5R_{11}\over l_p^3}$&\cr
\noalign{\hrule}
$(1^5,2^3,3,2,1,2)$&$Z^{aj_1\ldots j_4}$&$15$&${R_5R_8\ldots R_{11}\over l_p^6}$ &\cr
\noalign{\hrule}
$(1^5,2,3,4,5,3,1,3)$&$Z^{aj_1\ldots j_6,k_1}$&$6$&${R_5VR_{11}\over l_p^9}$ &\cr
\noalign{\hrule}
}
}\cr
Table 2.11 The string charge multiplet in $D=5$/The {\bf 27} of $E_6$ \cr}$$
In $D=6$ we look for the ${\bf 10}$ of $SO(5,5)$. This corresponds to roots of the form $(1^6,2, m_7 \ldots m_{11})$, which we list in table 2.12. 
$$\halign{\centerline{#} \cr
\vbox{\offinterlineskip
\halign{\strut \vrule \quad \hfil # \hfil\quad &\vrule \quad \hfil # \hfil\quad 
&\vrule \quad \hfil # \hfil\quad  &\vrule \quad \hfil # \hfil\quad &\vrule #
\cr
\noalign{\hrule}
$E_{12}$ root&Charge&Dimension of $SL(5)$ tensor&Mass&\cr 
\noalign{\hrule}
$(1^6,2^3,0,0,1)$&$Z^{a,j}$&$5$&${R_6R_{11}\over l_p^3}$&\cr
\noalign{\hrule}
$(1^6,2^2,3,2,1,2)$&$Z^{aj_1\ldots j_4}$&$5$&${R_6R_8\ldots R_{11}\over l_p^6}$&\cr
\noalign{\hrule}
}
}\cr
Table 2.12 The string charge multiplet in $D=6$/The {\bf 10} of $SO(5,5)$\cr}$$
The associated invariant mass squared for the representation in table 2.12 is
$${\cal M}^2_{\bf10} \sim {R_a^2V\over l_p^{9}}$$
In $D=7$ we look for the ${\bf 5}$ of $SL(5)$. This corresponds to roots of the form $(1^7,2, m_8 \ldots m_{11})$, which we list in table 2.13.
\eject
$$\halign{\centerline{#} \cr
\vbox{\offinterlineskip
\halign{\strut \vrule \quad \hfil # \hfil\quad &\vrule \quad \hfil # \hfil\quad 
&\vrule \quad \hfil # \hfil\quad  &\vrule \quad \hfil # \hfil\quad &\vrule #
\cr
\noalign{\hrule}
$E_{12}$ root&Charge&Dimension of SL(4) tensor&Mass&\cr 
\noalign{\hrule}
$(1^7,2^2,0,0,1)$&$Z^{a,j}$&$4$&${R_7R_{11}\over l_p^3}$&\cr
\noalign{\hrule}
$(1^7,2,3,2,1,2)$&$Z^{a,j_1\ldots j_4}$&$1$&${R_7V\over l_p^6}$&\cr
\noalign{\hrule}
}
}\cr
Table 2.13 The string charge multiplet in $D=7$/The {\bf 5} of $SL(5)$ \cr}$$
In $D=8$ we find the ${\bf (1,3)} $ of $SU(2)\times SU(3)$. This corresponds to roots of the form $(1^8,2, m_9 \ldots m_{11})$, and we find only the root with simple root coefficients $(1^8,2,0,0,1)$, corresponding to a charge $Z^{aj_1}$ of dimension $3$ and tension ${R_8R_{11}\over l_p^3}$. The appearance of the string multiplet in the $l_1$ representation is summarised for $3\leq D\leq 8$ in table 2.14.
$$\halign{\centerline{#} \cr
\vbox{\offinterlineskip
\halign{\strut \vrule \quad \hfil # \hfil\quad &\vrule \quad \hfil # \hfil\quad 
&\vrule \quad \hfil # \hfil\quad  &\vrule \quad \hfil # \hfil\quad &\vrule \quad \hfil # \hfil\quad &\vrule\quad \hfil # \hfil\quad &\vrule\quad \hfil # \hfil\quad &\vrule\quad \hfil # \hfil\quad &\vrule\quad \hfil # \hfil\quad &\vrule#
\cr
\noalign{\hrule}
D&$Z^a$&$Z^{aj}$&$Z^{aj_1j_2}$&$Z^{aj_1\ldots j_3}$&$Z^{aj_1\ldots j_4}$&$Z^{aj_1\ldots j_5}$&$Z^{aj_1\ldots j_6}$&$Z^{aj_1\ldots j_7}$&\cr 
\noalign{\hrule}
3&&$8^{(1)}$&&&$70^{(2)}$&&&$224^{(3)}$&\cr 
&&&$484^{(4)}$&&&$728^{(5)}$&&&\cr 
&$847^{(6)}$&&&$728^{(7)}$&&&$484^{(8)}$&&\cr 
&&$224^{(9)}$&&&$70^{(10)}$&&&$8^{(11)}$&\cr 
\noalign{\hrule}
4&&$7^{(1)}$&&&$35^{(2)}$&&&-&\cr 
&$49^{(3)}$&&&$35^{(4)}$&&&$7^{(5)}$&-&\cr 
\noalign{\hrule}
5&&$6^{(1)}$&&&$15^{(2)}$&&-&-&\cr 
&&$6^{(3)}$&&&&&-&-&\cr 
\noalign{\hrule}
6&&$5^{(1)}$&&&$5^{(2)}$&-&-&-&\cr 
\noalign{\hrule}
7&&$4^{(1)}$&&&-&-&-&-&\cr 
&$1^{(2)}$&&&&-&-&-&-&\cr 
\noalign{\hrule}
8&&$3^{(1)}$&&-&-&-&-&-&\cr 
\noalign{\hrule}
}}\cr
Table 2.14 The string charge multiplets in $3\leq D \leq 8$\cr}$$
In addition to the string and particle multiplets, we may also find multiplets associated with the membrane charge, a threebrane charge, a fourbrane charge and other charges up to that of a $D-1$-brane, as discussed in [8,9]. One might also find a fivebrane multiplet whose highest weight is $\mu_{D-5}\otimes 2\lambda_{11-D}$, indeed one may find multiplets corresponding to a variety of exotic charges whose interpretation is obscure, but are all a natural consequence of the conjectured eleven dimensional $E_{11}$ symmetry. In the next section we will consider the membrane multiplets in detail, and in section 2.6 we will find the remaining charge multiplets shown in table 2.2.
\medskip
{\bf 2.5 The membrane charge multiplet}
\medskip
Until this point we have focussed on reproducing the charge multiplets previously found from U-duality. The membrane charge multiplet has not been given in the literature previously. Here we treat the membrane charge in the same way as the particle and string charges and we apply our method to find the membrane multiplet in three, four, five, six, seven and eight dimensions.
For a membrane charge we take $p=2$ in equation (2.4) and look for solutions when $l=k=0$. Using equation (2.3) we look for roots of the form $(1^{(D-1)},2,3,m_{D+1},\ldots m_{11})$ in the tables of $E_{12}$ roots showing the $l_1$ representation of $E_{11}$ which are listed at length in [16]. $$\halign{\centerline{#} \cr
\vbox{\offinterlineskip
\halign{\strut \vrule \quad \hfil # \hfil\quad &\vrule \quad \hfil # \hfil\quad 
&\vrule \quad \hfil # \hfil\quad  &\vrule \quad \hfil # \hfil\quad &\vrule #
\cr
\noalign{\hrule}
Level&Charge&Dimension of &Mass&\cr 
&&$SL(8)$ tensor&&\cr 
\noalign{\hrule}
$1$&$Z^{ab}$&$1$&${R_2R_3\over l_p^3}$&\cr
\noalign{\hrule}
$2$&$Z^{abj_1\ldots j_3}$&$56$&${R_2R_3R_9\ldots R_{11}\over l_p^6}$&\cr
\noalign{\hrule}
$3$&$Z^{abj_1\ldots j_6}$&$28$&${R_2R_3R_6\ldots R_{11} \over l_p^9}$&\cr
$$&$Z^{abj_1\ldots j_5,k_1}$&$420$&${R_2R_3R_7\ldots R_{10}R_{11}^2 \over l_p^9}$&\cr
\noalign{\hrule}
$4$&$Z^{abj_1\ldots j_6,k_1\ldots k_3}$&$1344$&${R_2R_3VR_9\ldots R_{11}\over R_4R_5l_p^{12}}$ &\cr
$$&$Z^{abj_1\ldots j_7,k_1k_2}$&$216$&${R_2R_3 V R_{10}R_{11}\over R_4l_p^{12}}$&\cr
$$&$2Z^{abj_1\ldots j_8,k_1}$&$2\times8$&${R_2R_3VR_{11} \over l_p^9}$&\cr
$$&$2Z^{abj_1\ldots j_7,k_1,l_1}$&$280$&${R_2R_3VR_{11}^2 \over R_4l_p^9}$&\cr
\noalign{\hrule}
$5$&$Z^{abj_1\ldots j_6,k_1\ldots k_6}$&336&${R_2R_3 V^2 \over (R_4R_5)^2l_p^{15}}$&\cr
&$Z^{abj_1\ldots j_7,k_1\ldots k_5}$&$378$&${R_2R_3VR_7\ldots R_{11}\over R_4l_p^{15}}$&\cr
&$2Z^{abj_1\ldots j_8,k_1\ldots k_4}$&$2\times 70$&${R_2R_3VR_8\ldots R_{11}\over l_p^{15}}$&\cr
&$Z^{abj_1\ldots j_7,k_1\ldots k_4,l_1}$&$3584$&${R_2R_3VR_8\ldots R_{10}R_{11}^2\over R_4l_p^{15}}$&\cr
&$2Z^{abj_1\ldots j_8,k_1\ldots k_3,l_1}$&$2\times 378$&${R_2R_3VR_9R_{10}R_{11}^2\over l_p^{15}}$&\cr
\noalign{\hrule}
$6$&$2Z^{abj_1\ldots j_7}$&$2\times8$&${R_2R_3V^2\over R_4l_p^{18}}$ &\cr
$$&$Z^{abj_1\ldots j_7,k_1\ldots k_7,l_1}$&280&${R_2R_3V^2R_{11}\over R_4R_5l_p^{18}}$ &\cr
&$Z^{abj_1\ldots j_7,k_1\ldots k_6,l_1l_2}$&4200&${R_2R_3V^2R_{10}R_{11}\over R_4^2R_5l_p^{18}}$ &\cr
&$Z^{abj_1\ldots j_8,k_1\ldots k_4,l_1\ldots l_3}$&2352&${R_2R_3VR_8(R_9\ldots R_{11})^2\over l_p^{18}}$ &\cr
&$Z^{abj_1\ldots j_8,k_1\ldots k_5,l_1l_2}$&1344&${R_2R_3V^2R_{10}R_{11}\over R_4R_5R_6l_p^{18}}$ &\cr
&$4Z^{abj_1\ldots j_8,k_1\ldots k_6,l_1}$&$4\times 216$&${R_2R_3V^2R_{11}\over R_4R_5l_p^{18}}$ &\cr
&$Z^{abj_1\ldots j_8,k_1\ldots k_5,l_1,m_1}$&$1800$&${R_2R_3V^2R_{11}^2\over R_4R_5R_6l_p^{18}}$ &\cr
\noalign{\hrule}
$7$&$Z^{abj_1\ldots j_7,k_1\ldots k_7,l_1\ldots l_4}$&$2100$&${R_2R_3V^2R_8\ldots R_{11}\over R_4^2 l_p^{21}}$ &\cr
&$3Z^{abj_1\ldots j_8,k_1\ldots k_7,l_1\ldots l_3}$&$3\times420$&${R_2R_3V^2R_9R_{10}R_{11}\over R_4l_p^{21}}$ &\cr
&$3Z^{abj_1\ldots j_8,k_1\ldots k_8,l_1l_2}$&$3\times28$&${R_2R_3V^2R_{10}R_{11}\over l_p^{21}}$ &\cr
&$2Z^{abj_1\ldots j_8,k_1\ldots k_6,l_1\ldots l_4}$&$2\times 1512$&${R_2R_3V^2R_8R_9R_{10}R_{11}\over R_4R_5l_p^{21}}$&\cr
&$2Z^{abj_1\ldots j_8,k_1\ldots k_7,l_1\ldots l_2,m_1}$&$2\times1280$&${R_2R_3V^2R_{10}R_{11}^2\over R_4l_p^{21}}$&\cr
&$3Z^{abj_1\ldots j_8,k_1\ldots k_8,l_1,m_1}$&$3\times36$&${R_2R_3V^2R_{11}^2\over l_p^{21}}$&\cr
&$Z^{abj_1\ldots j_8,k_1\ldots k_6,l_1\ldots l_3,m_1}$&$8820$&${R_2R_3V^2R_9R_{10}R_{11}^2\over R_4R_5l_p^{21}}$&\cr
\noalign{\hrule}
$8$&$Z^{abj_1\ldots j_7,k_1\ldots k_7,l_1\ldots l_7}$&$120$&${R_2R_3V^3\over R_4^3 l_p^{24}}$ &\cr
&$3Z^{abj_1\ldots j_8,k_1\ldots k_7,l_1\ldots l_6}$&$3\times 168$&${R_2R_3V^3R_{11}\over R_4^2R_5 l_p^{24}}$ &\cr
&$3Z^{abj_1\ldots j_8,k_1\ldots k_8,l_1\ldots l_5}$&$3\times 56$&${R_2R_3V^3\over R_4R_5R_6l_p^{24}}$ &\cr
&$2Z^{abj_1\ldots j_8,k_1\ldots k_7,l_1\ldots l_5,m_1}$&$2\times 2800$&${R_2R_3V^3R_{10}R_{11}\over R_4^2R_5R_6l_p^{24}}$ &\cr
&$Z^{abj_1\ldots j_8,k_1\ldots k_8,l_1\ldots l_3,m_1m_2}$&$1008$&${R_2R_3V^2R_9R_{10} R_{11}^3\over l_p^{24}}$ &\cr
&$4Z^{abj_1\ldots j_8,k_1\ldots k_8,l_1\ldots l_4,m_1}$&$4\times 504$&${R_2R_3V^2R_8R_9R_{10} R_{11}^2\over l_p^{24}}$ &\cr
&$4Z^{abj_1\ldots j_8,k_1\ldots k_6,l_1\ldots l_6,m_1}$&$2520$&${R_2R_3V^3R_{11}\over R_4R_5l_p^{24}}$ &\cr
&$Z^{abj_1\ldots j_8,k_1\ldots k_7,l_1\ldots l_4,m_1m_2}$&$10584$&${R_2R_3V^3R_{10}R_{11}\over R_4^2R_5R_6R_7l_p^{24}}$ &\cr
&$Z^{abj_1\ldots j_8,k_1\ldots k_8,l_1\ldots l_3,m_1,n_1}$&$1512$&${R_2R_3V^3R_{10}R_{11}\over R_4^2R_5R_6R_7l_p^{24}}$ &\cr
\noalign{\hrule}
}
}\cr}$$
$$\halign{\centerline{#} \cr
\vbox{\offinterlineskip
\halign{\strut \vrule \quad \hfil # \hfil\quad &\vrule \quad \hfil # \hfil\quad 
&\vrule \quad \hfil # \hfil\quad  &\vrule \quad \hfil # \hfil\quad &\vrule #
\cr
\noalign{\hrule}
Level&Charge&Dimension of &Mass&\cr 
&&$SL(8)$ tensor&&\cr 
\noalign{\hrule}
$9$&$Z^{abj_1\ldots j_8,k_1\ldots k_8,l_1\ldots l_8,m_1\ldots m_8}$&$1$&${R_2R_3V^3\over l_p^{27}}$ &\cr
&$2Z^{abj_1\ldots j_8,k_1\ldots k_7,l_1\ldots l_7,m_1m_2}$&$2\times 945$&${R_2R_3V^3R_{9}R_{10} R_{11}\over R_4^2R_5l_p^{27}}$ &\cr
&$5Z^{abj_1\ldots j_8,k_1\ldots k_8,l_1\ldots l_7,m_1}$&$5\times 63$&${R_2R_3V^3R_{11}\over R_4l_p^{27}}$ &\cr
&$Z^{abj_1\ldots j_8,k_1\ldots k_8,l_1\ldots l_5,m_1\ldots m_3}$&$2352$&${R_2R_3V^3R_9R_{10}R_{11}\over R_4R_5R_6l_p^{27}}$&\cr
&$4Z^{abj_1\ldots j_8,k_1\ldots k_8,l_1\ldots l_6,m_1m_2}$&$4\times 720$&${R_2R_3V^3R_{10}R_{11}\over R_4R_5l_p^{27}}$&\cr
&$Z^{abj_1\ldots j_8,k_1\ldots k_7,l_1\ldots l_6,m_1\ldots m_3}$&$7680$&${R_2R_3V^3R_9R_{10}R_{11}\over R_4^2R_5l_p^{27}}$&\cr
&$2Z^{abj_1\ldots j_8,k_1\ldots k_7,l_1\ldots l_7,m_1m_2}$&$2\times 945$&${R_2R_3V^3R_{10}R_{11}\over R_4^2l_p^{27}}$&\cr
&$Z^{abj_1\ldots j_8,k_1\ldots k_8,l_1\ldots l_4,m_1\ldots m_4}$&$1764$&${R_2R_3V^4\over (R_4\ldots R_7)^2l_p^{27}}$&\cr
&$Z^{abj_1\ldots j_8,k_1\ldots k_8,l_1\ldots l_6,m_1,n_1}$&$7680$&${R_2R_3V^3R_{10}R_{11}^2\over R_4\ldots R_6l_p^{27}}$&\cr
\noalign{\hrule}
}
}\cr
Table 2.15 The membrane charge multiplet in $D=3$/The {\bf 147250} of $E_8$ \cr}$$
The membrane charge multiplet is a representation of $E_{11-D}$ with highest weight $\lambda_{11-D}$, in $D=3$ it is the ${\bf 147250}$ of $E_8$, in $D=4$ it is the ${\bf 912}$ of $E_7$ and so on. In $D=3$ the roots in the $l_1$ representation having coefficients of the form $(1^{(D-1)},2,3,m_{D+1},\ldots m_{11})$ identify the ${\bf 147250\oplus 30380 \oplus 3875 \oplus 248 \oplus 1}$ of $E_8$ from which we can identify the root vectors of the ${\bf 147250}$ and we list the charges in table 2.15, where the charges at levels ($m_{11}$) $10$ to $17$ which are the duals of those at levels $8$ to $1$ are not shown.

The associated invariant mass squared for the membrane charge multiplet in table 2.15 is
$${\cal M}^2_{\bf147250} \sim {(R_aR_b)^2V^6\over l_p^{54}}$$
$$\halign{\centerline{#} \cr
\vbox{\offinterlineskip
\halign{\strut \vrule \quad \hfil # \hfil\quad &\vrule \quad \hfil # \hfil\quad 
&\vrule \quad \hfil # \hfil\quad  &\vrule \quad \hfil # \hfil\quad &\vrule #
\cr
\noalign{\hrule}
Level&Charge&Dimension of $SL(7)$ tensor&Mass&\cr 
\noalign{\hrule}
$1$&$Z^{ab}$&$1$&${R_3R_4\over l_p^3}$&\cr
\noalign{\hrule}
$2$&$Z^{ab,j_1\ldots j_3}$&$35$&${R_3R_4R_9\ldots R_{11}\over l_p^6}$&\cr
\noalign{\hrule}
$3$&$Z^{ab,j_1\ldots j_6}$&$7$&${R_3R_4 V \over R_5l_p^9}$&\cr
$$&$Z^{ab,j_1\ldots j_5,k}$&$140$&${R_3R_4 VR_{11} \over R_5R_6l_p^9}$&\cr
\noalign{\hrule}
$4$&$Z^{ab,j_1\ldots j_6,k_1\ldots k_3}$&224&${R_3R_4 V^2R_9R_{10}R_{11}\over R_5^2R_6R_7R_8l_p^{12}}$ &\cr
$$&$Z^{ab,j_1\ldots j_7,k_1k_2}$&21&${R_3R_4 VR_{11}^2\over l_p^{12}}$ &\cr
$$&$Z^{ab,k_1,l_1}$&28&${R_3R_4 VR_{11}^2\over l_p^{12}}$ &\cr
\noalign{\hrule}
$5$&$Z^{ab,j_1\ldots j_6,k_1\ldots k_6}$&28&${R_3R_4 V^2\over R_5^2l_p^{15}}$ &\cr
$$&$Z^{ab,j_1\ldots j_7,k_1\ldots k_5}$&21&${R_3R_4 V^2\over R_5R_6l_p^{15}}$ &\cr
$$&$Z^{ab,j_1\ldots j_7,k_1\ldots k_4,l_1}$&224&${R_3R_4 V^2R_{11}\over R_5R_6R_7l_p^{15}}$ &\cr
\noalign{\hrule}
$6$&$Z^{ab,j_1\ldots j_7,k_1\ldots k_6,l_1l_2}$&$140$&${R_3R_4V^2R_{11}\over R_5l_p^{18}}$ &\cr
$$&$Z^{ab,j_1\ldots j_7,k_1\ldots k_7,l_1}$&$7$&${R_3R_4V^2R_{11}\over R_5l_p^{18}}$ &\cr
\noalign{\hrule}
$7$&$Z^{ab,j_1\ldots j_7,k_1\ldots k_7,l_1\ldots l_4}$&$35$&${R_3R_4V^3\over R_5R_6R_7l_p^{21}}$ &\cr
\noalign{\hrule}
$8$&$Z^{ab,j_1\ldots j_7,k_1\ldots k_7,l_1\ldots l_7}$&$1$&${R_3R_4V^3\over l_p^{24}}$ &\cr
\noalign{\hrule}
}
}\cr
Table 2.16 The membrane charge multiplet in $D=4$/The {\bf 912} of $E_7$ \cr}$$

In $D=4$ we look for roots of the form $(1^3,2,3,m_5,\ldots m_{11})$ and find the set of charges in table 2.16, which form the {\bf 912} of $E_7$. The associated invariant mass squared for the membrane charge multiplet in table 2.16 is
$${\cal M}^2_{\bf912} \sim {(R_aR_b)^2V^3\over l_p^{27}}$$
In $D=5$ we find the {\bf 78} of $E_6$ as shown in table 2.17.
$$\halign{\centerline{#} \cr
\vbox{\offinterlineskip
\halign{\strut \vrule \quad \hfil # \hfil\quad &\vrule \quad \hfil # \hfil\quad 
&\vrule \quad \hfil # \hfil\quad  &\vrule \quad \hfil # \hfil\quad &\vrule #
\cr
\noalign{\hrule}
Level&Charge&Dimension of $SL(5)$ tensor&Mass&\cr 
\noalign{\hrule}
$1$&$Z^{ab}$&$1$&${R_4R_5\over l_p^3}$&\cr
\noalign{\hrule}
$2$&$Z^{ab,j_1\ldots j_3}$&$20$&${R_4R_5R_9\ldots R_{11}\over l_p^6}$&\cr
\noalign{\hrule}
$3$&$Z^{ab,j_1\ldots j_5,k}$&$36$&${R_4R_5 VR_{11} \over R_6l_p^9}$&\cr
\noalign{\hrule}
$4$&$Z^{ab,j_1\ldots j_3}$&20&${R_4R_5 V(R_9\ldots R_{11})\over l_p^{12}}$ &\cr
\noalign{\hrule}
$5$&$Z^{ab}$&$1$&${R_4R_5 V^2\over l_p^{15}}$ &\cr
\noalign{\hrule}
}
}\cr
Table 2.17 The membrane charge multiplet in $D=5$/The ${\bf 78}$ of $E_6$ \cr}$$
The associated invariant mass squared for the membrane charge multiplet in table 2.17 is
$${\cal M}^2_{\bf 78} \sim {(R_aR_b)^2V^2\over l_p^{18}}$$
We can continue this process and find the ${\bf 1\oplus10\oplus5=16}$ of $SO(5,5)$ in $D=6$, the ${\bf 1\oplus 4=5}$ of $SL(5)$ in $D=7$ and the ${\bf 1\oplus1=(1.2)}$ of $SL(3)\otimes SL(2)$ in $D=8$. We summarise the appearance of the membrane multiplet charges from the $l_1$ representation of $E_{11}$ in table 2.18.
$$\halign{\centerline{#} \cr
\vbox{\offinterlineskip
\halign{\strut \vrule \quad \hfil # \hfil\quad &\vrule \quad \hfil # \hfil\quad 
&\vrule \quad \hfil # \hfil\quad  &\vrule \quad \hfil # \hfil\quad &\vrule \quad \hfil # \hfil\quad &\vrule\quad \hfil # \hfil\quad &\vrule\quad \hfil # \hfil\quad &\vrule\quad \hfil # \hfil\quad &\vrule\quad \hfil # \hfil\quad &\vrule#
\cr
\noalign{\hrule}
D&$Z^{ab}$&$Z^{abj}$&$Z^{abj_1j_2}$&$Z^{abj_1\ldots j_3}$&$Z^{abj_1\ldots j_4}$&$Z^{abj_1\ldots j_5}$&$Z^{abj_1\ldots j_6}$&$Z^{abj_1\ldots j_7}$&\cr 
\noalign{\hrule}
3&$1^{(1)}$&&&$56^{(2)}$&&&$420^{(3)}$&&\cr 
&&$1856^{(4)}$&&&$5152^{(5)}$&&&$11696^{(6)}$&\cr 
&&&$16408^{(7)}$&&&$23472^{(8)}$&&&\cr 
&$29128^{(9)}$&&&$23472^{(10)}$&&&$16408^{(11)}$&&\cr 
&&$11696^{(12)}$&&&$5152^{(13)}$&&&$1856^{(14)}$&\cr
&&&$420^{(15)}$&&&$56^{(16)}$&&&\cr
&$1^{(17)}$&&&&&&&&\cr
\noalign{\hrule}
4&$1^{(1)}$&&&$35^{(2)}$&&&$147^{(3)}$&-&\cr 
&&&$273^{(4)}$&&&$273^{(5)}$&&-&\cr 
&&$147^{(6)}$&&&$35^{(7)}$&&&-&\cr 
&$1^{(8)}$&&&&&&&&\cr
\noalign{\hrule}
5&$1^{(1)}$&&&$20^{(2)}$&&&-&-&\cr 
&$36^{(3)}$&&&$20^{(4)}$&&&-&-&\cr 
&$1^{(5)}$&&&&&&&&\cr
\noalign{\hrule}
6&$1^{(1)}$&&&$10^{(2)}$&&-&-&-&\cr 
&&$5^{(3)}$&&&&-&-&-&\cr 
\noalign{\hrule}
7&$1^{(1)}$&&&$4^{(2)}$&-&-&-&-&\cr 
\noalign{\hrule}
8&$1^{(1)}$&&&-&-&-&-&-&\cr 
&$1^{(2)}$&&&-&-&-&-&-&\cr 
\noalign{\hrule}
}}\cr
Table 2.18 The membrane charge multiplets in $3\leq D \leq 8$\cr}$$
\eject
{\bf 2.6 p-brane charges associated to general weights of $E_{11-D}$}

In the previous sections we considered representations of $A_{D-1}\otimes E_{11-D}$ in the $l_1$ representation of $E_{11}$. The representations of $A_{D-1}$ had a single set of antisymmetric indices and corresponded to p-brane charges, the representations of $E_{11-D}$ we considered all had a highest weight which was a single fundamental weight of $E_{11-D}$. Now we will continue to consider p-brane charges in the $A_{D-1}$ sub-algebra but we will generalise our considerations of $E_{11-D}$ to include those representations whose highest weight is a sum of fundamental weights. As we shall see, representations of $E_{11-D}$, with highest weight more general than a single fundamental weight, provide a straightforward constraint for the root coefficient $m_D$. By including the most general highest weight representations of $E_{11-D}$ we will complete table 2.2.

So far we have not considered the significance of blocks of $(11-D)$ indices appearing in the index structure of the generators in the $E_{11-D}$ sub-algebra. Previously we have restricted ourselves to the constraint that $l=0$ in equation (2.6), implying that we have ignored representations of $E_{11-D}$ which are identical up to the $\epsilon^{j_1\ldots j_{11-D}}$ tensor. However the appearance of a block of $(11-D)$ indices may be acted upon by the generators of the $A_{10}$ sub-algebra of $E_{11}$ so that a set of trivial internal indices may acquire spacetime indices and the remaining internal indices form part of a non-trivial representation of $E_{11-D}$. Let us consider a generic highest weight representation of $E_{11-D}$ labelled by $[p_1,p_2\ldots p_{11-D}]$. Putting this into equation (2.1), where instead of $\lambda_i$ we now have $\sum p_i\lambda_i$, we find, after making use of equation (2.2), 
$$\eqalign{
-n_j &= \sum p_i <\lambda_i,\lambda_j> - m_D<\lambda_1,\lambda_j>\cr
 &=\cases{\sum_{i\leq j}ip_i+j\sum_{i\geq j}^{8-D}p_i-m_D+{j\over D-2}[\sum_i^{8-D}ip_i+ 4p_{9-D}+2p_{10-D}+3p_{11-D}-m_D]&${j\leq 8-D}$\cr
{1\over D-2}[4\sum_{i\leq 8-D}ip_i+2(10-D)p_{9-D}+(10-D)p_{10-D}+2(8-D)p_{11-D}-4m_D] & $j=9-D$\cr
{1\over D-2}[2\sum_{i\leq 8-D}ip_i+(10-D)p_{9-D}+4p_{10-D}+(8-D)p_{11-D}-2m_D] & $j=10-D$\cr
{1\over D-2}[3\sum_{i\leq 8-D}ip_i+2(8-D)p_{9-D}+(8-D)p_{10-D}+(11-D)p_{11-D}-3m_D] & $j=11-D$}}
$$
We see that we have a general solution giving positive integer values for $n_j$ when,
$$m_D=\sum_{i=1}^{8-D}ip_i + 4p_{(9-D)}+2p_{(10-D)}+3p_{(11-D)}+h(D-2)\eqno{(2.7)}$$
Where $h$ is some positive integer, or zero. For this solution we find the simple root coefficients in $E_{11-D}$ are:
$$n_j =\cases{\sum_{i\geq j}^{8-D}(i-j)p_i+4p_{9-D}+2p_{10-D}+3p_{11-D}+hj+h(D-2) &${j\leq 8-D}$\cr
2p_{9-D}+p_{10-D}+2p_{11-D}+4h & $j=9-D$\cr
p_{9-D}+p_{11-D}+2h & $j=10-D$\cr
2p_{9-D}+p_{10-D}+p_{11-D}+3h & $j=11-D$}
$$
With these coefficients the full $E_{12}$ root is:
$$\eqalignno{\beta=&e_*+k(e_1+\ldots +e_{D-p})+(1+k)(e_{D-p+1}+\ldots +e_D)\cr &+\sum_{n=D+1}^{8-D}(h-\sum_{i=n}^{8-D}p_i)e_n+he_9+(h+p_{9-D})e_{10}+(h+p_{9-D}+p_{10-D})e_{11}}$$
Where $p_i$ are the Dynkin labels for the $E_{11-D}$ sub-algebra, and $p$ without an index indicates the corresponding p-brane charge multiplet. We can write the parameter, $h$, as:
$$h=\sum_{i=1}^{8-D}p_i+C$$
If we decompose the $E_{11-D}$ representation into $SL(11-D)$ tensor representations the number of $SL(11-D)$ indices appearing on generators at level $n_{11-D}$ in the decomposition, $\#$ is given by,
$$\#=3n_{11-D}-m_D=\sum_{i=1}^{10-D}(11-D-i)p_i +C(11-D)$$
So that $C$ controls the appearance of blocks of $(11-D)$ indices. We note that the lower bound on the number of indices, $\#$ occurs when $n_{11-D}$ and $p_1=m_D$, this implies that $C$ is bounded from below by $C\geq -m_D$.
Setting $k=0$ we find,
$$\eqalign{\beta^2=&1+p+\sum_{i=1}^{8-D}ip_i^2-2p_{9-D}^2-p_{11-D}^2+2\sum_{i\leq j}ip_ip_j-2p_{9-D}p_{10-D}-4p_{9-D}p_{11-D}-2p_{10-D}p_{11-D}\cr 
&-(D-2)h^2-2h\sum_{i=1}^{8-D}ip_i-8hp_{9-D}-4hp_{10-D}-6hp_{11-D}}\eqno{(2.8)}$$
Since $\beta^2=2,0,-2,\ldots$ equation (2.8) places constraints on which highest weights of $E_{11-D}$ are present in the $l_1$ representation. Let us see how useful this equation can be by looking for all the highest weight representations of $E_{11-D}$ appearing in the $l_1$ representation of $E_{11}$ for the particle, string and membrane charges. 

Consider the particle charge multiplet, having $p=0$, and $m_D=1$, and we will commence with the $D$-independent case where $h=0$. Earlier we found the particle charge multiplet associated to the first fundamental representation of $E_{11-D}$, whose highest weight is $\lambda_1$. From equation (2.7) we see that indeed $p_1=1$ is the only possible representation, and from equation (2.8) we find $\beta^2=2$ and so it is present in the $l_1$ representation. However if we now generalise our considerations to include the cases with $h> 0$ then from equation (2.7) we see that  in $D=3$ a new possible particle charge multiplet appears associated to a different highest weight of $E_{11-D}$.  In $D=3$ we can satisfy $m_D=1$ with ${p_i=0,h=1}$, and from equation (2.8) we see this corresponds to a root of $E_{12}$ with a squared length of zero. We now identify the complete multiplet associated to this particle charge by looking for roots having the form $(1^4,m_{D+1},\dots, m_{11})$. The highest weight root has $\beta^2=0$ and under the action of the $SL(11)$ sub-algebra the root length squared may be lowered or held constant, but not raised - due to the Serre relations. Consequently the highest weight representation may be found among the set of roots having the form $(1^4,m_{D+1},\dots, m_{11})$ and subject to $\beta^2\leq 0$.
The extra particle multiplet in $D=3$ is given in table 2.19.
$$\halign{\centerline{#} \cr
\vbox{\offinterlineskip
\halign{\strut \vrule \quad \hfil # \hfil\quad &\vrule \quad \hfil # \hfil\quad 
&\vrule \quad \hfil # \hfil\quad  &\vrule \quad \hfil # \hfil\quad &\vrule #
\cr
\noalign{\hrule}
$E_{12}$ root&Charge&Dimension of $SL(8)$ tensor&Mass&\cr 
\noalign{\hrule}
$(1^5,2,3,4,5,3,1,3)$&$Z$&$1$&${R_4\ldots R_{11}\over l_p^9}$&\cr
\noalign{\hrule}
}}\cr
Table 2.19 A second particle charge multiplet in $D=3$/The {\bf 1} of $E_8$ \cr}$$
From table 2.19 associated invariant mass squared for the trivial representation is:
$${\cal M}^2_{\bf 1}\sim {V^2\over l_p^{18}}$$
Let us repeat this process for the string charge multiplets, commencing as before with the case $h=0$. Putative string charge multiplets have $p=1$ and $m_D=2$. From equation (2.7) we find three possible highest weight representations:
$$(p_1=2),\quad (p_2=1),\quad (p_{10-D}=1)$$
However from equation (2.8) (with $h=0$) we notice that we face the restriction 
$$2 \geq 1+p+\sum_{i=1}^{8-D}ip_i^2-2p_{9-D}^2-p_{11-D}^2+2\sum_{i\leq j}ip_ip_j-2p_{9-D}p_{10-D}-4p_{9-D}p_{11-D}-2p_{10-D}p_{11-D}\eqno{(2.9)}$$
Bearing in mind that for the string $p+1=2$ we observe that the remaining positive terms must be balanced or outweighed by the negative terms. That is, any weight with $p_{i\leq 8-D} \neq 0$ must at least be accompanied by non-zero values of $p_{9-D}$ or $p_{11-D}$. Consequently the putative string charge representations with $(p_1=2)$ and $(p_2=1)$ do not appear in the $l_1$ representation. Thus the $(p_{10-D}=1)$ representation having highest weight $\lambda_{10-D}$ is the unique representation carrying the string charge in the $l_1$ representation  when $h=0$.
If we now consider solutions when $h\neq0$ satisfying $m_D=2=(p+1)$ in equation (2.7) we may find the remaining string charge multiplets in $D=3$. We find,
$$(h=1,p_1=1),\quad (h=2))$$
Where all the remaining $p_i$ are zero. From equation (2.8) we find that $\beta^2=0$ for the first case and $\beta^2=-2$ for the second case. So both of these are highest weights of string charge multiplets in $D=3$, they are the ${\bf 248}$ and the ${\bf 1}$ respectively. The string of roots in the representation may be found by searching amongst roots of the form $(1^3,2,m_{D+1},\dots, m_{11})$ in the $l_1$ representation such that $\beta^2\leq 0$ in the first case, and $\beta^2 \leq -2$ in the second case. The precise roots forming these multiplets are shown in tables 2.20 and 2.21.
$$\halign{\centerline{#} \cr
\vbox{\offinterlineskip
\halign{\strut \vrule \quad \hfil # \hfil\quad &\vrule \quad \hfil # \hfil\quad 
&\vrule \quad \hfil # \hfil\quad  &\vrule \quad \hfil # \hfil\quad &\vrule #
\cr
\noalign{\hrule}
$E_{12}$ root&Charge&Dimension of $SL(8)$ tensor&Mass&\cr 
\noalign{\hrule}
$(1^3,2^2,3,4,5,6,4,2,3)$&$Z^{aj_1\ldots j_7}$&$8$&${R_3V\over R_4l_p^9}$&\cr
\noalign{\hrule}
$(1^3,2,3,4,5,6,7,4,2,4)$&$Z^{aj_1\ldots j_8,k_1k_2}$&$28$&${R_3VR_{10}R_{11}\over l_p^{12}}$&\cr
\noalign{\hrule}
$(1^3,2,3,4,5,7,9,6,3,5)$&$Z^{aj_1\ldots j_8,k_1\ldots k_5}$&$56$&${R_3VR_7\ldots R_{11}\over l_p^{15}}$&\cr
\noalign{\hrule}
$(1^3,2,4,6,8,10,12,8,4,6)$&$Z^{aj_1\ldots j_8,k_1\ldots k_8}$&$1$&${R_3V^2\over l_p^{18}}$&\cr
$(1^3,2,3,5,7,9,11,7,3,6)$&$Z^{aj_1\ldots j_8,k_1\ldots k_7,l_1}$&$63$&${R_3V^2R_{11}\over R_4l_p^{18}}$&\cr
\noalign{\hrule}
$(1^3,2,4,6,8,10,12,8,4,7)$&$Z^{aj_1\ldots j_8,k_1\ldots k_8,l_1l_2l_3}$&$56$&${R_3V^2R_9\ldots R_{11}\over l_p^{21}}$&\cr
\noalign{\hrule}
$(1^3,2,4,6,9,12,15,10,5,8)$&$Z^{aj_1\ldots j_8,k_1\ldots k_8,l_1\ldots l_6}$&$28$&${R_3V^2R_6\ldots R_{11}\over l_p^{24}}$&\cr
\noalign{\hrule}
$(1^3,2,5,8,11,14,17,11,5,9)$&$Z^{aj_1\ldots j_8,k_1\ldots k_8,l_1\dots l_8,m_1}$&$8$&${R_3V^3R_{11}\over l_p^{27}}$&\cr
\noalign{\hrule}
}}\cr
Table 2.20 A second string charge multiplet in $D=3$/The {\bf 248} of $E_8$ \cr}$$
From table 2.20 we can identify dual generators in the ${\bf 248}$ representation and associate an invariant mass squared to the multiplet:
$${\cal M}^2_{\bf 248}\sim {R_a^2V^4\over l_p^{36}}$$
$$\halign{\centerline{#} \cr
\vbox{\offinterlineskip
\halign{\strut \vrule \quad \hfil # \hfil\quad &\vrule \quad \hfil # \hfil\quad 
&\vrule \quad \hfil # \hfil\quad  &\vrule \quad \hfil # \hfil\quad &\vrule #
\cr
\noalign{\hrule}
$E_{12}$ root&Charge&Dimension of $SL(8)$ tensor&Mass&\cr 
\noalign{\hrule}
$(1^3,2,4,6,8,10,12,8,4,6)$&$Z^{aj_1\ldots j_8,k_1\ldots k_8}$&$1$&${R_3V^2\over l_p^{18}}$&\cr
\noalign{\hrule}
}}\cr
Table 2.21 A third string charge multiplet in $D=3$/The {\bf 1} of $E_8$ \cr}$$
The invariant mass squared is:
$${\cal M}^2_{\bf 1}\sim {R_a^2V^4\over l_p^{36}}$$
We note that the invariant mass squared is the same over all three (the ${\bf 3875, 248, 1}$) string charge multiplets in three dimensions - the Casimir of the representations of the internal group is a property of the spacetime algebra.

We can carry out the same analysis for the possible membrane charge multiplets. With $p=2$, we look for $p_i$ giving $m_D=3$ with $h=0$ in equation (2.7). The possible highest weights have non-zero Dynkin labels:
$$(p_1=3),\quad (p_1=1,p_2=1),\quad (p_3=1),\quad (p_1=1,p_{10-D}=1),\quad (p_{11-D}=1)$$
For $p=2$ we must find a negative contribution to the squared root length in equation (2.9) coming from the $E_{11-D}$ Dynkin labels, $p_i$. This means that at least either $p_{9-D}$ or $p_{11-D}$ must be non-zero.  Consequently the highest weight representation carrying the membrane charge is unique in the $l_1$ representation, the $\bf 147250$, having $(p_{11-D}=1)$. When $h>0$ we find the following highest weights carrying spacetime membrane charges with $\beta^2=2,0,-2,\ldots$:
$$(h=1,p_2=1),\quad (h=1, p_{10-D}=1), \quad (h=2,p_1=1), \quad (h=3)$$
Their highest weights have root lengths such that $\beta^2\leq 0, -2, -4, -6$ and they correspond to the $\bf 30380,$ $\bf3875, 248$ and $\bf 1$ respectively, as indicated in table 2.2.

This process can be carried out to discover all the charge multiplets from the particle multiplet to the $D-1$-brane multiplet in D dimensions. The results of doing so are shown in table 2.2 for three, four, five, six, seven and eight spacetime dimensions. 

The reader will notice that in addition to the well-known charge multiplets there are additional representations of $E_{11-D}$ appearing in table 2.2. For example the particle charge multiplet in three dimensions is the $\bf{248}$ supplemented by a singlet $\bf{1}$, and the string multiplet is no longer just the $\bf{3875}$ but also the $\bf{248}$ and the $\bf{1}$ of $E_8$ and so on. The existence of some of these extra multiplets were in fact argued for in [9] see for example section 7.2  or appendix C therein. The authors of [9] considered bound states formed from the particle and string charge multiplets, given in the M-theory language, and then looked to see if all the bound states of D-branes in the IIA or IIB language appeared. By identifying the omissions and arguing in favour of completing the spectrum of bound states extra charge multiplets were conjectured. The examples of [9] were in D=5 where a singlet corresponding to a bound state of a $D6$ and a $D0$ brane in the IIA picture was predicted and in D=4 where an additional $\bf{56}$ and two singlets of $E_7$ were argued for to account for missing bound states of the $D7$ brane, now in the IIB picture. It is straightforward to confirm that the mass of the exotic states in these examples agrees with the masses associated to the matching additional representations shown in table 2.2. The natural appearance of additional charge multiplets from $E_{11}$, completing the spectrum of bound states in D=5 and D=4, is an interesting result tied up with the prediction that there exist further exotic states beyond those previously uncovered in the BPS spectrum but revealed fully here\footnote{$^3$}{We thank the reviewer for drawing our attention to the comments in this paragraph.}.

\medskip
{\bf 3 Mass and tension in toroidally compactified backgrounds}
\medskip 
The identity between the U-duality symmetries of M-theory and the action of the Weyl group of $E_n$ was achieved in [8,9] by making use of a formula that gave the tension encoded in a weight vector \footnote{$^4$}{See, for example, equation (3.13) on page 30 of [9]}. 

The tension formula was justified empirically but its origin appeared mysterious. The tension formula gave the known tension of p-branes, strings and particles corresponding to a weight vector in a $D+1$ dimensional vector space modulo the addition of the unique vector orthogonal to a $D$ dimensional (sub-)vector space associated with spacetime. The unique orthogonal vector was shown to correspond to Newton's gravitational constant. In this section we give the main result of this paper: an algebraic formula for deriving tensions from the $l_1$ charge algebra and show that it is consistent with the lower dimensional formula used in [6,7,8,9,10]. We apply the solution to find known brane and string tensions in M-theory, and the IIA and IIB theories, and retrospectively to reproduce the tensions of the brane charge multiplets derived in section 2. Furthermore we are able to interpret the most part of the charge algebra as being associated to KK-brane charges and offer a classification scheme for these exotic charges within the $l_1$ representation.
\medskip
{\bf 3.1 The truncated group element as a vielbein}
\medskip 
The group element of $l_1 \otimes_s E_{11}$ at low levels is,
$$g=\exp(x^\mu P_\mu)\exp({h_a}^b{K^a}_b)\exp({1\over 3!}A_{c_1c_2c_3}R^{c_1c_2c_3})\exp({1\over 6!}A_{d_1\ldots d_6}R^{d_1\ldots d_6})\ldots$$
Where the ellipsis indicates the exponentiation of further generators of both the $l_1$ representation of $E_{11}$ as well as its adjoint representation. The restriction of the group element to the Cartan subalgebra, leaving only the components ${K^a}_a$ of the ${K^a}_b$ generators non-zero, corresponds in the nonlinear realisation to considering a diagonalised vielbein and metric, 
$$g_{\mu\mu}={(e^{2h})_\mu}^a \eta_{aa}$$
We recall the success in reconstructing the $G^{+++}$ half-BPS solutions of the eleven dimensional, IIA and IIB supergravity theories [4] as well as the maximally oxidised supergravity theories [13] from the coefficients of the Cartan subalgebra in a truncated version of the group element given by,
$$g_{\beta}=\exp{(-{1\over \beta^2}\ln N H\cdot \beta)}\exp{((1-N)E_{\beta})}\eqno{(3.1)}$$
This group element encoded half-BPS solutions as a group element of $E_{11}$.
In this group element the part of a general root, $\beta$, occurring in the Cartan sub-algebra was singled out using the inner product of the root with the Cartan sub-algebra, $H\cdot \beta$. However in the present paper we work with the $l_1$ representation of $E_{11}$, or charge algebra, which is defined in the twelve-dimensional lattice of $E_{12}$. In this context it is natural to extend the form of the half-BPS group element of $E_{11}$ to $E_{12}$, where we allow the inner product  $H\cdot \beta$ to run over the twelve generators of the Cartan sub-algebra of $E_{12}$.

It will be fruitful to employ a change of notation at this stage in order to highlight the physical role played by the variables in the group element. There are two natural bases to work in, one uses the diagonalised generators, ${K^a}_a$, and the other uses the generators of the Cartan sub-algebra, $H_a$. We now make a change of notation and substitute $p_a$ for the field $h_a$ associated to the  generators, ${K^a}_a$. When we work in the basis of the Cartan subalgebra, $H_a$, we will label the fields pre-multiplying the $H_a$ by $q_a$.  

Let us consider a spacetime with a dimension compactified on a circle, the line element for such a direction becomes, 
$${(e^{2p_i})_\mu}^id\xi_i^2=dx_\mu^2$$
Where $\xi_i$ is a circular coordinate taking values in the range $[0,2\pi]$. So that if the radius of compactifcation is $R_i$ in the worldvolume, we find by integrating,
$$p_i=\ln{R_i \over l_p}\eqno{(3.2)}$$
Where $l_p$ is the Planck length, and appears so that the parameters $p_i$ are dimensionless. For the non-compact directions we set all the $p_i$ to the same constant [14].

In the setting of the $l_1$ representation we must also find an interpretation for $p_*$ associated to the additional coordinate. It was shown in [14] that the scaling of the Planck constant under T-duality was predicted by an $E_{11}$ symmetry. In that paper the non-compact $p_i$ were set equal to the same constant $C$ and the effect of the Weyl reflection in the root $\alpha_{11}$ was denoted by a primed index. It was shown that the non-compact parameters encoded the scaling of Minkowski metric of spacetime with respect to the eleven dimensional metric via,
$$C'-C=\ln{l_p\over l_p'} \eqno{(3.3)}$$
One way to satisfy this scaling is by setting $C=\ln({1\over l_p})=p_i$. We propose to treat the extra coordinate in the $l_1$ representation as such a non-compact parameter and we set:
$$p_*=\ln({1\over l_p})\eqno{(3.4)}$$ 
And, indeed, $$p_i=\ln({1\over l_p})$$ for all non-compact coordinates. This choice for the non-compact $p_i$ introduces dimensionful parameters into the group theory with the dimension of mass in Planck units. A consequence of $p_*$ being a massive parameter means that the fully-compactified theory will also have a dimensionful parameter. 

While this interpretation of the vierbein associated to the additional coordinate is motivated as a particularly simple solution of equation (3.3), it appears appears to be on the correct footing, having a comparable form to that of the compact parameters $p_i$ coordinates, and also being dependent only upon the Planck length. Ultimately we will rest upon the successful reproduction of the brane tensions as shown in the latter sections of this paper and on the correct tensions in the brane charge multiplets of section 2 to justify the choice of $p_*$.

Upon reduction to the ten dimensional IIA theory we may interpret the radius of the compact direction in terms of the string length,
$$p_{11}=\ln({R_{11}\over l_p})=\ln(g_s^{2\over 3})=\ln{({l_p\over l_s})^2}$$
Where we have used $R_{11}=g_sl_s$ and $l_p^3=g_sl_s^3$ which are well-known identifications but which may be independently derived solely from considering the $E_{11}$ algebra [15].

One can make similar speculations for the $l_1$ extension of  the $K_{27}$ algebra related to the twenty-six-dimensional bosonic string, for which the Dynkin diagram is shown below,
\medskip
\centerline{\epsfbox{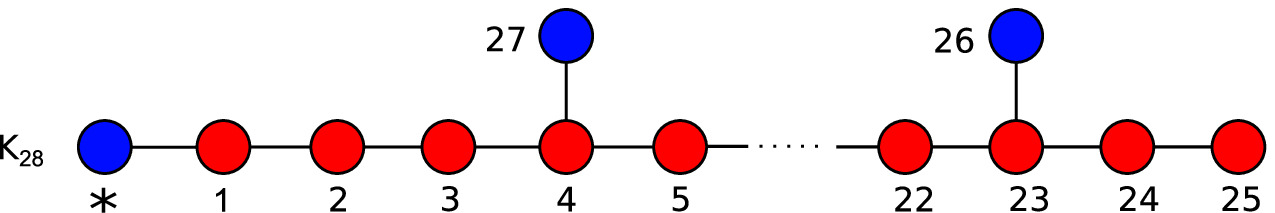}}
\medskip
In this case we have twenty-six $p_i$'s related to the spacetime coordinates, which we can interpret in a similar way as for $E_{11}$, i.e. $p_i=\ln{R_i\over l_p}$, and two more, $p_{27}$ and $p_*$. We would like to make a similar association with energy to the $*$ node, i.e. $p_*={1\over l_p}$, and we also have the dimensionful constant of string length which we associate with $p_{27}$, specifically one might take $p_{27}=({l_s\over l_p})^2$, and see if it leads to sensible and consistent results, but this is beyond the scope of the present work.
\medskip
{\bf 3.2 A tension formula}
\medskip
As we have mentioned the $l_1$ representation is conjectured to be the charge algebra of M-theory. For each brane solution in the $E_{11}$ adjoint algebra there is a corresponding brane charge in the $l_1$ representation. The $l_1$ representation is described equivalently by roots in the $E_{12}$ lattice or weights in the $E_{11}$ lattice. For each extremal half-BPS brane the brane charge is equal to the mass. 
We conjecture the following mass formula for an $E_{12}$ root, $\beta$, 
$$\eqalign{Z_{\beta}&={<\beta|\exp{(q^a\alpha_a {\cdot} H)}|\beta>\over <\beta|\beta>}\cr
&=\exp{(q^a\alpha_a {\cdot} \beta)}}\eqno{(3.5)}$$
Where $H_a\equiv \alpha_a\cdot H$ are generators of the $E_{12}$ Cartan subalgebra. It is the expectation value of the half-BPS brane solution group element of equation (3.1) generalised to the $E_{12}$ lattice. Where the Cartan sub-algebra now has twelve generators. We note that the generators which act as raising and lowering operators on $|\beta>$ are projected out as we take the same bra as ket and all that remains of the group element is the Cartan sub-algebra component.

We note here the effect of Weyl reflecting a root $\beta$ prior to applying the expression in equation (3.5). Recall that the Weyl reflections for a simply laced algebra act on the root lattice as:
$$S_a(\alpha_b)=\alpha_a-2{<\alpha_a,\alpha_b>\over <\alpha_a,\alpha_a>}\alpha_b=\alpha_a-A_{ab}\alpha_b$$
Where $A_{ab}$ is the Cartan matrix. Under this reflection $Z_{S_a(\beta)}$ is:
$$\eqalign{Z_{S_a(\beta)}&={<S_a(\beta)|\exp{(q^b\alpha_b {\cdot} H)}|S_a(\beta) >\over <\beta|\beta>}\cr
&=\exp{(q^b\alpha_b{\cdot} (S_a\beta))} \cr
&=Z_\beta\exp{(-q^bA_{ab}(\alpha_a\cdot \beta))}\cr
&=S_a(Z_\beta)}\eqno{(3.6)}$$
A Weyl reflection of a root in the $E_{12}$ lattice transforms the radii of the compact physical brane solution  and gives a correction to the mass formula for the original brane solution as shown. We will make use of this expression in the IIB scenario to find the change in the mass formula for known solutions which undergo an S-duality transformation.

We may write $\beta$ in terms of the fundamental weights, $\l_i$, of $E_{12}$,
$$\beta=\sum_{i=*,1}^{11} <\beta,\alpha_i>l_i\equiv \sum \hat{n}_il_i$$
Now, we have, 
$$\eqalignno{Z_{\beta}&=\exp{(q^a\alpha_a\cdot\beta)}\cr
&=\exp{(q^a\hat{n}_a)}}$$
The coordinates are now in the Chevalley basis, $H_a$, whose algebraic interpretation we understand but we would prefer to use a different basis, ${K^a}_a$, whose physical interpretation is clear [14]. We change basis to obtain an expression in terms of $p^a$, the coefficients of ${K^a}_a$, so we may interpret this as a mass formula using equations $(3.2)$ and $(3.3)$. Defining the transformation $\rho:{K^a}_a\rightarrow H_a$ and using,
$$H_a={K^a}_a-{K^{a+1}}_{a+1}\quad a={*,1,2,\ldots 10}$$
$$H_{11}=-{1\over 3}({K^*}_*+{K^1}_1+\ldots +{K^8}_8)+{2\over 3}({K^9}_9+{K^{10}}_{10}+{K^{11}}_{11})$$
Then,
$$-3(\rho^{-1})^T=\pmatrix{-2 & 1 & 1 & 1 & 1 & 1 & 1 & 1 & 1 & 1 & 1 & 1\cr 
-1 & -1 & 2 & 2 & 2 & 2 & 2 & 2 & 2 & 2 & 2 & 2\cr
0 & 0 & 0 & 3 & 3 & 3 & 3 & 3 & 3 & 3 & 3 & 3\cr
1 & 1 & 1 & 1 & 4 & 4 & 4 & 4 & 4 & 4 & 4 & 4\cr
2 & 2 & 2 & 2 & 2 & 5 & 5 & 5 & 5 & 5 & 5 & 5\cr
3 & 3 & 3 & 3 & 3 & 3 & 6 & 6 & 6 & 6 & 6 & 6\cr
4 & 4 & 4 & 4 & 4 & 4 & 4 & 7 & 7 & 7 & 7 & 7\cr
5 & 5 & 5 & 5 & 5 & 5 & 5 & 5 & 8 & 8 & 8 & 8\cr
6 & 6 & 6 & 6 & 6 & 6 & 6 & 6 & 6 & 9 & 9 & 9\cr
4 & 4 & 4 & 4 & 4 & 4 & 4 & 4 & 4 & 4 & 7 & 7\cr
2 & 2 & 2 & 2 & 2 & 2 & 2 & 2 & 2 & 2 & 2 & 5\cr
3 & 3 & 3 & 3 & 3 & 3 & 3 & 3 & 3 & 3 & 3 & 3
}
$$
Now,
$$q^a\alpha_a\cdot H = p^a\cdot\rho^{-1}\cdot\rho\cdot{K^a}_a=p^a\cdot\rho^{-1}\cdot H$$ Where we have suppressed the indices on the matrix $\rho$. For example we may read off, $$q_*=-{1\over3}(-2p_*+p_1+p_2+\ldots p_{11})$$
We now have,
$$\eqalign{Z_\beta&=\exp(p^a\cdot \rho^{-1}\cdot \hat{n}_a)\cr
&=\exp(\hat{n}^a\cdot(\rho^{-1})^T\cdot p_a)}\eqno{(3.7)}$$
For convenience we make a further basis transformation and express the fundamental weights of $E_{12}$, $l_i$, in terms of the 12-dimensional vector space basis, $e_i$, that we made use of in section 1,
$$\beta=\sum m^ie_i=\sum \hat{n}^jl_j$$
We observe that the basis transformation $R:l_i\rightarrow e_i$ is $R=(\rho)^T$ and $\hat{n}^i{\rho^{-1}}^T=m^i$, giving a more convenient form for $Z_\beta$ when substituted in (3.7),
$$Z_\beta=\exp(m\cdot p)$$
And making use of equations $(3.2)$ and $(3.4)$, we find a formula that we may interpret as a mass for toroidally compactified dimensions,
$$Z_\beta=({1\over l_p})^{m_*}\prod_{i=1}^{i=11}({R_i \over l_p})^{m_i}\eqno{(3.8)}$$
Where $\beta=\sum_{i=*,1}^{11}m_ie_i$. For the $l_1$ representation $m_*=1$. This formula gives the results found in [6,7,8,9,10] directly from the algebra without the need to work modulo the addition of an orthogonal vector corresponding to constants of the theory. In particular see equation (4.28f) on page 54 of the review [9] where the mass of the Kaluza-Klein mode is found modulo the addition of a vector orthogonal to the fundamental weights of the $E_{11-D}$ algebra. 

We observe, in passing, that in the review [9] there is an orthogonal vector which in the presentation here is played by the $y$ vector used in section 1 - i.e. it represents the part of the $l_1$ representation that is orthogonal to the roots of the $E_n$ algebra. In [9] this orthogonal vector encoded Newton's constant. In the present case $y$ too is a function of Newton's constant. The vector $y=e_*-{1\over2}(e_1+\ldots e_{11})$ has a mass, $Z_y$ where,
$$Z_y={1\over l_p}({l_p^{11}\over R_1\ldots R_{11}})^{1\over 2}={\sqrt{G_{11}}}=\kappa_{11}$$
Where $\kappa_{11}$ is the gravitational constant in eleven dimensions. Thus the orthogonal vector $y$ encodes the gravitational coupling constant, $\kappa_{11}$. 
\medskip
{\bf 3.3 Brane tensions from weights of the $l_1$ representation}
\medskip
Now we apply the mass equation $(3.8)$ to low level weights in the $l_1$ representation which correspond in the various decompositions detailed in section 1 to charges of M-theory, the IIA string theory and the IIB string theory. The process is simplified immensely by deriving the tension in a background where the brane solution is wrapping a torus. That is the charge, which is identified with the mass for the extremal solutions, is contained entirely on the surface of a torus.
\medskip
{\bf 3.3.1 Compactifications of M-theory}
\medskip
As described we compactify each p-brane solution on a p-torus in the eleven dimensional background. In this case the tension is derived from the mass by dividing by the p-torus volume. Let us look at specific examples occurring at low levels in the $l_1$. The relevant decomposition is given explicitly in section 1, and the algebra is split into representations of $A_{10}$ graded by the level $m_{11}$ given in equation (1.1). At level $m_{11}$ we must solve 
$$-A+1+11k=3m_{11}$$
With the exception of the translation generator (for which k=0) we will restrict ourselves to the solutions where $k=\sum p_i$, which corresponds to generators with no blocks of eleven antisymmetrised indices (i.e. trivial volume forms $\epsilon$), in which case we may rearrange this formula to get an expression in terms of the number of indices $\#\equiv \sum(11-i)p_i$,
$$\#=3m_{11}-1$$
At level $m_{11}=0$ the corresponding charge has $-1$ indices, which we may interpret as one contravariant index. The solution has $k=0$, $p_1=1$ and all other $p_i=0$ and corresponds to the charge $P_a$. The corresponding root from equation (1.2) and its mass according to equation (3.8) are
$$\beta_{pp}=e_*-e_1 \quad\quad Z_{\beta_{pp}}={1\over R_1}$$
Where we have singled out $R_1$ as the radius of a compact direction in which the $pp$-wave circulates. This is the mass of a KK-mode, or compactified $pp$-wave solution, in which momentum circulates around the compact direction. The radius $R_1$ has been singled out in this example, but there is a democracy of spacetime coordinates that can be seen under the Weyl reflections of the gravity line. The simple root we have used here is $\alpha_*$ but under the Weyl reflections of $A_{10}$ may be rotated into $\alpha_*+\alpha_1+\ldots +\alpha_i$ where $i\leq 10$, which have mass ${1\over R_i}$.  In what follows the results will be given modulo the $A_{10}$ Weyl reflections and particular radii will be given in the mass formulae but as in this example there are no inherently special spacetime directions particular to the solution.

At level $m_{11}=1$ we find a two index charge, associated to the $M2$ brane charge. Specifically we solve equation (1.1),
$$-A+11k=2$$
This has a solution with $p_9=1$ and all other $p_i=0$ (implying $k=\sum p_i=1$), such that by equation (1.3) $\beta^2=2$. The corresponding root, from equation (1.2), and its mass according to equation (3.8), are,
$$\beta_{M2}=e_*+e_{10}+e_{11}\quad\quad Z_{\beta_{M2}}={R_{10}R_{11}\over l_p^3}$$
We calculate the tension by dividing through by the torus volume that the brane is wrapping, for the $M2$ brane wrapped on a 2-torus in the $x^{10},x^{11}$ directions we divide by $V_2=(2\pi)^2R_{10}R_{11}$ giving,
$$T_{M2}\equiv {Z_{\beta_{M2}} \over V_2}={1\over (2\pi)^2l_p^3}$$
This is the tension [26] of the M2 brane.

At level $m_{11}=2$ we find the charge corresponding to the $M5$ brane, having $p_6=1$ and all other $p_i=0$ such that $\beta^2=2$, giving a root with mass,
$$\beta_{M5}=e_*+e_7+\ldots +e_{11}\quad\quad Z_{\beta_{M5}}={R_7R_8\ldots R_{11}\over l_p^6}$$
Dividing through by the surface area of a 5-torus, $V_5=(2\pi)^5R_7R_8\ldots R_{11}$, we find the tension [19] of the $M5$ brane,
$$T_{M5}\equiv {Z_{\beta_{M5}} \over V_5}={1\over (2\pi)^5l_p^6}$$

At level $m_{11}=3$ we find the charge corresponding to the dual graviton or $KK6$ monopole, having $p_4=p_7=1$ and all other $p_i=0$ such that $\beta^2=2$, giving a root with mass,
$$\beta_{KK6}=e_*+e_5+\ldots +e_{10}+2e_{11}\quad\quad Z_{\beta_{KK6}}={R_5R_6\ldots R_{10}R_{11}^2\over l_p^9}$$
Dividing through by the surface area of a 7-torus, $V_5=(2\pi)^7R_5R_8\ldots R_{11}$, we find the tension,
$$T_{KK6}\equiv {Z_{\beta_{KK6}} \over V_7}={R_{11}\over (2\pi)^7l_p^9}$$
The tension here is sensible only in the compact setting. Upon decompactifying $(R_{11}\rightarrow   \infty)$ the tension diverges. This is an example of the tension of a Kaluza-Klein brane, the higher-dimensional analogue of the Kaluza-Klein monopole. Let us define a KK-brane to be objects related to p-branes by U-duality transformations whose tension is dependent upon the radii of compactification of the background. The $KK6$ has a Taub-NUT fibration in the $R_{11}$ direction, in our notation. We will find that most tensions arising from the charge algebra will diverge when carried over to the non-compact setting and associated to KK-branes carrying a generalisation of the four-dimensional Taub-NUT charge (for a discussion of the higher-dimensional Taub-NUT charge see [22]).

Let us now turn away from specific cases and look at the mass given by the general $E_{12}$ root in the $l_1$ representation of $E_{11}$. By this we mean putting the coefficients of the solution given in equation (1.2) into equation (3.8). We find,
$$\eqalign{Z_\beta &={1\over l_p}\prod_{n=1}^{10}({R_n\over l_p})^{(k-\sum_{i=n}^{10}p_i)}({R_{11}\over l_p})^k \cr
&={ R_1^{k-(p_1+\ldots +p_{10})}R_2^{k-(p_2+\ldots +p_{10})}\ldots R_{10}^{k-p_{10}}R_{11}^k \over l_p^{3m_{11}}}}$$
To find the tension we now divide through by the volume of the relevant compact torus. And now we hit a snag. In our previous tension calculations we knew the charge and hence the brane solution we were considering ab initio. It was therefore clear that for a p-brane solution we could wrap the charge on a p-torus and then divide through by the volume of the p-torus to find the tension. In our approach to a generalised tension formula we do not know the particular solution and neither do we know relevant  p-torus volume to divide the mass formula by. In the general case we expect a monomial in the radii to remain even after we have divided the mass formula by the appropriate p-torus volume. We have already seen an example of this in the case of the KK6 or dual graviton tension, where a single power of $R_{11}$ remained in the expression for the tension.

We shall therefore insert a parameter, $\Omega$, that will keep track of all the remaining radii after the appropriate p-torus volume has been divided out. This parameter is introduced to keep our expressions compact and in any particular solution it will be straightforward to give $\Omega$ explicitly. Our next step is to divide the mass formula by the part of its numerator which corresponds to the unknown p-torus volume with each radius dressed up with $2\pi$. That is we divide by a volume $V_p\sim (2\pi R)^p$, with one $2\pi R_i$ for each radius that occurs, where,
$$\eqalign{V_p\Omega&\equiv(2\pi R_1)^{k-(p_1+\ldots +p_{10})}(2\pi R_2)^{k-(p_2+\ldots +p_{10})}\ldots (2\pi R_{10})^{k-p_{10}}(2\pi R_{11})^k \cr
&=(2\pi)^{3m_{11}-1}R_1^{k-(p_1+\ldots +p_{10})}R_2^{k-(p_2+\ldots +p_{10})}\ldots R_{10}^{k-p_{10}}R_{11}^k }$$
Returning to our general formula and dividing by the volume $V_p$ of the relevant p-torus we find the tension for all solutions of the $l_1$ representation is:
$$\eqalign{T_\beta &={\Omega\over (2\pi)^{3m_{11}-1}l_p^{3m_{11}}} \cr
&={\Omega\over (2\pi)^{\#}l_p^{\#+1}}}$$
Where we have used $\#\equiv -A+11k$, which for the special class of roots with $k=\sum_ip_i$ is the number of indices on the charge.  

We observe now that $\Omega$ does record a property of the solution, for whenever $\Omega$ is not a constant the solution has a divergent tension in the non-compact setting. In this case the tension has a form associated to KKp-branes - objects included by duality arguments into brane charge multiplets in the eleven dimensional theory but really associated to KK-waves and winding modes in lower dimensions. For example in the case of the $KK6$ brane tension considered earlier $\Omega = 2\pi R_{11}$. 
Indeed in [6,8,17,21] non-perturbative sets of higher-dimensional Kaluza-Klein branes have been found and their masses given for M-theory, IIA and IIB theories. The KK-branes of string theory found in [21] constitute a class of solutions derived from the D7 brane and have masses proportional to ${1\over g_s^3}$. Indeed objects whose mass, $Z$, is proportional to ${1\over g_s^n}$ where $n\geq 3$ have a non-vanishing gravitational field strength,$F\propto GZ\propto g_s^2Z$ in the weak coupling limit $(g_s\rightarrow 0)$. States with these masses do not therefore admit an asymptotically flat spacetime [6]. The weak coupling limit is not sensible for these states, and their presence in M-theory remains an unsolved puzzle. We will identify in the $l_1$ representation the states of this sort including those discussed in [21] and for the M-theory, IIA and IIB theories we will find highly non-perturbative masses proportional to arbitrary positive powers of ${1\over l_p}$ for M-theory and ${1\over g_s}$ for the string theories. 

In the $l_1$ representation of $E_{11}$ we find the following roots in the $E_{12}$ root lattice with associated mass as shown,
$$\eqalign{\beta_{WM_7}=e_*+2(e_4+\ldots +e_{10})+3e_{11} &\qquad Z_{WM_7}={(R_4\ldots R_{10})^2R_{11}^3 \over l_p^{18}}\cr
\beta_{M2_6}=e_*+(e_4+e_5)+2(e_6+\ldots e_{11}) &\qquad Z_{M2_6}={R_4R_5(R_6\ldots R_{11})^2 \over l_p^{15}}\cr
\beta_{M5_3}=e_*+(e_4+\ldots +e_8)+2(e_9+\ldots e_{11}) &\qquad Z_{M5_3}={R_4\ldots R_8(R_9\ldots R_{11})^2 \over l_p^{12}}
}$$
These roots have masses which agree with deformations of the fundamental states of M-theory that were found in [21]. These states were argued to exist in the eleven dimensional theory in order to explain the origin of ten-dimensional string theory KK-branes whose existence is inferred from U-duality transformations.  Following the notation of [21] the additional M-theory solutions are labelled $WM_7$, $M2_6$ and $M5_3$. It was recognised [21] that these charges appear in the superalgebra but the attempt to include them in the expansion of the anticommutator, $\{Q,\bar{Q} \}$, were artificial whereas the supercharges arise naturally in the $l_1$ representation of $E_{11}$.

$$\halign{\centerline{#} \cr
\vbox{\offinterlineskip
\halign{\strut \vrule \quad \hfil # \hfil\quad &\vrule \quad \hfil # \hfil\quad 
&\vrule \quad \hfil # \hfil\quad  &\vrule \quad \hfil # \hfil\quad &\vrule \quad \hfil # \hfil\quad &\vrule #
\cr
\noalign{\hrule}
Level&$Mp_i$&Mass&$E_{12}$ Root&Root length&\cr 
($m_{11}$)&&&($e_i$ basis)&squared&\cr 
\noalign{\hrule}
$3$&$M6_1$&${R_5R_6R_7R_8R_9R_{10}R_{11}^2\over l_p^{9}}$&$(1,0,0,0,0,1,1,1,1,1,1,2)$&$2$ & \cr 
\noalign{\hrule}$4$&$M5_3$&${R_4R_5R_6R_7R_8R_9^2R_{10}^2R_{11}^2\over l_p^{12}}$&$(1,0,0,0,1,1,1,1,1,2,2,2)$&$2$ & \cr 
$4$&$M7_2$&${R_3R_4R_5R_6R_7R_8R_9R_{10}^2R_{11}^2\over l_p^{12}}$&$(1,0,0,1,1,1,1,1,1,1,2,2)$&$0$ & \cr 
$4$&$M9_1$&${R_2R_3R_4R_5R_6R_7R_8R_9R_{10}R_{11}^2\over l_p^{12}}$&$(1,0,1,1,1,1,1,1,1,1,1,2)$&$-2$ & \cr 
\noalign{\hrule}$5$&$M8_3$&${R_1R_2R_3R_4R_5R_6R_7R_8R_9^2R_{10}^2R_{11}^2\over l_p^{15}}$&$(1,1,1,1,1,1,1,1,1,2,2,2)$&$-4$ & \cr 
$5$&$M2_6$&${R_4R_5R_6^2R_7^2R_8^2R_9^2R_{10}^2R_{11}^2\over l_p^{15}}$&$(1,0,0,0,1,1,2,2,2,2,2,2)$&$2$ & \cr 
$5$&$M4_5$&${R_3R_4R_5R_6R_7^2R_8^2R_9^2R_{10}^2R_{11}^2\over l_p^{15}}$&$(1,0,0,1,1,1,1,2,2,2,2,2)$&$0$ & \cr 
$5$&$M6_3$&${R_2R_3R_4R_5R_6R_7R_8^2R_9^2R_{10}^2R_{11}^2\over l_p^{15}}$&$(1,0,1,1,1,1,1,1,2,2,2,2)$&$-2$ & \cr 
\noalign{\hrule}$6$&$M5_6$&${R_1R_2R_3R_4R_5R_6^2R_7^2R_8^2R_9^2R_{10}^2R_{11}^2\over l_p^{18}}$&$(1,1,1,1,1,1,2,2,2,2,2,2)$&$-6$ & \cr 
$6$&$M1_8$&${R_3R_4^2R_5^2R_6^2R_7^2R_8^2R_9^2R_{10}^2R_{11}^2\over l_p^{18}}$&$(1,0,0,1,2,2,2,2,2,2,2,2)$&$-2$ & \cr 
$6$&$M3_7$&${R_2R_3R_4R_5^2R_6^2R_7^2R_8^2R_9^2R_{10}^2R_{11}^2\over l_p^{18}}$&$(1,0,1,1,1,2,2,2,2,2,2,2)$&$-4$ & \cr 
\noalign{\hrule}$7$&$M2_9$&${R_1R_2R_3^2R_4^2R_5^2R_6^2R_7^2R_8^2R_9^2R_{10}^2R_{11}^2\over l_p^{21}}$&$(1,1,1,2,2,2,2,2,2,2,2,2)$&$-10$ & \cr 
$7$&$M0_{10}$&${R_2^2R_3^2R_4^2R_5^2R_6^2R_7^2R_8^2R_9^2R_{10}^2R_{11}^2\over l_p^{21}}$&$(1,0,2,2,2,2,2,2,2,2,2,2)$&$-8$ & \cr 
\noalign{\hrule}
}
}\cr
Table 3.1 The full set of M-theory $Mp_i$ branes from the $l_1$ representation of $E_{11}$ \cr}$$

We emphasise that these are a small set of all the KK-branes of M-theory the full set of which is contained in the $l_1$ representation of $E_{11}$. Indeed it seems the role of the variable $k$ used in section 1.1 in the decomposition of the $l_1$ representation classifies the different types of KK-branes that appear. From equation (1.2), or from the expression for the mass, $Z_\beta$, one can see that the dependence on the eleventh radius $R_{11}$ is controlled by $k$. The KK-branes have masses with powers of the radii greater than one, and since $k$ is also the greatest radial power appearing in the mass formula one could use $k$ to classify finite sets of KK-branes. For example in M-theory one could look for all the KK-branes with masses quadratic in the radii by listing the roots with $k=2$. From table A.1, one finds the following, given explicitly in table 3.1: $M6_1, M5_3, M7_2, M9_1, M8_3, M2_6, M4_5, M6_4, M5_6, M1_8, M3_7, M2_9$ where the script number refers to the number of linear radii and the subscript to the number of radii which are squared in the mass formula.
\medskip 
{\bf 3.3.2 IIA supergravity}
\medskip 
To study the ten-dimensional IIA theory the $l_1$ algebra is decomposed into representations of the $A_9$ subalgebra formed from the roots numbered 1 to 9, in the $E_{12}$ Dynkin diagrams shown in the introduction. The decomposition was carried out in section 1 and a general root given in equation (1.4). We now find the formula for the mass this general root as given by equation (3.8),
$$\eqalign{Z_\beta &={1\over l_p}\prod_{n=1}^{9}({R_n\over l_p})^{(k-\sum_{i=n}^{9}p_i)}({R_{10}\over l_p})^k({R_{11}\over l_p})^{(m_{11}-m_{10})} \cr
&={{ R_1^{k-(p_1+\ldots +p_{9})}R_2^{k-(p_2+\ldots +p_{9})}\ldots R_{9}^{k-p_{9}}R_{10}^kR_{11}^{{m_{11}-m_{10}}}} \over l_p^{3m_{11}}}}$$
To find a convenient notation we divide through by the appropriate volume $V_p$ as before to remove the radial dependencies, i.e. we divide by $$V_p\equiv{(2\pi)^{10k-A}\over \Omega}R_1^{k-(p_1+\ldots +p_{9})}R_2^{k-(p_2+\ldots +p_{9})}\ldots R_{9}^{k-p_{9}}R_{10}^k$$ giving, 
$$\eqalign{T_\beta &={\Omega R_{11}^{m_{11}-m_{10}}\over (2\pi)^{10k-A}l_p^{3m_{11}}} \cr
&={\Omega\over (2\pi)^{\#}g_s^{m_{10}}l_s^{\#+1}}}$$
Where we have made use of the identities $R_{11}=g_sl_s$ and $l_p^3=g_sl_s^3$. As before we have defined $\#\equiv 10k-A$ and in the case where $k=\sum_i p_i$, $\#$ is the number of indices on the corresponding generator in the algebra.
We now may pose the question: which roots in the algebra have tensions with a single string coupling constant in them. Immediately we see we are looking for the roots such that $m_{10}=1$. Now $\#=2m_{11}+m_{10}-1=2m_{11}$. Therefore we note that the charges we will find matching this criteria will have an even number of indices if they exist in the algebra. Specifically $m_{10}=1$ gives $q=-k+1$ and so $2m_{11}=-A+10k$. The cases where $k=\sum_ip_i$ and where only a single $p_i$ is nonzero and equal to one we have $k=1$ and we must solve $2m_{11}=10-A\geq 0$ which has solutions for $A=0,2,4,6,8$ corresponding to the D10, D8, D6, D4, D2 brane solutions. For each Dp-brane we have $p_{(10-p)}=1$ with all other $p_i=0$. Therefore for these cases, we have $\#=p$ where $p$ is even and less than ten, we find,
$$T_{\beta_{Dp}} ={1\over (2\pi)^pg_sl_s^{p+1}}$$
Which is the tension formula for the RR D-brane charges of the IIA theory. 
The D0-brane charge occurs when all $p_i=0$ so that $\#=0$, consequently,
$$\beta_{D0}=e_*-e_{11} \quad\quad T_{\beta_{D0}}={1\over g_sl_s}\equiv{1\over g_s\sqrt{\alpha'}}$$
Note that the result is precisely given by by the tension formula from $\beta_{D0}$ and not modulo an additional orthogonal vector as was the case for the KK mode of [9]. 
The fundamental string charge is listed at level $(1,1)$ in table A.3 in the appendix - it's highest weight generator is associated to the root $\beta_{F1}=e_*-e_{11}$ and its tension is:
$$T_{\beta_{F1}}={1\over 2\pi l_s^2}$$
One can also identify the tension of the $NS5$ brane, whose charge is associated to the root $\beta_{NS5}=e_*+e_6+e_7+e_8+e_9+e_{11}$ appearing at level $(3,1)$ in table A3. The corresponding tension is:
$$T_{\beta_{NS5}}={1\over {(2\pi)}^5g_s^2 l_s^6}$$
Amongst the weights in the $l_1$ representation of $E_{11}$ one can use the tension formula to identify the roots corresponding to the KK-branes of the IIA theory given in [21], namely the $D0_7, D2_5, D4_3$ and $D6_1$ which are all derived from the D7 brane by U-duality transformations. Using table A.2 in the appendix we may identify:
$$\eqalign{\beta_{D0_7}=e_*+2(e_4+\ldots +e_{10})+3e_{11} &\qquad Z_{D0_7}={(R_4\ldots R_{10})^2 \over g_s^3l_s^{15}}\cr
\beta_{D2_5}=e_*+(e_4+e_5)+2(e_6+\ldots e_{11}) &\qquad Z_{D2_5}={R_4R_5(R_6\ldots R_{10})^2 \over g_s^3l_s^{13}}\cr
\beta_{D4_3}=e_*+(e_4+\ldots +e_7)+2(e_8+\ldots e_{10})+e_{11} &\qquad Z_{D4_3}={R_4\ldots R_7(R_8\ldots R_{10})^2 \over g_s^3l_s^{11}}\cr
\beta_{D6_1}=e_*+(e_4+\ldots +e_9)+2e_{10} &\qquad Z_{D6_1}={R_4\ldots R_9(R_{10})^2 \over g_s^3l_s^9}
}$$
One can use the parameter $k$ to classify the KK-brane solutions; in the IIA decomposition $k$ is the coefficient of $e_{10}$. The full set of $Dp_i$ branes, where $i$ indicates the number of directions with a Taub-NUT fibration, corresponds to all the weights in the $l_1$ representation having $k=2$, shown in table 3.2. 

$$\halign{\centerline{#} \cr
\vbox{\offinterlineskip
\halign{\strut \vrule \quad \hfil # \hfil\quad &\vrule \quad \hfil # \hfil\quad 
&\vrule \quad \hfil # \hfil\quad  &\vrule \quad \hfil # \hfil\quad &\vrule \quad \hfil # \hfil\quad &\vrule #
\cr
\noalign{\hrule}
Level&$Dp_i$&Mass&$E_{12}$ Root&Root length&\cr 
($m_{10},m_{11}$)&&&($e_i$ basis)&squared&\cr 
\noalign{\hrule}$(2,3)$&$D5_1$&${R_{5}R_{6}R_{7}R_{8}R_{9}R_{10}^{2}\over g_s^{2}l_s^{8}}$&$(1,0,0,0,0,1,1,1,1,1,2,1)$&$2$ & \cr 
\noalign{\hrule}$(3,3)$&$D6_1$&${R_{4}R_{5}R_{6}R_{7}R_{8}R_{9}R_{10}^{2}\over g_s^{3}l_s^{9}}$&$(1,0,0,0,1,1,1,1,1,1,2,0)$&$2$ & \cr 
\noalign{\hrule}$(2,4)$&$D5_2$&${R_{4}R_{5}R_{6}R_{7}R_{8}R_{9}^{2}R_{10}^{2}\over g_s^{2}l_s^{10}}$&$(1,0,0,0,1,1,1,1,1,2,2,2)$&$2$ & \cr 
$(2,4)$&$D7_1$&${R_{3}R_{4}R_{5}R_{6}R_{7}R_{8}R_{9}R_{10}^{2}\over g_s^{2}l_s^{10}}$&$(1,0,0,1,1,1,1,1,1,1,2,2)$&$0$ & \cr 
\noalign{\hrule}$(3,4)$&$D4_3$&${R_{4}R_{5}R_{6}R_{7}R_{8}^{2}R_{9}^{2}R_{10}^{2}\over g_s^{3}l_s^{11}}$&$(1,0,0,0,1,1,1,1,2,2,2,1)$&$2$ & \cr 
$(3,4)$&$D6_2$&${R_{3}R_{4}R_{5}R_{6}R_{7}R_{8}R_{9}^{2}R_{10}^{2}\over g_s^{3}l_s^{11}}$&$(1,0,0,1,1,1,1,1,1,2,2,1)$&$0$ & \cr 
$(3,4)$&$D8_1$&${R_{2}R_{3}R_{4}R_{5}R_{6}R_{7}R_{8}R_{9}R_{10}^{2}\over g_s^{3}l_s^{11}}$&$(1,0,1,1,1,1,1,1,1,1,2,1)$&$-2$ & \cr 
\noalign{\hrule} $(4,4)$&$D9_1$&${R_{1}R_{2}R_{3}R_{4}R_{5}R_{6}R_{7}R_{8}R_{9}R_{10}^{2}\over g_s^{4}l_s^{12}}$&$(1,1,1,1,1,1,1,1,1,1,2,0)$&$-2$ & \cr 
$(4,4)$&$D5_3$&${R_{3}R_{4}R_{5}R_{6}R_{7}R_{8}^{2}R_{9}^{2}R_{10}^{2}\over g_s^{4}l_s^{12}}$&$(1,0,0,1,1,1,1,1,2,2,2,0)$&$2$ & \cr 
$(4,4)$&$D7_2$&${R_{2}R_{3}R_{4}R_{5}R_{6}R_{7}R_{8}R_{9}^{2}R_{10}^{2}\over g_s^{4}l_s^{12}}$&$(1,0,1,1,1,1,1,1,1,2,2,0)$&$0$ & \cr 
\noalign{\hrule}$(5,4)$&$D8_2$&${R_{1}R_{2}R_{3}R_{4}R_{5}R_{6}R_{7}R_{8}R_{9}^{2}R_{10}^{2}\over g_s^{5}l_s^{13}}$&$(1,1,1,1,1,1,1,1,1,2,2,-1)$&$2$ & \cr 
\noalign{\hrule}$(2,5)$&$D9_1$&${R_{1}R_{2}R_{3}R_{4}R_{5}R_{6}R_{7}R_{8}R_{9}R_{10}^{2}\over g_s^{2}l_s^{12}}$&$(1,1,1,1,1,1,1,1,1,1,2,3)$&$-2$ & \cr 
$(2,5)$&$D5_3$&${R_{3}R_{4}R_{5}R_{6}R_{7}R_{8}^{2}R_{9}^{2}R_{10}^{2}\over g_s^{2}l_s^{12}}$&$(1,0,0,1,1,1,1,1,2,2,2,3)$&$2$ & \cr 
$(2,5)$&$D7_2$&${R_{2}R_{3}R_{4}R_{5}R_{6}R_{7}R_{8}R_{9}^{2}R_{10}^{2}\over g_s^{2}l_s^{12}}$&$(1,0,1,1,1,1,1,1,1,2,2,3)$&$0$ & \cr 
\noalign{\hrule}$(3,5)$&$D8_2$&${R_{1}R_{2}R_{3}R_{4}R_{5}R_{6}R_{7}R_{8}R_{9}^{2}R_{10}^{2}\over g_s^{3}l_s^{13}}$&$(1,1,1,1,1,1,1,1,1,2,2,2)$&$-4$ & \cr 
$(3,5)$&$D2_5$&${R_{4}R_{5}R_{6}^{2}R_{7}^{2}R_{8}^{2}R_{9}^{2}R_{10}^{2}\over g_s^{3}l_s^{13}}$&$(1,0,0,0,1,1,2,2,2,2,2,2)$&$2$ & \cr 
$(3,5)$&$D4_4$&${R_{3}R_{4}R_{5}R_{6}R_{7}^{2}R_{8}^{2}R_{9}^{2}R_{10}^{2}\over g_s^{3}l_s^{13}}$&$(1,0,0,1,1,1,1,2,2,2,2,2)$&$0$ & \cr 
$(3,5)$&$D6_3$&${R_{2}R_{3}R_{4}R_{5}R_{6}R_{7}R_{8}^{2}R_{9}^{2}R_{10}^{2}\over g_s^{3}l_s^{13}}$&$(1,0,1,1,1,1,1,1,2,2,2,2)$&$-2$ & \cr 
\noalign{\hrule}$(4,5)$&$D7_3$&${R_{1}R_{2}R_{3}R_{4}R_{5}R_{6}R_{7}R_{8}^{2}R_{9}^{2}R_{10}^{2}\over g_s^{4}l_s^{14}}$&$(1,1,1,1,1,1,1,1,2,2,2,1)$&$-4$ & \cr 
$(4,5)$&$D1_6$&${R_{4}R_{5}^{2}R_{6}^{2}R_{7}^{2}R_{8}^{2}R_{9}^{2}R_{10}^{2}\over g_s^{4}l_s^{14}}$&$(1,0,0,0,1,2,2,2,2,2,2,1)$&$2$ & \cr 
$(4,5)$&$D3_5$&${R_{3}R_{4}R_{5}R_{6}^{2}R_{7}^{2}R_{8}^{2}R_{9}^{2}R_{10}^{2}\over g_s^{4}l_s^{14}}$&$(1,0,0,1,1,1,2,2,2,2,2,1)$&$0$ & \cr 
$(4,5)$&$D5_4$&${R_{2}R_{3}R_{4}R_{5}R_{6}R_{7}^{2}R_{8}^{2}R_{9}^{2}R_{10}^{2}\over g_s^{4}l_s^{14}}$&$(1,0,1,1,1,1,1,2,2,2,2,1)$&$-2$ & \cr 
\noalign{\hrule}$(5,5)$&$D6_4$&${R_{1}R_{2}R_{3}R_{4}R_{5}R_{6}R_{7}^{2}R_{8}^{2}R_{9}^{2}R_{10}^{2}\over g_s^{5}l_s^{15}}$&$(1,1,1,1,1,1,1,2,2,2,2,0)$&$-2$ & \cr 
$(5,5)$&$D2_6$&${R_{3}R_{4}R_{5}^{2}R_{6}^{2}R_{7}^{2}R_{8}^{2}R_{9}^{2}R_{10}^{2}\over g_s^{5}l_s^{15}}$&$(1,0,0,1,1,2,2,2,2,2,2,0)$&$2$ & \cr 
$(5,5)$&$D4_5$&${R_{2}R_{3}R_{4}R_{5}R_{6}^{2}R_{7}^{2}R_{8}^{2}R_{9}^{2}R_{10}^{2}\over g_s^{5}l_s^{15}}$&$(1,0,1,1,1,1,2,2,2,2,2,0)$&$0$ & \cr 
\noalign{\hrule}$(6,5)$&$D5_5$&${R_{1}R_{2}R_{3}R_{4}R_{5}R_{6}^{2}R_{7}^{2}R_{8}^{2}R_{9}^{2}R_{10}^{2}\over g_s^{6}l_s^{16}}$&$(1,1,1,1,1,1,2,2,2,2,2,-1)$&$2$ & \cr 
\noalign{\hrule}
$(2,6)$&$D7_3$&${R_{1}R_{2}R_{3}R_{4}R_{5}R_{6}R_{7}R_{8}^{2}R_{9}^{2}R_{10}^{2}\over g_s^{2}l_s^{14}}$&$(1,1,1,1,1,1,1,1,2,2,2,4)$&$0$ & \cr 
$(2,6)$&$D5_4$&${R_{2}R_{3}R_{4}R_{5}R_{6}R_{7}^{2}R_{8}^{2}R_{9}^{2}R_{10}^{2}\over g_s^{2}l_s^{14}}$&$(1,0,1,1,1,1,1,2,2,2,2,4)$&$2$ & \cr 
\noalign{\hrule}$(3,6)$&$D6_4$&${R_{1}R_{2}R_{3}R_{4}R_{5}R_{6}R_{7}^{2}R_{8}^{2}R_{9}^{2}R_{10}^{2}\over g_s^{3}l_s^{15}}$&$(1,1,1,1,1,1,1,2,2,2,2,3)$&$-4$ & \cr 
$(3,6)$&$D0_7$&${R_{4}^{2}R_{5}^{2}R_{6}^{2}R_{7}^{2}R_{8}^{2}R_{9}^{2}R_{10}^{2}\over g_s^{3}l_s^{15}}$&$(1,0,0,0,2,2,2,2,2,2,2,3)$&$2$ & \cr 
$(3,6)$&$D2_6$&${R_{3}R_{4}R_{5}^{2}R_{6}^{2}R_{7}^{2}R_{8}^{2}R_{9}^{2}R_{10}^{2}\over g_s^{3}l_s^{15}}$&$(1,0,0,1,1,2,2,2,2,2,2,3)$&$0$ & \cr 
$(3,6)$&$D4_5$&${R_{2}R_{3}R_{4}R_{5}R_{6}^{2}R_{7}^{2}R_{8}^{2}R_{9}^{2}R_{10}^{2}\over g_s^{3}l_s^{15}}$&$(1,0,1,1,1,1,2,2,2,2,2,3)$&$-2$ & \cr
\noalign{\hrule}$(4,6)$&$D4_6$&${R_{1}R_{2}R_{3}R_{4}R_{5}R_{6}^{2}R_{7}^{2}R_{8}^{2}R_{9}^{2}R_{10}^{2}\over g_s^{4}l_s^{16}}$&$(1,1,1,1,1,1,2,2,2,2,2,2)$&$-6$ & \cr 
$(4,6)$&$D1_7$&${R_{3}R_{4}^{2}R_{5}^{2}R_{6}^{2}R_{7}^{2}R_{8}^{2}R_{9}^{2}R_{10}^{2}\over g_s^{4}l_s^{16}}$&$(1,0,0,1,2,2,2,2,2,2,2,2)$&$-2$ & \cr 
$(4,6)$&$D3_6$&${R_{2}R_{3}R_{4}R_{5}^{2}R_{6}^{2}R_{7}^{2}R_{8}^{2}R_{9}^{2}R_{10}^{2}\over g_s^{4}l_s^{16}}$&$(1,0,1,1,1,2,2,2,2,2,2,2)$&$-4$ & \cr 
\noalign{\hrule}
}
}\cr
\cr}$$
$$\halign{\centerline{#} \cr
\vbox{\offinterlineskip
\halign{\strut \vrule \quad \hfil # \hfil\quad &\vrule \quad \hfil # \hfil\quad 
&\vrule \quad \hfil # \hfil\quad  &\vrule \quad \hfil # \hfil\quad &\vrule \quad \hfil # \hfil\quad &\vrule #
\cr
\noalign{\hrule}
Level&$Dp_i$&Mass&$E_{12}$ Root&Root length&\cr 
($m_{10},m_{11}$)&&&($e_i$ basis)&squared&\cr
\noalign{\hrule}$(5,6)$&$D4_6$&${R_{1}R_{2}R_{3}R_{4}R_{5}^{2}R_{6}^{2}R_{7}^{2}R_{8}^{2}R_{9}^{2}R_{10}^{2}\over g_s^{5}l_s^{17}}$&$(1,1,1,1,1,2,2,2,2,2,2,1)$&$-6$ & \cr 
$(5,6)$&$D0_8$&${R_{3}^{2}R_{4}^{2}R_{5}^{2}R_{6}^{2}R_{7}^{2}R_{8}^{2}R_{9}^{2}R_{10}^{2}\over g_s^{5}l_s^{17}}$&$(1,0,0,2,2,2,2,2,2,2,2,1)$&$-2$ & \cr 
$(5,6)$&$D2_7$&${R_{2}R_{3}R_{4}^{2}R_{5}^{2}R_{6}^{2}R_{7}^{2}R_{8}^{2}R_{9}^{2}R_{10}^{2}\over g_s^{5}l_s^{17}}$&$(1,0,1,1,2,2,2,2,2,2,2,1)$&$-4$ & \cr 
\noalign{\hrule}$(6,6)$&$D3_7$&${R_{1}R_{2}R_{3}R_{4}^{2}R_{5}^{2}R_{6}^{2}R_{7}^{2}R_{8}^{2}R_{9}^{2}R_{10}^{2}\over g_s^{6}l_s^{18}}$&$(1,1,1,1,2,2,2,2,2,2,2,0)$&$-4$ & \cr 
$(6,6)$&$D1_8$&${R_{2}R_{3}^{2}R_{4}^{2}R_{5}^{2}R_{6}^{2}R_{7}^{2}R_{8}^{2}R_{9}^{2}R_{10}^{2}\over g_s^{6}l_s^{18}}$&$(1,0,1,2,2,2,2,2,2,2,2,0)$&$-2$ & \cr 
\noalign{\hrule}
$(7,6)$&$D2_8$&${R_{1}R_{2}R_{3}^{2}R_{4}^{2}R_{5}^{2}R_{6}^{2}R_{7}^{2}R_{8}^{2}R_{9}^{2}R_{10}^{2}\over g_s^{7}l_s^{19}}$&$(1,1,1,2,2,2,2,2,2,2,2,-1)$&$0$ & \cr 
$(7,6)$&$D0_9$&${R_{2}^{2}R_{3}^{2}R_{4}^{2}R_{5}^{2}R_{6}^{2}R_{7}^{2}R_{8}^{2}R_{9}^{2}R_{10}^{2}\over g_s^{7}l_s^{19}}$&$(1,0,2,2,2,2,2,2,2,2,2,-1)$&$2$ & \cr 
\noalign{\hrule}$(2,7)$&$D5_5$&${R_{1}R_{2}R_{3}R_{4}R_{5}R_{6}^{2}R_{7}^{2}R_{8}^{2}R_{9}^{2}R_{10}^{2}\over g_s^{2}l_s^{16}}$&$(1,1,1,1,1,1,2,2,2,2,2,5)$&$2$ & \cr 
\noalign{\hrule}$(3,7)$&$D4_6$&${R_{1}R_{2}R_{3}R_{4}R_{5}^{2}R_{6}^{2}R_{7}^{2}R_{8}^{2}R_{9}^{2}R_{10}^{2}\over g_s^{3}l_s^{17}}$&$(1,1,1,1,1,2,2,2,2,2,2,4)$&$-4$ & \cr 
$(3,7)$&$D0_8$&${R_{3}^{2}R_{4}^{2}R_{5}^{2}R_{6}^{2}R_{7}^{2}R_{8}^{2}R_{9}^{2}R_{10}^{2}\over g_s^{3}l_s^{17}}$&$(1,0,0,2,2,2,2,2,2,2,2,4)$&$0$ & \cr 
$(3,7)$&$D2_7$&${R_{2}R_{3}R_{4}^{2}R_{5}^{2}R_{6}^{2}R_{7}^{2}R_{8}^{2}R_{9}^{2}R_{10}^{2}\over g_s^{3}l_s^{17}}$&$(1,0,1,1,2,2,2,2,2,2,2,4)$&$-2$ & \cr 
\noalign{\hrule}$(4,7)$&$D3_7$&${R_{1}R_{2}R_{3}R_{4}^{2}R_{5}^{2}R_{6}^{2}R_{7}^{2}R_{8}^{2}R_{9}^{2}R_{10}^{2}\over g_s^{4}l_s^{18}}$&$(1,1,1,1,2,2,2,2,2,2,2,3)$&$-8$ & \cr 
$(4,7)$&$D1_8$&${R_{2}R_{3}^{2}R_{4}^{2}R_{5}^{2}R_{6}^{2}R_{7}^{2}R_{8}^{2}R_{9}^{2}R_{10}^{2}\over g_s^{4}l_s^{18}}$&$(1,0,1,2,2,2,2,2,2,2,2,3)$&$-6$ & \cr 
\noalign{\hrule}$(5,7)$&$D2_8$&${R_{1}R_{2}R_{3}^{2}R_{4}^{2}R_{5}^{2}R_{6}^{2}R_{7}^{2}R_{8}^{2}R_{9}^{2}R_{10}^{2}\over g_s^{5}l_s^{19}}$&$(1,1,1,2,2,2,2,2,2,2,2,2)$&$-10$ & \cr 
$(5,7)$&$D0_9$&${R_{2}^{2}R_{3}^{2}R_{4}^{2}R_{5}^{2}R_{6}^{2}R_{7}^{2}R_{8}^{2}R_{9}^{2}R_{10}^{2}\over g_s^{5}l_s^{19}}$&$(1,0,2,2,2,2,2,2,2,2,2,2)$&$-8$ & \cr 
\noalign{\hrule}$(6,7)$&$D1_9$&${R_{1}R_{2}^{2}R_{3}^{2}R_{4}^{2}R_{5}^{2}R_{6}^{2}R_{7}^{2}R_{8}^{2}R_{9}^{2}R_{10}^{2}\over g_s^{6}l_s^{20}}$&$(1,1,2,2,2,2,2,2,2,2,2,1)$&$-10$ & \cr 
\noalign{\hrule}$(7,7)$&$D0_{10}$&${R_{1}^{2}R_{2}^{2}R_{3}^{2}R_{4}^{2}R_{5}^{2}R_{6}^{2}R_{7}^{2}R_{8}^{2}R_{9}^{2}R_{10}^{2}\over g_s^{7}l_s^{21}}$&$(1,2,2,2,2,2,2,2,2,2,2,0)$&$-8$ & \cr 
\noalign{\hrule}$(3,8)$&$D2_8$&${R_{1}R_{2}R_{3}^{2}R_{4}^{2}R_{5}^{2}R_{6}^{2}R_{7}^{2}R_{8}^{2}R_{9}^{2}R_{10}^{2}\over g_s^{3}l_s^{19}}$&$(1,1,1,2,2,2,2,2,2,2,2,5)$&$-4$ & \cr 
$(3,8)$&$D0_9$&${R_{2}^{2}R_{3}^{2}R_{4}^{2}R_{5}^{2}R_{6}^{2}R_{7}^{2}R_{8}^{2}R_{9}^{2}R_{10}^{2}\over g_s^{3}l_s^{19}}$&$(1,0,2,2,2,2,2,2,2,2,2,5)$&$-2$ & \cr 
\noalign{\hrule}$(4,8)$&$D1_9$&${R_{1}R_{2}^{2}R_{3}^{2}R_{4}^{2}R_{5}^{2}R_{6}^{2}R_{7}^{2}R_{8}^{2}R_{9}^{2}R_{10}^{2}\over g_s^{4}l_s^{20}}$&$(1,1,2,2,2,2,2,2,2,2,2,4)$&$-10$ & \cr 
\noalign{\hrule}$(5,8)$&$D0_{10}$&${R_{1}^{2}R_{2}^{2}R_{3}^{2}R_{4}^{2}R_{5}^{2}R_{6}^{2}R_{7}^{2}R_{8}^{2}R_{9}^{2}R_{10}^{2}\over g_s^{5}l_s^{21}}$&$(1,2,2,2,2,2,2,2,2,2,2,3)$&$-14$ & \cr 
\noalign{\hrule}$(3,9)$&$D0_{10}$&${R_{1}^{2}R_{2}^{2}R_{3}^{2}R_{4}^{2}R_{5}^{2}R_{6}^{2}R_{7}^{2}R_{8}^{2}R_{9}^{2}R_{10}^{2}\over g_s^{3}l_s^{21}}$&$(1,2,2,2,2,2,2,2,2,2,2,6)$&$-4$ & \cr 
\noalign{\hrule}
}
}\cr
Table 3.2 The full set of IIA $Dp_i$ branes from the $l_1$ representation of $E_{11}$ \cr}$$

\medskip 
{\bf 3.3.3 IIB supergravity}
\medskip 
We repeat the process applied to the IIA case but making use of the IIB variables. The decomposition was carried out in section 1 and a general root given in equation (1.5) which we use with equation (3.3) to find the mass,
$$\eqalign{Z_\beta &={1\over l_p}\prod_{n=1}^{9}({R_n\over l_p})^{(k-\sum_{i=n}^{8,11}p_i)}({R_{10}\over l_p})^{l-k}({R_{11}\over l_p})^{(m_9-k-l)} \cr
&={{ R_1^{k-(p_1+\ldots +p_{8}+p_{11})}R_2^{k-(p_2+\ldots +p_{8}+p_{11})}\ldots R_{9}^{k-p_{11}}R_{10}^{l-k}R_{11}^{m_9-k-l}} \over l_p^{3m_{11}}}}$$
To find a convenient notation for the tension we divide through by the volume to remove the radial dependencies, i.e. we divide by 
$$V_p\equiv{(2\pi)^{(10k-A)}\over \Omega}R_1^{k-(p_1+\ldots +p_{8}+p_{11})}R_2^{k-(p_2+\ldots +p_{8}+p_{11})}\ldots R_{9}^{k-p_{11}}\hat{R}_{10}^k$$ This contains the volume of the compact spacetime torus as a factor and as before we record remaining powers of $2\pi R$ in a factor, $\Omega$. We have used the notation $\hat{R}_{10}\equiv {\l_p^3\over R_{10}R_{11}}$ [15] so that, 
$$T_\beta ={\Omega \over (2\pi)^{2m_9-1}\hat{g}_s^ll_s^{2m_9}}$$ Where we have used the IIB parameters $l_s^2={l_p^3\over R_{11}}$ and $\hat{g}_s={R_{11}\over R_{10}}$. If, as before, we define $\#\equiv 10k-A=2m_9-1$ then we have,
$$T_\beta ={\Omega \over (2\pi)^{\#}\hat{g}_s^ll_s^{\#+1}}$$
As before, in the case where $k=\sum_i p_i$, $\#$ is the number of indices on the corresponding generator in the algebra. We now may single out the tensions with a single $\hat{g}^s$ appearing.
In this case we are looking for roots with $l=m_{10}=1$. Now $\#=2m_9-1$. Therefore we note that the charges we will find matching this criteria will have an odd number of indices if they exist in the algebra. Specifically where $k=\sum_ip_i$ and where only a single $p_i$ is nonzero and equal to one we have $k=1$ and we must solve $2m_{9}=11-A\geq 0$ which has solutions for $A=1,3,5,7,9$ corresponding to the D9, D7, D5, D3, D1 brane solutions. Excluding the D1 brane we have for each Dp-brane, $p_{(10-p)}=1$ with all other $p_i=0$. For the D1 brane we have $p_{11}=1$ and all other $p_i=0$. Therefore for these cases, we have $\#=p$ where $p$ is odd and less than ten, we find,
$$T_{\beta_{Dp}} ={1\over (2\pi)^pg_sl_s^{p+1}}$$
Which is the tension formula for the RR D-brane charges of the IIB theory. Furthermore we may apply the S-duality transformation to the IIB theory which corresponds to the Weyl reflection in the plane perpendicular to the $\alpha_{10}$ root vector:
$$S_{10}\beta=\beta-(2l-m_9)\alpha_{10}$$
That is $l\rightarrow m_9-l$, leaving $\beta^2$ in equation (1.6) unaltered, as expected, but altering the tension formula to:
$$T'_{\beta_{Dp}} ={1\over (2\pi)^pg_s^{m_9-1}l_s^{p+1}}$$
This formula could also be obtained using equation (3.6) for the Weyl reflection of the mass formula. Explicitly we have,
$$Z'_{\beta_{Dp}}\equiv Z_{S_{10}(\beta_{D_p})}=Z_{\beta_{D_p}}\exp{(-(-q^9+2q^{10})(\alpha_{10}\cdot \beta_{D_p}))}$$
Using the parameters of solution given in section 1.3 we have,
$$\alpha_{10}\cdot \beta =-m_9+2m_{10}=-q$$
We recall that in general $q$ is an integer taking values less than or equal to the coefficient, $m_9$, but in the particular set of solutions we are considering here $q=m_9-2$. Let us translate the $-q^9+2q^{10}$ into the coordinates $p^a$, using the transformation matrix $(\rho^{-1})^{T}$ from section 3.2 we read off,
$$\eqalign{q^9&=-{1\over 3}(4(p_*+p_1+\ldots p_9)+7(p_{10}+p_{11}))\cr
q^{10}&=-{1\over 3}(2(p_*+p_1+\ldots p_{10})+5p_{11}))}$$
So that,
$$-q^9+2q^{10}=p_{10}-p_{11}=\ln({R_{10}\over R_{11}})=\ln{(1\over  g_s)}$$
Substituting back into our expression for $Z_{S_{10}(\beta)}$ an dividing by $V_p\sim (2\pi R)^p$ to find the tension we obtain,
$$T_{S_{10}(\beta_{D_p})}=T_{\beta_{D_p}}({1\over  g_s})^q={1\over (2\pi)^pg_s^{1+q}l_s^{p+1}}$$
In the case we have been concerned with the parameters take specific values, $l=1$ and so $q=m_9-2$ and the expression is identical to the S-dual tension derived previously.

We may use this form of the tension to write down the tensions of the S-dual states to the $Dp$-branes of the IIB theory, explicitly the $F1$ string, the $NS5$ brane, the $S7$ brane and the $S9$ brane. Notice that the tension of the D3 brane is mapped to itself by S-duality. Explicitly we find, 
$$\eqalignno{T_{\beta_{F1}}&={1\over 2\pi l_s^2} \cr
T_{\beta_{NS5}}&={1\over {(2\pi)}^5g_s^2l_s^6} \cr
T_{\beta_{S7}}&={1\over {(2\pi)}^7g_s^3l_s^8} \cr
T_{\beta_{S9}}&={1\over {(2\pi)}^9g_s^4l_s^{10}}}$$

In addition there are two further 9-brane states making up the charge $Z^{a_1\ldots a_9(\alpha\beta\gamma)}$, these are S-dual to each other and have tensions:
$$\eqalignno{T_{\beta_{9}}&={1\over {(2\pi)}^9 g_s^2l_s^{10}} \cr
T_{\beta_{9}}&={1\over {(2\pi)}^9 g_s^3l_s^{10}}}$$

Let us also identify the weights in the $l_1$ representation corresponding to the KK-branes of the IIB theory given in [21], namely the $D1_6, D3_4$ and $D5_2$ which are all derived from the D7 brane by U-duality transformations. From table A.3 in the appendix we may read off:
$$\eqalign{\beta_{D1_6}=e_*+2(e_4+\ldots +e_{10})+3e_{11} &\qquad Z_{D0_7}={(R_4\ldots R_{10})^2 \over g_s^3l_s^{15}}\cr
\beta_{D3_4}=e_*+(e_4+e_5)+2(e_6+\ldots e_{11}) &\qquad Z_{D2_5}={R_4R_5(R_6\ldots R_{10})^2 \over g_s^3l_s^{13}}\cr
\beta_{D5_2}=e_*+(e_4+\ldots +e_7)+2(e_8+\ldots e_{10})+e_{11} &\qquad Z_{D4_3}={R_4\ldots R_7(R_8\ldots R_{10})^2 \over g_s^3l_s^{11}}
}$$
One can use the parameter $k$ to classify the KK-brane solutions; in the IIB decomposition $m_{10}-k$ is the coefficient of $e_{10}$. The full set of $Dp_i$ branes, where $i$ indicates the number of directions with a Taub-NUT fibration, corresponds to all the weights in the $l_1$ representation having $k=2$ and is shown in table 3.3. 
$$\halign{\centerline{#} \cr
\vbox{\offinterlineskip
\halign{\strut \vrule \quad \hfil # \hfil\quad &\vrule \quad \hfil # \hfil\quad 
&\vrule \quad \hfil # \hfil\quad  &\vrule \quad \hfil # \hfil\quad &\vrule \quad \hfil # \hfil\quad &\vrule #
\cr
\noalign{\hrule}
Level&$Dp_i$&Mass&$E_{12}$ Root&Root length&\cr 
($m_{9},m_{10}$)&&&($e_i$ basis)&squared&\cr 
\noalign{\hrule}$(4,2)$&$D5_{1}$&${R_{5}R_{6}R_{7}R_{8}R_{9}\hat{R}_{10}^{2}\over g_s^{2}l_s^{8}}$&$(1,0,0,0,0,1,1,1,1,1,0,0)$&$2$ & \cr 
\noalign{\hrule}$(5,2)$&$D5_{2}$&${R_{4}R_{5}R_{6}R_{7}R_{8}R_{9}^{2}\hat{R}_{10}^{2}\over g_s^{2}l_s^{10}}$&$(1,0,0,0,1,1,1,1,1,2,0,1)$&$2$ & \cr 
$(5,2)$&$D7_{1}$&${R_{3}R_{4}R_{5}R_{6}R_{7}R_{8}R_{9}\hat{R}_{10}^{2}\over g_s^{2}l_s^{10}}$&$(1,0,0,1,1,1,1,1,1,1,0,1)$&$0$ & \cr 
\noalign{\hrule}$(5,3)$&$D5_{2}$&${R_{4}R_{5}R_{6}R_{7}R_{8}R_{9}^{2}\hat{R}_{10}^{2}\over g_s^{3}l_s^{10}}$&$(1,0,0,0,1,1,1,1,1,2,1,0)$&$2$ & \cr 
$(5,3)$&$D7_{1}$&${R_{3}R_{4}R_{5}R_{6}R_{7}R_{8}R_{9}\hat{R}_{10}^{2}\over g_s^{3}l_s^{10}}$&$(1,0,0,1,1,1,1,1,1,1,1,0)$&$0$ & \cr 
\noalign{\hrule}$(6,2)$&$D9_{1}$&${R_{1}R_{2}R_{3}R_{4}R_{5}R_{6}R_{7}R_{8}R_{9}\hat{R}_{10}^{2}\over g_s^{2}l_s^{12}}$&$(1,1,1,1,1,1,1,1,1,1,0,2)$&$-2$ & \cr 
$(6,2)$&$D5_{3}$&${R_{3}R_{4}R_{5}R_{6}R_{7}R_{8}^{2}R_{9}^{2}\hat{R}_{10}^{2}\over g_s^{2}l_s^{12}}$&$(1,0,0,1,1,1,1,1,2,2,0,2)$&$2$ & \cr 
$(6,2)$&$D7_{2}$&${R_{2}R_{3}R_{4}R_{5}R_{6}R_{7}R_{8}R_{9}^{2}\hat{R}_{10}^{2}\over g_s^{2}l_s^{12}}$&$(1,0,1,1,1,1,1,1,1,2,0,2)$&$0$ & \cr 
\noalign{\hrule}$(6,3)$&$D9_{1}$&${R_{1}R_{2}R_{3}R_{4}R_{5}R_{6}R_{7}R_{8}R_{9}\hat{R}_{10}^{2}\over g_s^{3}l_s^{12}}$&$(1,1,1,1,1,1,1,1,1,1,1,1)$&$-4$ & \cr 
$(6,3)$&$D3_{4}$&${R_{4}R_{5}R_{6}R_{7}^{2}R_{8}^{2}R_{9}^{2}\hat{R}_{10}^{2}\over g_s^{3}l_s^{12}}$&$(1,0,0,0,1,1,1,2,2,2,1,1)$&$2$ & \cr 
$(6,3)$&$D5_{3}$&${R_{3}R_{4}R_{5}R_{6}R_{7}R_{8}^{2}R_{9}^{2}\hat{R}_{10}^{2}\over g_s^{3}l_s^{12}}$&$(1,0,0,1,1,1,1,1,2,2,1,1)$&$0$ & \cr 
$(6,3)$&$D7_{2}$&${R_{2}R_{3}R_{4}R_{5}R_{6}R_{7}R_{8}R_{9}^{2}\hat{R}_{10}^{2}\over g_s^{3}l_s^{12}}$&$(1,0,1,1,1,1,1,1,1,2,1,1)$&$-2$ & \cr 
\noalign{\hrule}$(6,4)$&$D9_{1}$&${R_{1}R_{2}R_{3}R_{4}R_{5}R_{6}R_{7}R_{8}R_{9}\hat{R}_{10}^{2}\over g_s^{4}l_s^{12}}$&$(1,1,1,1,1,1,1,1,1,1,2,0)$&$-2$ & \cr 
$(6,4)$&$D5_{3}$&${R_{3}R_{4}R_{5}R_{6}R_{7}R_{8}^{2}R_{9}^{2}\hat{R}_{10}^{2}\over g_s^{4}l_s^{12}}$&$(1,0,0,1,1,1,1,1,2,2,2,0)$&$2$ & \cr 
$(6,4)$&$D7_{2}$&${R_{2}R_{3}R_{4}R_{5}R_{6}R_{7}R_{8}R_{9}^{2}\hat{R}_{10}^{2}\over g_s^{4}l_s^{12}}$&$(1,0,1,1,1,1,1,1,1,2,2,0)$&$0$ & \cr 
\noalign{\hrule}$(7,2)$&$D7_{3}$&${R_{1}R_{2}R_{3}R_{4}R_{5}R_{6}R_{7}R_{8}^{2}R_{9}^{2}\hat{R}_{10}^{2}\over g_s^{2}l_s^{14}}$&$(1,1,1,1,1,1,1,1,2,2,0,3)$&$0$ & \cr 
$(7,2)$&$D5_{4}$&${R_{2}R_{3}R_{4}R_{5}R_{6}R_{7}^{2}R_{8}^{2}R_{9}^{2}\hat{R}_{10}^{2}\over g_s^{2}l_s^{14}}$&$(1,0,1,1,1,1,1,2,2,2,0,3)$&$2$ & \cr 
\noalign{\hrule}$(7,3)$&$D7_{3}$&${R_{1}R_{2}R_{3}R_{4}R_{5}R_{6}R_{7}R_{8}^{2}R_{9}^{2}\hat{R}_{10}^{2}\over g_s^{3}l_s^{14}}$&$(1,1,1,1,1,1,1,1,2,2,1,2)$&$-4$ & \cr 
$(7,3)$&$D1_{6}$&${R_{4}R_{5}^{2}R_{6}^{2}R_{7}^{2}R_{8}^{2}R_{9}^{2}\hat{R}_{10}^{2}\over g_s^{3}l_s^{14}}$&$(1,0,0,0,1,2,2,2,2,2,1,2)$&$2$ & \cr 
$(7,3)$&$D3_{5}$&${R_{3}R_{4}R_{5}R_{6}^{2}R_{7}^{2}R_{8}^{2}R_{9}^{2}\hat{R}_{10}^{2}\over g_s^{3}l_s^{14}}$&$(1,0,0,1,1,1,2,2,2,2,1,2)$&$0$ & \cr 
$(7,3)$&$D5_{4}$&${R_{2}R_{3}R_{4}R_{5}R_{6}R_{7}^{2}R_{8}^{2}R_{9}^{2}\hat{R}_{10}^{2}\over g_s^{3}l_s^{14}}$&$(1,0,1,1,1,1,1,2,2,2,1,2)$&$-2$ & \cr 
\noalign{\hrule}$(7,4)$&$D7_{3}$&${R_{1}R_{2}R_{3}R_{4}R_{5}R_{6}R_{7}R_{8}^{2}R_{9}^{2}\hat{R}_{10}^{2}\over g_s^{4}l_s^{14}}$&$(1,1,1,1,1,1,1,1,2,2,2,1)$&$-4$ & \cr 
$(7,4)$&$D1_{6}$&${R_{4}R_{5}^{2}R_{6}^{2}R_{7}^{2}R_{8}^{2}R_{9}^{2}\hat{R}_{10}^{2}\over g_s^{4}l_s^{14}}$&$(1,0,0,0,1,2,2,2,2,2,2,1)$&$2$ & \cr 
$(7,4)$&$D3_{5}$&${R_{3}R_{4}R_{5}R_{6}^{2}R_{7}^{2}R_{8}^{2}R_{9}^{2}\hat{R}_{10}^{2}\over g_s^{4}l_s^{14}}$&$(1,0,0,1,1,1,2,2,2,2,2,1)$&$0$ & \cr 
$(7,4)$&$D5_{4}$&${R_{2}R_{3}R_{4}R_{5}R_{6}R_{7}^{2}R_{8}^{2}R_{9}^{2}\hat{R}_{10}^{2}\over g_s^{4}l_s^{14}}$&$(1,0,1,1,1,1,1,2,2,2,2,1)$&$-2$ & \cr 
\noalign{\hrule}$(7,5)$&$D7_{3}$&${R_{1}R_{2}R_{3}R_{4}R_{5}R_{6}R_{7}R_{8}^{2}R_{9}^{2}\hat{R}_{10}^{2}\over g_s^{5}l_s^{14}}$&$(1,1,1,1,1,1,1,1,2,2,3,0)$&$0$ & \cr 
$(7,5)$&$D5_{4}$&${R_{2}R_{3}R_{4}R_{5}R_{6}R_{7}^{2}R_{8}^{2}R_{9}^{2}\hat{R}_{10}^{2}\over g_s^{5}l_s^{14}}$&$(1,0,1,1,1,1,1,2,2,2,3,0)$&$2$ & \cr 
\noalign{\hrule}$(8,2)$&$D5_{5}$&${R_{1}R_{2}R_{3}R_{4}R_{5}R_{6}^{2}R_{7}^{2}R_{8}^{2}R_{9}^{2}\hat{R}_{10}^{2}\over g_s^{2}l_s^{16}}$&$(1,1,1,1,1,1,2,2,2,2,0,4)$&$2$ & \cr 
\noalign{\hrule}$(8,3)$&$D5_{5}$&${R_{1}R_{2}R_{3}R_{4}R_{5}R_{6}^{2}R_{7}^{2}R_{8}^{2}R_{9}^{2}\hat{R}_{10}^{2}\over g_s^{3}l_s^{16}}$&$(1,1,1,1,1,1,2,2,2,2,1,3)$&$-4$ & \cr 
$(8,3)$&$D1_{7}$&${R_{3}R_{4}^{2}R_{5}^{2}R_{6}^{2}R_{7}^{2}R_{8}^{2}R_{9}^{2}\hat{R}_{10}^{2}\over g_s^{3}l_s^{16}}$&$(1,0,0,1,2,2,2,2,2,2,1,3)$&$0$ & \cr 
$(8,3)$&$D3_{6}$&${R_{2}R_{3}R_{4}R_{5}^{2}R_{6}^{2}R_{7}^{2}R_{8}^{2}R_{9}^{2}\hat{R}_{10}^{2}\over g_s^{3}l_s^{16}}$&$(1,0,1,1,1,2,2,2,2,2,1,3)$&$-2$ & \cr 
\noalign{\hrule}$(8,4)$&$D5_{5}$&${R_{1}R_{2}R_{3}R_{4}R_{5}R_{6}^{2}R_{7}^{2}R_{8}^{2}R_{9}^{2}\hat{R}_{10}^{2}\over g_s^{4}l_s^{16}}$&$(1,1,1,1,1,1,2,2,2,2,2,2)$&$-6$ & \cr 
$(8,4)$&$D1_{7}$&${R_{3}R_{4}^{2}R_{5}^{2}R_{6}^{2}R_{7}^{2}R_{8}^{2}R_{9}^{2}\hat{R}_{10}^{2}\over g_s^{4}l_s^{16}}$&$(1,0,0,1,2,2,2,2,2,2,2,2)$&$-2$ & \cr 
$(8,4)$&$D3_{6}$&${R_{2}R_{3}R_{4}R_{5}^{2}R_{6}^{2}R_{7}^{2}R_{8}^{2}R_{9}^{2}\hat{R}_{10}^{2}\over g_s^{4}l_s^{16}}$&$(1,0,1,1,1,2,2,2,2,2,2,2)$&$-4$ & \cr 
\noalign{\hrule}
}
}\cr}$$

$$\halign{\centerline{#} \cr
\vbox{\offinterlineskip
\halign{\strut \vrule \quad \hfil # \hfil\quad &\vrule \quad \hfil # \hfil\quad 
&\vrule \quad \hfil # \hfil\quad  &\vrule \quad \hfil # \hfil\quad &\vrule \quad \hfil # \hfil\quad &\vrule #
\cr
\noalign{\hrule}
Level&$Dp_i$&Mass&$E_{12}$ Root&Root length&\cr 
($m_{9},m_{10}$)&&&($e_i$ basis)&squared&\cr 
\noalign{\hrule}
$(8,5)$&$D5_{5}$&${R_{1}R_{2}R_{3}R_{4}R_{5}R_{6}^{2}R_{7}^{2}R_{8}^{2}R_{9}^{2}\hat{R}_{10}^{2}\over g_s^{5}l_s^{16}}$&$(1,1,1,1,1,1,2,2,2,2,3,1)$&$-4$ & \cr 
$(8,5)$&$D1_{7}$&${R_{3}R_{4}^{2}R_{5}^{2}R_{6}^{2}R_{7}^{2}R_{8}^{2}R_{9}^{2}\hat{R}_{10}^{2}\over g_s^{5}l_s^{16}}$&$(1,0,0,1,2,2,2,2,2,2,3,1)$&$0$ & \cr 
$(8,5)$&$D3_{6}$&${R_{2}R_{3}R_{4}R_{5}^{2}R_{6}^{2}R_{7}^{2}R_{8}^{2}R_{9}^{2}\hat{R}_{10}^{2}\over g_s^{5}l_s^{16}}$&$(1,0,1,1,1,2,2,2,2,2,3,1)$&$-2$ & \cr 
\noalign{\hrule}$(8,6)$&$D5_{5}$&${R_{1}R_{2}R_{3}R_{4}R_{5}R_{6}^{2}R_{7}^{2}R_{8}^{2}R_{9}^{2}\hat{R}_{10}^{2}\over g_s^{6}l_s^{16}}$&$(1,1,1,1,1,1,2,2,2,2,4,0)$&$2$ & \cr 
\noalign{\hrule}$(9,3)$&$D3_{7}$&${R_{1}R_{2}R_{3}R_{4}^{2}R_{5}^{2}R_{6}^{2}R_{7}^{2}R_{8}^{2}R_{9}^{2}\hat{R}_{10}^{2}\over g_s^{3}l_s^{18}}$&$(1,1,1,1,2,2,2,2,2,2,1,4)$&$-4$ & \cr 
$(9,3)$&$D1_{8}$&${R_{2}R_{3}^{2}R_{4}^{2}R_{5}^{2}R_{6}^{2}R_{7}^{2}R_{8}^{2}R_{9}^{2}\hat{R}_{10}^{2}\over g_s^{3}l_s^{18}}$&$(1,0,1,2,2,2,2,2,2,2,1,4)$&$-2$ & \cr 
\noalign{\hrule}$(9,4)$&$D3_{7}$&${R_{1}R_{2}R_{3}R_{4}^{2}R_{5}^{2}R_{6}^{2}R_{7}^{2}R_{8}^{2}R_{9}^{2}\hat{R}_{10}^{2}\over g_s^{4}l_s^{18}}$&$(1,1,1,1,2,2,2,2,2,2,2,3)$&$-8$ & \cr 
$(9,4)$&$D1_{8}$&${R_{2}R_{3}^{2}R_{4}^{2}R_{5}^{2}R_{6}^{2}R_{7}^{2}R_{8}^{2}R_{9}^{2}\hat{R}_{10}^{2}\over g_s^{4}l_s^{18}}$&$(1,0,1,2,2,2,2,2,2,2,2,3)$&$-6$ & \cr 
\noalign{\hrule}$(9,5)$&$D3_{7}$&${R_{1}R_{2}R_{3}R_{4}^{2}R_{5}^{2}R_{6}^{2}R_{7}^{2}R_{8}^{2}R_{9}^{2}\hat{R}_{10}^{2}\over g_s^{5}l_s^{18}}$&$(1,1,1,1,2,2,2,2,2,2,3,2)$&$-8$ & \cr 
$(9,5)$&$D1_{8}$&${R_{2}R_{3}^{2}R_{4}^{2}R_{5}^{2}R_{6}^{2}R_{7}^{2}R_{8}^{2}R_{9}^{2}\hat{R}_{10}^{2}\over g_s^{5}l_s^{18}}$&$(1,0,1,2,2,2,2,2,2,2,3,2)$&$-6$ & \cr 
\noalign{\hrule}$(9,6)$&$D3_{7}$&${R_{1}R_{2}R_{3}R_{4}^{2}R_{5}^{2}R_{6}^{2}R_{7}^{2}R_{8}^{2}R_{9}^{2}\hat{R}_{10}^{2}\over g_s^{6}l_s^{18}}$&$(1,1,1,1,2,2,2,2,2,2,4,1)$&$-4$ & \cr 
$(9,6)$&$D1_{8}$&${R_{2}R_{3}^{2}R_{4}^{2}R_{5}^{2}R_{6}^{2}R_{7}^{2}R_{8}^{2}R_{9}^{2}\hat{R}_{10}^{2}\over g_s^{6}l_s^{18}}$&$(1,0,1,2,2,2,2,2,2,2,4,1)$&$-2$ & \cr 
\noalign{\hrule}$(10,3)$&$D1_{9}$&${R_{1}R_{2}^{2}R_{3}^{2}R_{4}^{2}R_{5}^{2}R_{6}^{2}R_{7}^{2}R_{8}^{2}R_{9}^{2}\hat{R}_{10}^{2}\over g_s^{3}l_s^{20}}$&$(1,1,2,2,2,2,2,2,2,2,1,5)$&$-4$ & \cr 
\noalign{\hrule}$(10,4)$&$D1_{9}$&${R_{1}R_{2}^{2}R_{3}^{2}R_{4}^{2}R_{5}^{2}R_{6}^{2}R_{7}^{2}R_{8}^{2}R_{9}^{2}\hat{R}_{10}^{2}\over g_s^{4}l_s^{20}}$&$(1,1,2,2,2,2,2,2,2,2,2,4)$&$-10$ & \cr 
\noalign{\hrule}$(10,5)$&$D1_{9}$&${R_{1}R_{2}^{2}R_{3}^{2}R_{4}^{2}R_{5}^{2}R_{6}^{2}R_{7}^{2}R_{8}^{2}R_{9}^{2}\hat{R}_{10}^{2}\over g_s^{5}l_s^{20}}$&$(1,1,2,2,2,2,2,2,2,2,3,3)$&$-12$ & \cr 
\noalign{\hrule}$(10,6)$&$D1_{9}$&${R_{1}R_{2}^{2}R_{3}^{2}R_{4}^{2}R_{5}^{2}R_{6}^{2}R_{7}^{2}R_{8}^{2}R_{9}^{2}\hat{R}_{10}^{2}\over g_s^{6}l_s^{20}}$&$(1,1,2,2,2,2,2,2,2,2,4,2)$&$-10$ & \cr 
\noalign{\hrule}$(10,7)$&$D1_{9}$&${R_{1}R_{2}^{2}R_{3}^{2}R_{4}^{2}R_{5}^{2}R_{6}^{2}R_{7}^{2}R_{8}^{2}R_{9}^{2}\hat{R}_{10}^{2}\over g_s^{7}l_s^{20}}$&$(1,1,2,2,2,2,2,2,2,2,5,1)$&$-4$ & \cr 
\noalign{\hrule}
}
}\cr
Table 3.3 The full set of IIB $Dp_i$ branes from the $l_1$ representation of $E_{11}$ \cr}$$
\medskip
{\bf Conclusion}
\medskip
In this paper we have identified the U-duality brane charge multiplets within the $l_1$ representation and given explicitly the weights associated to the particle, string and membrane multiplets when D=3,4,5,6,7,8 - this is a generalisation of the results of [11]. We have also introduced a tension formula that associates a tension to each root in the $E_{12}$ root lattice, the lattice natural to the $l_1$ representation of $E_{11}$. The tension formula can be readily extended to other infinite dimensional algebras. The formula was constructed by introducing a dimensionful parameter into the $l_1$ representation of $E_{11}$, associated to the extra node (denoted in this paper with a "*") of the $E_{12}$ Dynkin diagram that differentiates it from the $E_{11}$ diagram. The tension formula reproduced the tensions of the pp-wave, the M2-brane, the M5-brane and the KK6 monopole of M-theory from the associated weights in the $l_1$ algebra. Furthermore all the tensions of the $Dp$-branes of IIA and IIB superstring theories were also found, together with the correct powers of the string coupling constant, $g_s$, and the string length, $l_s$. The tension formula was also applied to all the states in the particle and string charge multiplets for comparison with previously known results. The dimensionful parameter introduced was found to correctly reproduce all the known masses of the U-duality brane charge multiplets presented in [6,7,8,9,10]. The formula was then applied to the charge multiplet of the membrane and the corresponding masses were given. It would be illuminating to analyse the content of other $G^{+++}$ algebras using the same construction. This would be especially interesting for the pure gravitational Kac-Moody algebra $A_{D-3}^{+++}$ and the algebra associated to the bosonic string, $K_{27}\equiv D_{24}^{+++}$.

One consequence of the tension formula is the observation that almost all the content of $E_{11}$ is associated to KK-branes (or monopoles). This interpretation is based on the observation that the tension of KK-branes is divergent when the spacetime is decompactified. Given the $l_1$ representation of $E_{11}$ we can calculate for any of the charges an associated mass and tension. The tension is found by dividing the mass by a volume, $V_p\sim (2\pi R)^p$. For branes the tension is independent of a radius of compactification, in the cases where the tension remains dependent on the compact radii the associated solution is a KK-brane. A familiar example is generalisation of the Taub-NUT solution to eleven dimensions, which is also called the KK6-brane and is associated to the dual gravity field. The charge conserved by the dual gravity field appears in the $l_1$ representation at level three in the algebraic decomposition. The root in the $E_{12}$ lattice is:
$$\eqalignno{\beta_{KK6} & \equiv \alpha_*+\alpha_1+\ldots \alpha_4 +2\alpha_5 +3\alpha_6 +4\alpha_7+5\alpha_8+3\alpha_9+\alpha_{10}+3\alpha_{11} \cr
& = (e_5+\ldots e_{10} +2e_{11})}$$
Using the formula of equation (3.5)  the mass of this root is found. 
$$Z_{\beta_{KK6}}={R_5\ldots R_{10}R^2_{11}\over l_p^9}$$
For convenience we analyse the solution when it is compactified on a seven-torus and the tension of the solution may be found by dividing the mass by $(2\pi)^7R_5\ldots R_{11}$:
$$T_{\beta_{KK6}}={R_{11}\over (2\pi)^7l_p^9}$$
Evidently this result does not carry naturally back into the uncompactified eleven dimensional spacetime and diverges when $R_{11}$ is decompactified. The diverging tension in uncompactified spacetime is the signature of a KK-brane. For most of the weights of $E_{11}$ appearing in the $l_1$ representation, in fact all those associated to a mixed symmetry field, the tension found by using equation (3.5) is divergent in the non-compact spacetime. In this sense much of the $l_1$ representation, being composed mostly of mixed symmetry fields, is associated to KK-brane charges. One may say that KK-brane charges are the rule and their vanishing in the case of the $M2$, $M5$ and pp-wave charges are the exceptions. 

We have been able to classify the KK-brane charges that appear in the $l_1$ algebra since they are directly related to the parameter, $k$, appearing in the algebraic decomposition of section 1 of this paper. We recall that this parameter played a twofold role of counting the number of blocks of antisymmetric indices as well as controlling the blocks of eleven antisymmetric indices, or volume forms $\epsilon$, appearing in the generators of the $l_1$ representation of $E_{11}$. The KK-branes may be labelled by the powers of the radii appearing in the mass formula. The most studied class of KK-branes, labelled $Dp_i$ in the literature, have a mass which is quadratic in the spatial radii and may be labelled by two integers corresponding to the number of linear, $p$, and squared radii, $i$, respectively. In tables 3.1, 3.2 and 3.3 we list the full set of such $Dp_i$ brane charges in the $l_1$ representation of $E_{11}$ relevant to M-theory, the IIA theory and the IIB theory including the charges of KK-branes previously found by U-duality transformations of the D7 brane charge in [21]. The role played by KK-branes is unclear. A simple interpretation of the KK-brane charges of M-theory is that they give an eleven dimensional origin to the $Dp$-brane charges ($p>5$) of IIA and IIB string theory upon dimensional reduction as well as other KK-brane charges. However, whether they play a more important part than simply book-keeping the branes, and also winding and KK-modes that appear from duality arguments in lower dimensional theories remains to be seen. One may hope that the extra KK-brane tower of states, being non-perturbative, may reveal significant details about the kinematics and dynamics of the $E_{11}$ fields. More simply, since the prototype KK-brane is the dual graviton their further investigation may shed more light on the dual gravity theories.

An interesting infinite class of roots in the adjoint of $E_{11}$ which corresponds to generators with no blocks of ten or eleven antisymmetrised indices has recently been completely found [20] and it was highlighted that $E_{11}$ contained all the dualised versions of the tensors of massless dualised supergravity. The fields in this class of roots have associated generators taking the form,
$${K^{a^1_1\ldots a^1_9, \ldots a^n_1\ldots a^n_9,b}}_c,R^{a^1_1\ldots a^1_9, \ldots a^n_1\ldots a^n_9, j_1j_2j_3},R^{a^1_1\ldots a^1_9, \ldots a^n_1\ldots a^n_9, j_1\ldots j_6},R^{a^1_1\ldots a^1_9, \ldots a^n_1\ldots a^n_9, j_1\ldots j_8,k}$$
Where $n\geq 0$. At first sight these generators, being all representations of the little group in eleven dimensions, $SO(9)$, appear to indicate all possible massless dual solutions in the algebra, but if we consider the charges associated to these roots in the $l_1$ representation we find that they have different "masses", namely,
$${V^n \over l_p^{9n}R_{11}},{V^n R_{10}R_{11}\over l_p^{9n+3}},{V^n R_7R_8R_9R_{10}R_{11}\over l_p^{9n+6}}, {V^n R_5 \ldots R_{10}R_{11}^2 \over l_p^{9n+9}}$$
Where $V=R_3\ldots R_{11}$. For the case where $n=0$ three of the tensions, given by division by $(2\pi)^9V$, are well-defined and the fourth, which is the $KK6$ brane tension diverges. However for the other roots in this class $(n>0)$ the tensions all diverge in the non-compact setting which may indicate some difference in their nature. Using the techniques of [4,13] one can write down a line element associated to a half-BPS brane solution for each of the dual roots in the adjoint of $E_{11}$. The set of roots are:
$$\eqalign{\beta_{pp*}&=\alpha_{10}+n\beta_0\cr
\beta_{M2*}&=\alpha_{11}+n\beta_0\cr
\beta_{M5*}&=\alpha_6+2\alpha_7+3\alpha_8+2\alpha_9+\alpha_{10}+2\alpha_{11}+n\beta_0\cr
\beta_{KK6*}&=\alpha_4+2\alpha_5+3\alpha_6+4\alpha_7+5\alpha_8+3\alpha_9+\alpha_{10}+3\alpha_{11}+n\beta_0}$$
Where,
$$\beta_0\equiv \alpha_3+2\alpha_4+3\alpha_5+4\alpha_6+5\alpha_7+6\alpha_8+4\alpha_9+2\alpha_{10}+3\alpha_{11}$$
is the root controlling the blocks of nine antisymmetric indices in the dual generators. For reference,
$$Z_{\beta_0}={R_3\ldots R_{11}\over l_p^9}={R_1R_2R_3\over G_{11}}$$
The line elements corresponding to the set of dual roots are,
$$\eqalign{ds^2_{pp*}&=(1+K)^nd\bar{x}_2^2-(1-K)(dt^2)+(1+K)dy^2-2Kdtdy+d\Sigma_7^2\cr
ds^2_{M2*}&=N^{(n+{1\over 3})}(d\bar{x}_2^2)+N^{1\over 3}(d\bar{x}_6^2)+N^{-{2\over 3}}(d\bar{y}_2^2-dt^2)\cr
ds^2_{M5*}&=N^{(n+{2\over 3})}(d\bar{x}_2^2)+N^{2\over 3}(d\bar{x}_3^2)+N^{-{1\over 3}}(d\bar{y}_5^2-dt^2)\cr
ds^2_{KK6*}&=N^{(n+1)}(d\bar{x}_2^2)+N(dy^2)+N^{-1}(-dt^2)+d\Sigma_7^2
}$$
Where $d\Sigma_7^2$ denotes a seven dimensional Euclidean line element. These volume elements are dependent on $n$, the number of blocks of nine antisymmetric indices appearing in the generator of the associated dual roots. It would be interesting to study this class of roots further.
 
\medskip 
{\bf Acknowledgments}
P.P.C. would like to thank the Scuola Normale Superiore di Pisa for their hospitality during various stages of this work and to thank Per Sundell, Carlo Iazeolla, Nicolas Boulanger, Augusto Sagnotti and Chris Hull for conversations and inspiration during the preparation of this paper. P.P.C. is funded by EU Superstring Network by the research contract MRTN-CT-2004-512194 and INFN, via MIUR-PRIN contract 2003-023852. The research of P.W. is supported by a PPARC rolling grant PP/C5071745/1 and the EU Marie Curie research training network grant HPRN-CT-2000-00122. 

\bigskip
\bigskip
\bigskip
\bigskip
\bigskip
\bigskip\bigskip
\bigskip
\bigskip
\bigskip
\bigskip
\bigskip
\eject

\medskip
{\bf Appendix A - Low level weights in the $l_1$ representation of $E_{11}$ relevant to 11D, IIA and IIB SuGra}
$$\halign{\centerline{#} \cr
\vbox{\offinterlineskip
\halign{\strut \vrule \quad \hfil # \hfil\quad &\vrule \quad \hfil # \hfil\quad 
&\vrule \quad \hfil # \hfil\quad  &\vrule \quad \hfil # \hfil\quad &\vrule \quad \hfil # \hfil\quad &\vrule #
\cr
\noalign{\hrule}
Level&$A_{10}$ weights&$E_{12}$ Root&$E_{12}$ Root&Root length&\cr 
($m_{11}$)&&($\alpha_i$ basis)&($e_i$ basis)&squared&\cr 
\noalign{\hrule}
$0$&$[1,0,0,0,0,0,0,0,0,0]$& $(1,0,0,0,0,0,0,0,0,0,0,0)$&$(1,-1,0,0,0,0,0,0,0,0,0,0)$&$2$ & \cr 
\noalign{\hrule}$1$&$[0,0,0,0,0,0,0,0,1,0]$& $(1,1,1,1,1,1,1,1,1,0,0,1)$&$(1,0,0,0,0,0,0,0,0,0,1,1)$&$2$ & \cr 
\noalign{\hrule}$2$&$[0,0,0,0,0,1,0,0,0,0]$& $(1,1,1,1,1,1,1,2,3,2,1,2)$&$(1,0,0,0,0,0,0,1,1,1,1,1)$&$2$ & \cr 
\noalign{\hrule}$3$&$[0,0,1,0,0,0,0,0,0,0]$& $(1,1,1,1,2,3,4,5,6,4,2,3)$&$(1,0,0,0,1,1,1,1,1,1,1,1)$&$0$ & \cr 
$3$&$[0,0,0,1,0,0,0,0,0,1]$& $(1,1,1,1,1,2,3,4,5,3,1,3)$&$(1,0,0,0,0,1,1,1,1,1,1,2)$&$2$ & \cr 
\noalign{\hrule}$4$&$[0,0,1,0,0,0,0,1,0,0]$& $(1,1,1,1,2,3,4,5,6,4,2,4)$&$(1,0,0,0,1,1,1,1,1,2,2,2)$&$2$ & \cr 
$4$&$[0,1,0,0,0,0,0,0,1,0]$& $(1,1,1,2,3,4,5,6,7,4,2,4)$&$(1,0,0,1,1,1,1,1,1,1,2,2)$&$0$ & \cr 
$4$&$[1,0,0,0,0,0,0,0,0,1]$& $(1,1,2,3,4,5,6,7,8,5,2,4)$&$(1,0,1,1,1,1,1,1,1,1,1,2)$&$-2$ & \cr 
$4$&$[0,1,0,0,0,0,0,0,0,2]$& $(1,1,1,2,3,4,5,6,7,4,1,4)$&$(1,0,0,1,1,1,1,1,1,1,1,3)$&$2$ & \cr 
\noalign{\hrule}$5$&$[0,0,0,0,0,0,0,1,0,0]$& $(1,2,3,4,5,6,7,8,9,6,3,5)$&$(1,1,1,1,1,1,1,1,1,2,2,2)$&$-4$ & \cr 
$5$&$[0,0,1,0,1,0,0,0,0,0]$& $(1,1,1,1,2,3,5,7,9,6,3,5)$&$(1,0,0,0,1,1,2,2,2,2,2,2)$&$2$ & \cr 
$5$&$[0,1,0,0,0,1,0,0,0,0]$& $(1,1,1,2,3,4,5,7,9,6,3,5)$&$(1,0,0,1,1,1,1,2,2,2,2,2)$&$0$ & \cr 
$5$&$[1,0,0,0,0,0,1,0,0,0]$& $(1,1,2,3,4,5,6,7,9,6,3,5)$&$(1,0,1,1,1,1,1,1,2,2,2,2)$&$-2$ & \cr 
$5$&$[0,0,0,0,0,0,0,0,1,1]$& $(1,2,3,4,5,6,7,8,9,5,2,5)$&$(1,1,1,1,1,1,1,1,1,1,2,3)$&$-2$ & \cr 
$5$&$[0,1,0,0,0,0,1,0,0,1]$& $(1,1,1,2,3,4,5,6,8,5,2,5)$&$(1,0,0,1,1,1,1,1,2,2,2,3)$&$2$ & \cr 
$5$&$[1,0,0,0,0,0,0,0,2,0]$& $(1,1,2,3,4,5,6,7,8,4,2,5)$&$(1,0,1,1,1,1,1,1,1,1,3,3)$&$2$ & \cr 
$5$&$[1,0,0,0,0,0,0,1,0,1]$& $(1,1,2,3,4,5,6,7,8,5,2,5)$&$(1,0,1,1,1,1,1,1,1,2,2,3)$&$0$ & \cr 
$5$&$[0,0,0,0,0,0,0,0,0,3]$& $(1,2,3,4,5,6,7,8,9,5,1,5)$&$(1,1,1,1,1,1,1,1,1,1,1,4)$&$2$ & \cr 
\noalign{\hrule}$6$&$[0,0,0,0,1,0,0,0,0,0]$& $(1,2,3,4,5,6,8,10,12,8,4,6)$&$(1,1,1,1,1,1,2,2,2,2,2,2)$&$-6$ & \cr 
$6$&$[0,1,1,0,0,0,0,0,0,0]$& $(1,1,1,2,4,6,8,10,12,8,4,6)$&$(1,0,0,1,2,2,2,2,2,2,2,2)$&$-2$ & \cr 
$6$&$[1,0,0,1,0,0,0,0,0,0]$& $(1,1,2,3,4,6,8,10,12,8,4,6)$&$(1,0,1,1,1,2,2,2,2,2,2,2)$&$-4$ & \cr 
$6$&$[0,0,0,0,0,0,0,2,0,0]$& $(1,2,3,4,5,6,7,8,9,6,3,6)$&$(1,1,1,1,1,1,1,1,1,3,3,3)$&$0$ & \cr 
$6$&$[0,0,0,0,0,0,1,0,1,0]$& $(1,2,3,4,5,6,7,8,10,6,3,6)$&$(1,1,1,1,1,1,1,1,2,2,3,3)$&$-2$ & \cr 
$6$&$[0,0,0,0,0,1,0,0,0,1]$& $(1,2,3,4,5,6,7,9,11,7,3,6)$&$(1,1,1,1,1,1,1,2,2,2,2,3)$&$-4$ & \cr 
$6$&$[0,0,2,0,0,0,0,0,0,1]$& $(1,1,1,1,3,5,7,9,11,7,3,6)$&$(1,0,0,0,2,2,2,2,2,2,2,3)$&$2$ & \cr 
$6$&$[0,1,0,0,1,0,0,0,1,0]$& $(1,1,1,2,3,4,6,8,10,6,3,6)$&$(1,0,0,1,1,1,2,2,2,2,3,3)$&$2$ & \cr 
$6$&$[0,1,0,1,0,0,0,0,0,1]$& $(1,1,1,2,3,5,7,9,11,7,3,6)$&$(1,0,0,1,1,2,2,2,2,2,2,3)$&$0$ & \cr 
$6$&$[1,0,0,0,0,0,1,1,0,0]$& $(1,1,2,3,4,5,6,7,9,6,3,6)$&$(1,0,1,1,1,1,1,1,2,3,3,3)$&$2$ & \cr 
$6$&$[1,0,0,0,0,1,0,0,1,0]$& $(1,1,2,3,4,5,6,8,10,6,3,6)$&$(1,0,1,1,1,1,1,2,2,2,3,3)$&$0$ & \cr 
$6$&$[1,0,0,0,1,0,0,0,0,1]$& $(1,1,2,3,4,5,7,9,11,7,3,6)$&$(1,0,1,1,1,1,2,2,2,2,2,3)$&$-2$ & \cr 
$6$&$[0,0,0,0,0,0,0,1,1,1]$& $(1,2,3,4,5,6,7,8,9,5,2,6)$&$(1,1,1,1,1,1,1,1,1,2,3,4)$&$2$ & \cr 
$6$&$[0,0,0,0,0,0,1,0,0,2]$& $(1,2,3,4,5,6,7,8,10,6,2,6)$&$(1,1,1,1,1,1,1,1,2,2,2,4)$&$0$ & \cr 
$6$&$[1,0,0,0,0,1,0,0,0,2]$& $(1,1,2,3,4,5,6,8,10,6,2,6)$&$(1,0,1,1,1,1,1,2,2,2,2,4)$&$2$ & \cr 
\noalign{\hrule}$7$&$[0,1,0,0,0,0,0,0,0,0]$& $(1,2,3,5,7,9,11,13,15,10,5,7)$&$(1,1,1,2,2,2,2,2,2,2,2,2)$&$-10$ & \cr 
$7$&$[2,0,0,0,0,0,0,0,0,0]$& $(1,1,3,5,7,9,11,13,15,10,5,7)$&$(1,0,2,2,2,2,2,2,2,2,2,2)$&$-8$ & \cr 
$7$&$[0,0,0,0,0,1,1,0,0,0]$& $(1,2,3,4,5,6,7,9,12,8,4,7)$&$(1,1,1,1,1,1,1,2,3,3,3,3)$&$-2$ & \cr 
$7$&$[0,0,0,0,1,0,0,1,0,0]$& $(1,2,3,4,5,6,8,10,12,8,4,7)$&$(1,1,1,1,1,1,2,2,2,3,3,3)$&$-4$ & \cr 
$7$&$[0,0,0,1,0,0,0,0,1,0]$& $(1,2,3,4,5,7,9,11,13,8,4,7)$&$(1,1,1,1,1,2,2,2,2,2,3,3)$&$-6$ & \cr 
$7$&$[0,0,1,0,0,0,0,0,0,1]$& $(1,2,3,4,6,8,10,12,14,9,4,7)$&$(1,1,1,1,2,2,2,2,2,2,2,3)$&$-8$ & \cr 
$7$&$[0,1,0,1,0,0,1,0,0,0]$& $(1,1,1,2,3,5,7,9,12,8,4,7)$&$(1,0,0,1,1,2,2,2,3,3,3,3)$&$2$ & \cr 
$7$&$[0,1,1,0,0,0,0,1,0,0]$& $(1,1,1,2,4,6,8,10,12,8,4,7)$&$(1,0,0,1,2,2,2,2,2,3,3,3)$&$0$ & \cr 
$7$&$[0,2,0,0,0,0,0,0,1,0]$& $(1,1,1,3,5,7,9,11,13,8,4,7)$&$(1,0,0,2,2,2,2,2,2,2,3,3)$&$-2$ & \cr 
$7$&$[1,0,0,0,0,2,0,0,0,0]$& $(1,1,2,3,4,5,6,9,12,8,4,7)$&$(1,0,1,1,1,1,1,3,3,3,3,3)$&$2$ & \cr 
$7$&$[1,0,0,0,1,0,1,0,0,0]$& $(1,1,2,3,4,5,7,9,12,8,4,7)$&$(1,0,1,1,1,1,2,2,3,3,3,3)$&$0$ & \cr 
$7$&$[1,0,0,1,0,0,0,1,0,0]$& $(1,1,2,3,4,6,8,10,12,8,4,7)$&$(1,0,1,1,1,2,2,2,2,3,3,3)$&$-2$ & \cr 
$7$&$[1,0,1,0,0,0,0,0,1,0]$& $(1,1,2,3,5,7,9,11,13,8,4,7)$&$(1,0,1,1,2,2,2,2,2,2,3,3)$&$-4$ & \cr 
$7$&$[1,1,0,0,0,0,0,0,0,1]$& $(1,1,2,4,6,8,10,12,14,9,4,7)$&$(1,0,1,2,2,2,2,2,2,2,2,3)$&$-6$ & \cr 
\noalign{\hrule}}
}\cr
Table A.1 Low level weights in the {\it 11D supergravity} decomposition of the $l_1$ representation of $E_{11}$ (continued)\cr}$$
$$\halign{\centerline{#} \cr
\vbox{\offinterlineskip
\halign{\strut \vrule \quad \hfil # \hfil\quad &\vrule \quad \hfil # \hfil\quad 
&\vrule \quad \hfil # \hfil\quad  &\vrule \quad \hfil # \hfil\quad &\vrule \quad \hfil # \hfil\quad &\vrule #
\cr
\noalign{\hrule}
Level&$A_{10}$ weights&$E_{12}$ Root&$E_{12}$ Root&Root length&\cr 
($m_{11}$)&&($\alpha_i$ basis)&($e_i$ basis)&squared&\cr 
\noalign{\hrule}
$7$&$[0,0,0,0,0,0,2,0,0,1]$& $(1,2,3,4,5,6,7,8,11,7,3,7)$&$(1,1,1,1,1,1,1,1,3,3,3,4)$&$2$ & \cr 
$7$&$[0,0,0,0,0,1,0,0,2,0]$& $(1,2,3,4,5,6,7,9,11,6,3,7)$&$(1,1,1,1,1,1,1,2,2,2,4,4)$&$2$ & \cr 
$7$&$[0,0,0,0,0,1,0,1,0,1]$& $(1,2,3,4,5,6,7,9,11,7,3,7)$&$(1,1,1,1,1,1,1,2,2,3,3,4)$&$0$ & \cr 
$7$&$[0,0,0,0,1,0,0,0,1,1]$& $(1,2,3,4,5,6,8,10,12,7,3,7)$&$(1,1,1,1,1,1,2,2,2,2,3,4)$&$-2$ & \cr 
$7$&$[0,0,0,1,0,0,0,0,0,2]$& $(1,2,3,4,5,7,9,11,13,8,3,7)$&$(1,1,1,1,1,2,2,2,2,2,2,4)$&$-4$ & \cr 
$7$&$[0,1,1,0,0,0,0,0,1,1]$& $(1,1,1,2,4,6,8,10,12,7,3,7)$&$(1,0,0,1,2,2,2,2,2,2,3,4)$&$2$ & \cr 
$7$&$[0,2,0,0,0,0,0,0,0,2]$& $(1,1,1,3,5,7,9,11,13,8,3,7)$&$(1,0,0,2,2,2,2,2,2,2,2,4)$&$0$ & \cr 
$7$&$[1,0,0,0,1,0,0,1,0,1]$& $(1,1,2,3,4,5,7,9,11,7,3,7)$&$(1,0,1,1,1,1,2,2,2,3,3,4)$&$2$ & \cr 
$7$&$[1,0,0,1,0,0,0,0,1,1]$& $(1,1,2,3,4,6,8,10,12,7,3,7)$&$(1,0,1,1,1,2,2,2,2,2,3,4)$&$0$ & \cr 
$7$&$[1,0,1,0,0,0,0,0,0,2]$& $(1,1,2,3,5,7,9,11,13,8,3,7)$&$(1,0,1,1,2,2,2,2,2,2,2,4)$&$-2$ & \cr 
$7$&$[0,0,0,0,1,0,0,0,0,3]$& $(1,2,3,4,5,6,8,10,12,7,2,7)$&$(1,1,1,1,1,1,2,2,2,2,2,5)$&$2$ & \cr
\noalign{\hrule}
$8$&$[0,0,0,0,0,0,0,0,0,1]$& $(1,3,5,7,9,11,13,15,17,11,5,8)$&$(1,2,2,2,2,2,2,2,2,2,2,3)$&$-14$ & \cr 
$8$&$[0,0,0,0,2,0,0,0,0,0]$& $(1,2,3,4,5,6,9,12,15,10,5,8)$&$(1,1,1,1,1,1,3,3,3,3,3,3)$&$-4$ & \cr 
$8$&$[0,0,0,1,0,1,0,0,0,0]$& $(1,2,3,4,5,7,9,12,15,10,5,8)$&$(1,1,1,1,1,2,2,3,3,3,3,3)$&$-6$ & \cr 
$8$&$[0,0,1,0,0,0,1,0,0,0]$& $(1,2,3,4,6,8,10,12,15,10,5,8)$&$(1,1,1,1,2,2,2,2,3,3,3,3)$&$-8$ & \cr 
$8$&$[0,1,0,0,0,0,0,1,0,0]$& $(1,2,3,5,7,9,11,13,15,10,5,8)$&$(1,1,1,2,2,2,2,2,2,3,3,3)$&$-10$ & \cr 
$8$&$[0,1,0,2,0,0,0,0,0,0]$& $(1,1,1,2,3,6,9,12,15,10,5,8)$&$(1,0,0,1,1,3,3,3,3,3,3,3)$&$2$ & \cr 
$8$&$[0,1,1,0,1,0,0,0,0,0]$& $(1,1,1,2,4,6,9,12,15,10,5,8)$&$(1,0,0,1,2,2,3,3,3,3,3,3)$&$0$ & \cr 
$8$&$[0,2,0,0,0,1,0,0,0,0]$& $(1,1,1,3,5,7,9,12,15,10,5,8)$&$(1,0,0,2,2,2,2,3,3,3,3,3)$&$-2$ & \cr 
$8$&$[1,0,0,0,0,0,0,0,1,0]$& $(1,2,4,6,8,10,12,14,16,10,5,8)$&$(1,1,2,2,2,2,2,2,2,2,3,3)$&$-12$ & \cr 
$8$&$[1,0,0,1,1,0,0,0,0,0]$& $(1,1,2,3,4,6,9,12,15,10,5,8)$&$(1,0,1,1,1,2,3,3,3,3,3,3)$&$-2$ & \cr 
$8$&$[1,0,1,0,0,1,0,0,0,0]$& $(1,1,2,3,5,7,9,12,15,10,5,8)$&$(1,0,1,1,2,2,2,3,3,3,3,3)$&$-4$ & \cr 
$8$&$[1,1,0,0,0,0,1,0,0,0]$& $(1,1,2,4,6,8,10,12,15,10,5,8)$&$(1,0,1,2,2,2,2,2,3,3,3,3)$&$-6$ & \cr 
$8$&$[2,0,0,0,0,0,0,1,0,0]$& $(1,1,3,5,7,9,11,13,15,10,5,8)$&$(1,0,2,2,2,2,2,2,2,3,3,3)$&$-8$ & \cr 
$8$&$[0,0,0,0,0,2,0,0,1,0]$& $(1,2,3,4,5,6,7,10,13,8,4,8)$&$(1,1,1,1,1,1,1,3,3,3,4,4)$&$2$ & \cr 
$8$&$[0,0,0,0,1,0,0,2,0,0]$& $(1,2,3,4,5,6,8,10,12,8,4,8)$&$(1,1,1,1,1,1,2,2,2,4,4,4)$&$2$ & \cr 
$8$&$[0,0,0,0,1,0,1,0,1,0]$& $(1,2,3,4,5,6,8,10,13,8,4,8)$&$(1,1,1,1,1,1,2,2,3,3,4,4)$&$0$ & \cr 
$8$&$[0,0,0,0,1,1,0,0,0,1]$& $(1,2,3,4,5,6,8,11,14,9,4,8)$&$(1,1,1,1,1,1,2,3,3,3,3,4)$&$-2$ & \cr 
$8$&$[0,0,0,1,0,0,0,1,1,0]$& $(1,2,3,4,5,7,9,11,13,8,4,8)$&$(1,1,1,1,1,2,2,2,2,3,4,4)$&$-2$ & \cr 
$8$&$[0,0,0,1,0,0,1,0,0,1]$& $(1,2,3,4,5,7,9,11,14,9,4,8)$&$(1,1,1,1,1,2,2,2,3,3,3,4)$&$-4$ & \cr 
$8$&$[0,0,1,0,0,0,0,0,2,0]$& $(1,2,3,4,6,8,10,12,14,8,4,8)$&$(1,1,1,1,2,2,2,2,2,2,4,4)$&$-4$ & \cr 
$8$&$[0,0,1,0,0,0,0,1,0,1]$& $(1,2,3,4,6,8,10,12,14,9,4,8)$&$(1,1,1,1,2,2,2,2,2,3,3,4)$&$-6$ & \cr 
$8$&$[0,1,0,0,0,0,0,0,1,1]$& $(1,2,3,5,7,9,11,13,15,9,4,8)$&$(1,1,1,2,2,2,2,2,2,2,3,4)$&$-8$ & \cr 
$8$&$[0,1,1,0,0,1,0,0,0,1]$& $(1,1,1,2,4,6,8,11,14,9,4,8)$&$(1,0,0,1,2,2,2,3,3,3,3,4)$&$2$ & \cr 
$8$&$[0,2,0,0,0,0,0,1,1,0]$& $(1,1,1,3,5,7,9,11,13,8,4,8)$&$(1,0,0,2,2,2,2,2,2,3,4,4)$&$2$ & \cr 
$8$&$[0,2,0,0,0,0,1,0,0,1]$& $(1,1,1,3,5,7,9,11,14,9,4,8)$&$(1,0,0,2,2,2,2,2,3,3,3,4)$&$0$ & \cr 
$8$&$[1,0,0,0,0,0,0,0,0,2]$& $(1,2,4,6,8,10,12,14,16,10,4,8)$&$(1,1,2,2,2,2,2,2,2,2,2,4)$&$-10$ & \cr 
$8$&$[1,0,0,0,2,0,0,0,0,1]$& $(1,1,2,3,4,5,8,11,14,9,4,8)$&$(1,0,1,1,1,1,3,3,3,3,3,4)$&$2$ & \cr 
$8$&$[1,0,0,1,0,0,1,0,1,0]$& $(1,1,2,3,4,6,8,10,13,8,4,8)$&$(1,0,1,1,1,2,2,2,3,3,4,4)$&$2$ & \cr 
$8$&$[1,0,0,1,0,1,0,0,0,1]$& $(1,1,2,3,4,6,8,11,14,9,4,8)$&$(1,0,1,1,1,2,2,3,3,3,3,4)$&$0$ & \cr 
$8$&$[1,0,1,0,0,0,0,1,1,0]$& $(1,1,2,3,5,7,9,11,13,8,4,8)$&$(1,0,1,1,2,2,2,2,2,3,4,4)$&$0$ & \cr 
$8$&$[1,0,1,0,0,0,1,0,0,1]$& $(1,1,2,3,5,7,9,11,14,9,4,8)$&$(1,0,1,1,2,2,2,2,3,3,3,4)$&$-2$ & \cr 
$8$&$[1,1,0,0,0,0,0,0,2,0]$& $(1,1,2,4,6,8,10,12,14,8,4,8)$&$(1,0,1,2,2,2,2,2,2,2,4,4)$&$-2$ & \cr 
$8$&$[1,1,0,0,0,0,0,1,0,1]$& $(1,1,2,4,6,8,10,12,14,9,4,8)$&$(1,0,1,2,2,2,2,2,2,3,3,4)$&$-4$ & \cr 
$8$&$[2,0,0,0,0,0,0,0,1,1]$& $(1,1,3,5,7,9,11,13,15,9,4,8)$&$(1,0,2,2,2,2,2,2,2,2,3,4)$&$-6$ & \cr 
$8$&$[0,0,0,0,1,0,1,0,0,2]$& $(1,2,3,4,5,6,8,10,13,8,3,8)$&$(1,1,1,1,1,1,2,2,3,3,3,5)$&$2$ & \cr 
$8$&$[0,0,0,1,0,0,0,0,2,1]$& $(1,2,3,4,5,7,9,11,13,7,3,8)$&$(1,1,1,1,1,2,2,2,2,2,4,5)$&$2$ & \cr 
$8$&$[0,0,0,1,0,0,0,1,0,2]$& $(1,2,3,4,5,7,9,11,13,8,3,8)$&$(1,1,1,1,1,2,2,2,2,3,3,5)$&$0$ & \cr 
$8$&$[0,0,1,0,0,0,0,0,1,2]$& $(1,2,3,4,6,8,10,12,14,8,3,8)$&$(1,1,1,1,2,2,2,2,2,2,3,5)$&$-2$ & \cr 
\noalign{\hrule}}
}\cr
Table A.1 Low level weights in the  {\it 11D supergravity} decomposition of the $l_1$ representation of $E_{11}$ \cr}$$
$$\halign{\centerline{#} \cr
\vbox{\offinterlineskip
\halign{\strut \vrule \quad \hfil # \hfil\quad &\vrule \quad \hfil # \hfil\quad 
&\vrule \quad \hfil # \hfil\quad  &\vrule \quad \hfil # \hfil\quad &\vrule \quad \hfil # \hfil\quad &\vrule #
\cr
\noalign{\hrule}
Level&$A_{9}$ weights&$E_{12}$ Root&$E_{12}$ Root&Root length&\cr 
($m_{10},m_{11}$)&&($\alpha_i$ basis)&($e_i$ basis)&squared&\cr 
\noalign{\hrule}
$(0,0)$&$[1,0,0,0,0,0,0,0,0]$& $(1,0,0,0,0,0,0,0,0,0,0,0)$&$(1,-1,0,0,0,0,0,0,0,0,0,0)$&$2$ & \cr 
\noalign{\hrule}$(1,0)$&$[0,0,0,0,0,0,0,0,0]$& $(1,1,1,1,1,1,1,1,1,1,1,0)$&$(1,0,0,0,0,0,0,0,0,0,0,-1)$&$2$ & \cr 
\noalign{\hrule}\noalign{\hrule}$(0,1)$&$[0,0,0,0,0,0,0,0,1]$& $(1,1,1,1,1,1,1,1,1,0,0,1)$&$(1,0,0,0,0,0,0,0,0,0,1,1)$&$2$ & \cr 
\noalign{\hrule}$(1,1)$&$[0,0,0,0,0,0,0,1,0]$& $(1,1,1,1,1,1,1,1,1,1,1,1)$&$(1,0,0,0,0,0,0,0,0,1,1,0)$&$2$ & \cr 
\noalign{\hrule}$(1,2)$&$[0,0,0,0,0,1,0,0,0]$& $(1,1,1,1,1,1,1,2,3,2,1,2)$&$(1,0,0,0,0,0,0,1,1,1,1,1)$&$2$ & \cr 
\noalign{\hrule}$(2,2)$&$[0,0,0,0,1,0,0,0,0]$& $(1,1,1,1,1,1,2,3,4,3,2,2)$&$(1,0,0,0,0,0,1,1,1,1,1,0)$&$2$ & \cr 
\noalign{\hrule}\noalign{\hrule}$(1,3)$&$[0,0,0,1,0,0,0,0,0]$& $(1,1,1,1,1,2,3,4,5,3,1,3)$&$(1,0,0,0,0,1,1,1,1,1,1,2)$&$2$ & \cr 
\noalign{\hrule}$(2,3)$&$[0,0,1,0,0,0,0,0,0]$& $(1,1,1,1,2,3,4,5,6,4,2,3)$&$(1,0,0,0,1,1,1,1,1,1,1,1)$&$0$ & \cr 
$(2,3)$&$[0,0,0,1,0,0,0,0,1]$& $(1,1,1,1,1,2,3,4,5,3,2,3)$&$(1,0,0,0,0,1,1,1,1,1,2,1)$&$2$ & \cr 
\noalign{\hrule}$(3,3)$&$[0,1,0,0,0,0,0,0,0]$& $(1,1,1,2,3,4,5,6,7,5,3,3)$&$(1,0,0,1,1,1,1,1,1,1,1,0)$&$0$ & \cr 
$(3,3)$&$[0,0,1,0,0,0,0,0,1]$& $(1,1,1,1,2,3,4,5,6,4,3,3)$&$(1,0,0,0,1,1,1,1,1,1,2,0)$&$2$ & \cr 
\noalign{\hrule}$(4,3)$&$[1,0,0,0,0,0,0,0,0]$& $(1,1,2,3,4,5,6,7,8,6,4,3)$&$(1,0,1,1,1,1,1,1,1,1,1,-1)$&$2$ & \cr 
\noalign{\hrule}\noalign{\hrule}$(1,4)$&$[0,1,0,0,0,0,0,0,0]$& $(1,1,1,2,3,4,5,6,7,4,1,4)$&$(1,0,0,1,1,1,1,1,1,1,1,3)$&$2$ & \cr 
\noalign{\hrule}$(2,4)$&$[1,0,0,0,0,0,0,0,0]$& $(1,1,2,3,4,5,6,7,8,5,2,4)$&$(1,0,1,1,1,1,1,1,1,1,1,2)$&$-2$ & \cr 
$(2,4)$&$[0,0,1,0,0,0,0,1,0]$& $(1,1,1,1,2,3,4,5,6,4,2,4)$&$(1,0,0,0,1,1,1,1,1,2,2,2)$&$2$ & \cr 
$(2,4)$&$[0,1,0,0,0,0,0,0,1]$& $(1,1,1,2,3,4,5,6,7,4,2,4)$&$(1,0,0,1,1,1,1,1,1,1,2,2)$&$0$ & \cr 
\noalign{\hrule}$(3,4)$&$[0,0,0,0,0,0,0,0,0]$& $(1,2,3,4,5,6,7,8,9,6,3,4)$&$(1,1,1,1,1,1,1,1,1,1,1,1)$&$-4$ & \cr 
$(3,4)$&$[0,0,1,0,0,0,1,0,0]$& $(1,1,1,1,2,3,4,5,7,5,3,4)$&$(1,0,0,0,1,1,1,1,2,2,2,1)$&$2$ & \cr 
$(3,4)$&$[0,1,0,0,0,0,0,1,0]$& $(1,1,1,2,3,4,5,6,7,5,3,4)$&$(1,0,0,1,1,1,1,1,1,2,2,1)$&$0$ & \cr 
$(3,4)$&$[1,0,0,0,0,0,0,0,1]$& $(1,1,2,3,4,5,6,7,8,5,3,4)$&$(1,0,1,1,1,1,1,1,1,1,2,1)$&$-2$ & \cr 
$(3,4)$&$[0,1,0,0,0,0,0,0,2]$& $(1,1,1,2,3,4,5,6,7,4,3,4)$&$(1,0,0,1,1,1,1,1,1,1,3,1)$&$2$ & \cr 
\noalign{\hrule}$(4,4)$&$[0,0,0,0,0,0,0,0,1]$& $(1,2,3,4,5,6,7,8,9,6,4,4)$&$(1,1,1,1,1,1,1,1,1,1,2,0)$&$-2$ & \cr 
$(4,4)$&$[0,1,0,0,0,0,1,0,0]$& $(1,1,1,2,3,4,5,6,8,6,4,4)$&$(1,0,0,1,1,1,1,1,2,2,2,0)$&$2$ & \cr 
$(4,4)$&$[1,0,0,0,0,0,0,1,0]$& $(1,1,2,3,4,5,6,7,8,6,4,4)$&$(1,0,1,1,1,1,1,1,1,2,2,0)$&$0$ & \cr 
$(4,4)$&$[1,0,0,0,0,0,0,0,2]$& $(1,1,2,3,4,5,6,7,8,5,4,4)$&$(1,0,1,1,1,1,1,1,1,1,3,0)$&$2$ & \cr 
\noalign{\hrule}$(5,4)$&$[0,0,0,0,0,0,0,1,0]$& $(1,2,3,4,5,6,7,8,9,7,5,4)$&$(1,1,1,1,1,1,1,1,1,2,2,-1)$&$2$ & \cr 
\noalign{\hrule}\noalign{\hrule}$(1,5)$&$[0,0,0,0,0,0,0,0,0]$& $(1,2,3,4,5,6,7,8,9,5,1,5)$&$(1,1,1,1,1,1,1,1,1,1,1,4)$&$2$ & \cr 
\noalign{\hrule}$(2,5)$&$[0,0,0,0,0,0,0,0,1]$& $(1,2,3,4,5,6,7,8,9,5,2,5)$&$(1,1,1,1,1,1,1,1,1,1,2,3)$&$-2$ & \cr 
$(2,5)$&$[0,1,0,0,0,0,1,0,0]$& $(1,1,1,2,3,4,5,6,8,5,2,5)$&$(1,0,0,1,1,1,1,1,2,2,2,3)$&$2$ & \cr 
$(2,5)$&$[1,0,0,0,0,0,0,1,0]$& $(1,1,2,3,4,5,6,7,8,5,2,5)$&$(1,0,1,1,1,1,1,1,1,2,2,3)$&$0$ & \cr 
$(2,5)$&$[1,0,0,0,0,0,0,0,2]$& $(1,1,2,3,4,5,6,7,8,4,2,5)$&$(1,0,1,1,1,1,1,1,1,1,3,3)$&$2$ & \cr 
\noalign{\hrule}$(3,5)$&$[0,0,0,0,0,0,0,1,0]$& $(1,2,3,4,5,6,7,8,9,6,3,5)$&$(1,1,1,1,1,1,1,1,1,2,2,2)$&$-4$ & \cr 
$(3,5)$&$[0,0,1,0,1,0,0,0,0]$& $(1,1,1,1,2,3,5,7,9,6,3,5)$&$(1,0,0,0,1,1,2,2,2,2,2,2)$&$2$ & \cr 
$(3,5)$&$[0,1,0,0,0,1,0,0,0]$& $(1,1,1,2,3,4,5,7,9,6,3,5)$&$(1,0,0,1,1,1,1,2,2,2,2,2)$&$0$ & \cr 
$(3,5)$&$[1,0,0,0,0,0,1,0,0]$& $(1,1,2,3,4,5,6,7,9,6,3,5)$&$(1,0,1,1,1,1,1,1,2,2,2,2)$&$-2$ & \cr 
$(3,5)$&$[0,0,0,0,0,0,0,0,2]$& $(1,2,3,4,5,6,7,8,9,5,3,5)$&$(1,1,1,1,1,1,1,1,1,1,3,2)$&$-2$ & \cr 
$(3,5)$&$[0,1,0,0,0,0,1,0,1]$& $(1,1,1,2,3,4,5,6,8,5,3,5)$&$(1,0,0,1,1,1,1,1,2,2,3,2)$&$2$ & \cr 
$(3,5)$&$[1,0,0,0,0,0,0,1,1]$& $(1,1,2,3,4,5,6,7,8,5,3,5)$&$(1,0,1,1,1,1,1,1,1,2,3,2)$&$0$ & \cr 
\noalign{\hrule}$(4,5)$&$[0,0,0,0,0,0,1,0,0]$& $(1,2,3,4,5,6,7,8,10,7,4,5)$&$(1,1,1,1,1,1,1,1,2,2,2,1)$&$-4$ & \cr 
$(4,5)$&$[0,0,1,1,0,0,0,0,0]$& $(1,1,1,1,2,4,6,8,10,7,4,5)$&$(1,0,0,0,1,2,2,2,2,2,2,1)$&$2$ & \cr 
$(4,5)$&$[0,1,0,0,1,0,0,0,0]$& $(1,1,1,2,3,4,6,8,10,7,4,5)$&$(1,0,0,1,1,1,2,2,2,2,2,1)$&$0$ & \cr 
$(4,5)$&$[1,0,0,0,0,1,0,0,0]$& $(1,1,2,3,4,5,6,8,10,7,4,5)$&$(1,0,1,1,1,1,1,2,2,2,2,1)$&$-2$ & \cr 
$(4,5)$&$[0,0,0,0,0,0,0,1,1]$& $(1,2,3,4,5,6,7,8,9,6,4,5)$&$(1,1,1,1,1,1,1,1,1,2,3,1)$&$-2$ & \cr 
$(4,5)$&$[0,1,0,0,0,1,0,0,1]$& $(1,1,1,2,3,4,5,7,9,6,4,5)$&$(1,0,0,1,1,1,1,2,2,2,3,1)$&$2$ & \cr 
$(4,5)$&$[1,0,0,0,0,0,0,2,0]$& $(1,1,2,3,4,5,6,7,8,6,4,5)$&$(1,0,1,1,1,1,1,1,1,3,3,1)$&$2$ & \cr 
$(4,5)$&$[1,0,0,0,0,0,1,0,1]$& $(1,1,2,3,4,5,6,7,9,6,4,5)$&$(1,0,1,1,1,1,1,1,2,2,3,1)$&$0$ & \cr 
$(4,5)$&$[0,0,0,0,0,0,0,0,3]$& $(1,2,3,4,5,6,7,8,9,5,4,5)$&$(1,1,1,1,1,1,1,1,1,1,4,1)$&$2$ & \cr 
\noalign{\hrule}
}
}\cr
Table A.2 Low level weights in the {\it IIA supergravity} decomposition of the $l_1$ representation of $E_{11} (continued)$ \cr}$$
$$\halign{\centerline{#} \cr
\vbox{\offinterlineskip
\halign{\strut \vrule \quad \hfil # \hfil\quad &\vrule \quad \hfil # \hfil\quad 
&\vrule \quad \hfil # \hfil\quad  &\vrule \quad \hfil # \hfil\quad &\vrule \quad \hfil # \hfil\quad &\vrule #
\cr
\noalign{\hrule}
Level&$A_{9}$ weights&$E_{12}$ Root&$E_{12}$ Root&Root length&\cr 
($m_{10},m_{11}$)&&($\alpha_i$ basis)&($e_i$ basis)&squared&\cr 
\noalign{\hrule}
$(5,5)$&$[0,0,0,0,0,1,0,0,0]$& $(1,2,3,4,5,6,7,9,11,8,5,5)$&$(1,1,1,1,1,1,1,2,2,2,2,0)$&$-2$ & \cr 
$(5,5)$&$[0,1,0,1,0,0,0,0,0]$& $(1,1,1,2,3,5,7,9,11,8,5,5)$&$(1,0,0,1,1,2,2,2,2,2,2,0)$&$2$ & \cr 
$(5,5)$&$[1,0,0,0,1,0,0,0,0]$& $(1,1,2,3,4,5,7,9,11,8,5,5)$&$(1,0,1,1,1,1,2,2,2,2,2,0)$&$0$ & \cr 
$(5,5)$&$[0,0,0,0,0,0,0,2,0]$& $(1,2,3,4,5,6,7,8,9,7,5,5)$&$(1,1,1,1,1,1,1,1,1,3,3,0)$&$2$ & \cr 
$(5,5)$&$[0,0,0,0,0,0,1,0,1]$& $(1,2,3,4,5,6,7,8,10,7,5,5)$&$(1,1,1,1,1,1,1,1,2,2,3,0)$&$0$ & \cr 
$(5,5)$&$[1,0,0,0,0,1,0,0,1]$& $(1,1,2,3,4,5,6,8,10,7,5,5)$&$(1,0,1,1,1,1,1,2,2,2,3,0)$&$2$ & \cr 
\noalign{\hrule}$(6,5)$&$[0,0,0,0,1,0,0,0,0]$& $(1,2,3,4,5,6,8,10,12,9,6,5)$&$(1,1,1,1,1,1,2,2,2,2,2,-1)$&$2$ & \cr 
\noalign{\hrule}\noalign{\hrule}$(2,6)$&$[0,0,0,0,0,0,1,0,0]$& $(1,2,3,4,5,6,7,8,10,6,2,6)$&$(1,1,1,1,1,1,1,1,2,2,2,4)$&$0$ & \cr 
$(2,6)$&$[1,0,0,0,0,1,0,0,0]$& $(1,1,2,3,4,5,6,8,10,6,2,6)$&$(1,0,1,1,1,1,1,2,2,2,2,4)$&$2$ & \cr 
$(2,6)$&$[0,0,0,0,0,0,0,1,1]$& $(1,2,3,4,5,6,7,8,9,5,2,6)$&$(1,1,1,1,1,1,1,1,1,2,3,4)$&$2$ & \cr 
\noalign{\hrule}$(3,6)$&$[0,0,0,0,0,1,0,0,0]$& $(1,2,3,4,5,6,7,9,11,7,3,6)$&$(1,1,1,1,1,1,1,2,2,2,2,3)$&$-4$ & \cr 
$(3,6)$&$[0,0,2,0,0,0,0,0,0]$& $(1,1,1,1,3,5,7,9,11,7,3,6)$&$(1,0,0,0,2,2,2,2,2,2,2,3)$&$2$ & \cr 
$(3,6)$&$[0,1,0,1,0,0,0,0,0]$& $(1,1,1,2,3,5,7,9,11,7,3,6)$&$(1,0,0,1,1,2,2,2,2,2,2,3)$&$0$ & \cr 
$(3,6)$&$[1,0,0,0,1,0,0,0,0]$& $(1,1,2,3,4,5,7,9,11,7,3,6)$&$(1,0,1,1,1,1,2,2,2,2,2,3)$&$-2$ & \cr 
$(3,6)$&$[0,0,0,0,0,0,0,2,0]$& $(1,2,3,4,5,6,7,8,9,6,3,6)$&$(1,1,1,1,1,1,1,1,1,3,3,3)$&$0$ & \cr 
$(3,6)$&$[0,0,0,0,0,0,1,0,1]$& $(1,2,3,4,5,6,7,8,10,6,3,6)$&$(1,1,1,1,1,1,1,1,2,2,3,3)$&$-2$ & \cr 
$(3,6)$&$[0,1,0,0,1,0,0,0,1]$& $(1,1,1,2,3,4,6,8,10,6,3,6)$&$(1,0,0,1,1,1,2,2,2,2,3,3)$&$2$ & \cr 
$(3,6)$&$[1,0,0,0,0,0,1,1,0]$& $(1,1,2,3,4,5,6,7,9,6,3,6)$&$(1,0,1,1,1,1,1,1,2,3,3,3)$&$2$ & \cr 
$(3,6)$&$[1,0,0,0,0,1,0,0,1]$& $(1,1,2,3,4,5,6,8,10,6,3,6)$&$(1,0,1,1,1,1,1,2,2,2,3,3)$&$0$ & \cr 
$(3,6)$&$[0,0,0,0,0,0,0,1,2]$& $(1,2,3,4,5,6,7,8,9,5,3,6)$&$(1,1,1,1,1,1,1,1,1,2,4,3)$&$2$ & \cr 
\noalign{\hrule}$(4,6)$&$[0,0,0,0,1,0,0,0,0]$& $(1,2,3,4,5,6,8,10,12,8,4,6)$&$(1,1,1,1,1,1,2,2,2,2,2,2)$&$-6$ & \cr 
$(4,6)$&$[0,1,1,0,0,0,0,0,0]$& $(1,1,1,2,4,6,8,10,12,8,4,6)$&$(1,0,0,1,2,2,2,2,2,2,2,2)$&$-2$ & \cr 
$(4,6)$&$[1,0,0,1,0,0,0,0,0]$& $(1,1,2,3,4,6,8,10,12,8,4,6)$&$(1,0,1,1,1,2,2,2,2,2,2,2)$&$-4$ & \cr 
$(4,6)$&$[0,0,0,0,0,0,1,1,0]$& $(1,2,3,4,5,6,7,8,10,7,4,6)$&$(1,1,1,1,1,1,1,1,2,3,3,2)$&$-2$ & \cr 
$(4,6)$&$[0,0,0,0,0,1,0,0,1]$& $(1,2,3,4,5,6,7,9,11,7,4,6)$&$(1,1,1,1,1,1,1,2,2,2,3,2)$&$-4$ & \cr 
$(4,6)$&$[0,0,2,0,0,0,0,0,1]$& $(1,1,1,1,3,5,7,9,11,7,4,6)$&$(1,0,0,0,2,2,2,2,2,2,3,2)$&$2$ & \cr 
$(4,6)$&$[0,1,0,0,1,0,0,1,0]$& $(1,1,1,2,3,4,6,8,10,7,4,6)$&$(1,0,0,1,1,1,2,2,2,3,3,2)$&$2$ & \cr 
$(4,6)$&$[0,1,0,1,0,0,0,0,1]$& $(1,1,1,2,3,5,7,9,11,7,4,6)$&$(1,0,0,1,1,2,2,2,2,2,3,2)$&$0$ & \cr 
$(4,6)$&$[1,0,0,0,0,0,2,0,0]$& $(1,1,2,3,4,5,6,7,10,7,4,6)$&$(1,0,1,1,1,1,1,1,3,3,3,2)$&$2$ & \cr 
$(4,6)$&$[1,0,0,0,0,1,0,1,0]$& $(1,1,2,3,4,5,6,8,10,7,4,6)$&$(1,0,1,1,1,1,1,2,2,3,3,2)$&$0$ & \cr 
$(4,6)$&$[1,0,0,0,1,0,0,0,1]$& $(1,1,2,3,4,5,7,9,11,7,4,6)$&$(1,0,1,1,1,1,2,2,2,2,3,2)$&$-2$ & \cr 
$(4,6)$&$[0,0,0,0,0,0,0,2,1]$& $(1,2,3,4,5,6,7,8,9,6,4,6)$&$(1,1,1,1,1,1,1,1,1,3,4,2)$&$2$ & \cr 
$(4,6)$&$[0,0,0,0,0,0,1,0,2]$& $(1,2,3,4,5,6,7,8,10,6,4,6)$&$(1,1,1,1,1,1,1,1,2,2,4,2)$&$0$ & \cr 
$(4,6)$&$[1,0,0,0,0,1,0,0,2]$& $(1,1,2,3,4,5,6,8,10,6,4,6)$&$(1,0,1,1,1,1,1,2,2,2,4,2)$&$2$ & \cr 
\noalign{\hrule}$(5,6)$&$[0,0,0,1,0,0,0,0,0]$& $(1,2,3,4,5,7,9,11,13,9,5,6)$&$(1,1,1,1,1,2,2,2,2,2,2,1)$&$-6$ & \cr 
$(5,6)$&$[0,2,0,0,0,0,0,0,0]$& $(1,1,1,3,5,7,9,11,13,9,5,6)$&$(1,0,0,2,2,2,2,2,2,2,2,1)$&$-2$ & \cr 
$(5,6)$&$[1,0,1,0,0,0,0,0,0]$& $(1,1,2,3,5,7,9,11,13,9,5,6)$&$(1,0,1,1,2,2,2,2,2,2,2,1)$&$-4$ & \cr 
$(5,6)$&$[0,0,0,0,0,0,2,0,0]$& $(1,2,3,4,5,6,7,8,11,8,5,6)$&$(1,1,1,1,1,1,1,1,3,3,3,1)$&$0$ & \cr 
$(5,6)$&$[0,0,0,0,0,1,0,1,0]$& $(1,2,3,4,5,6,7,9,11,8,5,6)$&$(1,1,1,1,1,1,1,2,2,3,3,1)$&$-2$ & \cr 
$(5,6)$&$[0,0,0,0,1,0,0,0,1]$& $(1,2,3,4,5,6,8,10,12,8,5,6)$&$(1,1,1,1,1,1,2,2,2,2,3,1)$&$-4$ & \cr 
$(5,6)$&$[0,1,0,1,0,0,0,1,0]$& $(1,1,1,2,3,5,7,9,11,8,5,6)$&$(1,0,0,1,1,2,2,2,2,3,3,1)$&$2$ & \cr 
$(5,6)$&$[0,1,1,0,0,0,0,0,1]$& $(1,1,1,2,4,6,8,10,12,8,5,6)$&$(1,0,0,1,2,2,2,2,2,2,3,1)$&$0$ & \cr 
$(5,6)$&$[1,0,0,0,0,1,1,0,0]$& $(1,1,2,3,4,5,6,8,11,8,5,6)$&$(1,0,1,1,1,1,1,2,3,3,3,1)$&$2$ & \cr 
$(5,6)$&$[1,0,0,0,1,0,0,1,0]$& $(1,1,2,3,4,5,7,9,11,8,5,6)$&$(1,0,1,1,1,1,2,2,2,3,3,1)$&$0$ & \cr 
$(5,6)$&$[1,0,0,1,0,0,0,0,1]$& $(1,1,2,3,4,6,8,10,12,8,5,6)$&$(1,0,1,1,1,2,2,2,2,2,3,1)$&$-2$ & \cr 
$(5,6)$&$[0,0,0,0,0,0,1,1,1]$& $(1,2,3,4,5,6,7,8,10,7,5,6)$&$(1,1,1,1,1,1,1,1,2,3,4,1)$&$2$ & \cr 
$(5,6)$&$[0,0,0,0,0,1,0,0,2]$& $(1,2,3,4,5,6,7,9,11,7,5,6)$&$(1,1,1,1,1,1,1,2,2,2,4,1)$&$0$ & \cr 
$(5,6)$&$[1,0,0,0,1,0,0,0,2]$& $(1,1,2,3,4,5,7,9,11,7,5,6)$&$(1,0,1,1,1,1,2,2,2,2,4,1)$&$2$ & \cr 
\noalign{\hrule}$(6,6)$&$[0,0,1,0,0,0,0,0,0]$& $(1,2,3,4,6,8,10,12,14,10,6,6)$&$(1,1,1,1,2,2,2,2,2,2,2,0)$&$-4$ & \cr 
\noalign{\hrule}
}
}\cr
Table A.2 Low level weights in the {\it IIA supergravity} of the $l_1$ representation of $E_{11}$ (continued)\cr}$$
$$\halign{\centerline{#} \cr
\vbox{\offinterlineskip
\halign{\strut \vrule \quad \hfil # \hfil\quad &\vrule \quad \hfil # \hfil\quad 
&\vrule \quad \hfil # \hfil\quad  &\vrule \quad \hfil # \hfil\quad &\vrule \quad \hfil # \hfil\quad &\vrule #
\cr
\noalign{\hrule}
Level&$A_{9}$ weights&$E_{12}$ Root&$E_{12}$ Root&Root length&\cr 
($m_{10},m_{11}$)&&($\alpha_i$ basis)&($e_i$ basis)&squared&\cr 
\noalign{\hrule}
$(6,6)$&$[1,1,0,0,0,0,0,0,0]$& $(1,1,2,4,6,8,10,12,14,10,6,6)$&$(1,0,1,2,2,2,2,2,2,2,2,0)$&$-2$ & \cr 
$(6,6)$&$[0,0,0,0,0,1,1,0,0]$& $(1,2,3,4,5,6,7,9,12,9,6,6)$&$(1,1,1,1,1,1,1,2,3,3,3,0)$&$2$ & \cr 
$(6,6)$&$[0,0,0,0,1,0,0,1,0]$& $(1,2,3,4,5,6,8,10,12,9,6,6)$&$(1,1,1,1,1,1,2,2,2,3,3,0)$&$0$ & \cr 
$(6,6)$&$[0,0,0,1,0,0,0,0,1]$& $(1,2,3,4,5,7,9,11,13,9,6,6)$&$(1,1,1,1,1,2,2,2,2,2,3,0)$&$-2$ & \cr 
$(6,6)$&$[0,2,0,0,0,0,0,0,1]$& $(1,1,1,3,5,7,9,11,13,9,6,6)$&$(1,0,0,2,2,2,2,2,2,2,3,0)$&$2$ & \cr 
$(6,6)$&$[1,0,0,1,0,0,0,1,0]$& $(1,1,2,3,4,6,8,10,12,9,6,6)$&$(1,0,1,1,1,2,2,2,2,3,3,0)$&$2$ & \cr 
$(6,6)$&$[1,0,1,0,0,0,0,0,1]$& $(1,1,2,3,5,7,9,11,13,9,6,6)$&$(1,0,1,1,2,2,2,2,2,2,3,0)$&$0$ & \cr 
$(6,6)$&$[0,0,0,0,1,0,0,0,2]$& $(1,2,3,4,5,6,8,10,12,8,6,6)$&$(1,1,1,1,1,1,2,2,2,2,4,0)$&$2$ & \cr 
\noalign{\hrule}\noalign{\hrule}$(2,7)$&$[0,0,0,0,1,0,0,0,0]$& $(1,2,3,4,5,6,8,10,12,7,2,7)$&$(1,1,1,1,1,1,2,2,2,2,2,5)$&$2$ & \cr 
\noalign{\hrule}$(3,7)$&$[0,0,0,1,0,0,0,0,0]$& $(1,2,3,4,5,7,9,11,13,8,3,7)$&$(1,1,1,1,1,2,2,2,2,2,2,4)$&$-4$ & \cr 
$(3,7)$&$[0,2,0,0,0,0,0,0,0]$& $(1,1,1,3,5,7,9,11,13,8,3,7)$&$(1,0,0,2,2,2,2,2,2,2,2,4)$&$0$ & \cr 
$(3,7)$&$[1,0,1,0,0,0,0,0,0]$& $(1,1,2,3,5,7,9,11,13,8,3,7)$&$(1,0,1,1,2,2,2,2,2,2,2,4)$&$-2$ & \cr 
$(3,7)$&$[0,0,0,0,0,0,2,0,0]$& $(1,2,3,4,5,6,7,8,11,7,3,7)$&$(1,1,1,1,1,1,1,1,3,3,3,4)$&$2$ & \cr 
$(3,7)$&$[0,0,0,0,0,1,0,1,0]$& $(1,2,3,4,5,6,7,9,11,7,3,7)$&$(1,1,1,1,1,1,1,2,2,3,3,4)$&$0$ & \cr 
$(3,7)$&$[0,0,0,0,1,0,0,0,1]$& $(1,2,3,4,5,6,8,10,12,7,3,7)$&$(1,1,1,1,1,1,2,2,2,2,3,4)$&$-2$ & \cr 
$(3,7)$&$[0,1,1,0,0,0,0,0,1]$& $(1,1,1,2,4,6,8,10,12,7,3,7)$&$(1,0,0,1,2,2,2,2,2,2,3,4)$&$2$ & \cr 
$(3,7)$&$[1,0,0,0,1,0,0,1,0]$& $(1,1,2,3,4,5,7,9,11,7,3,7)$&$(1,0,1,1,1,1,2,2,2,3,3,4)$&$2$ & \cr 
$(3,7)$&$[1,0,0,1,0,0,0,0,1]$& $(1,1,2,3,4,6,8,10,12,7,3,7)$&$(1,0,1,1,1,2,2,2,2,2,3,4)$&$0$ & \cr 
$(3,7)$&$[0,0,0,0,0,1,0,0,2]$& $(1,2,3,4,5,6,7,9,11,6,3,7)$&$(1,1,1,1,1,1,1,2,2,2,4,4)$&$2$ & \cr 
\noalign{\hrule}$(4,7)$&$[0,0,1,0,0,0,0,0,0]$& $(1,2,3,4,6,8,10,12,14,9,4,7)$&$(1,1,1,1,2,2,2,2,2,2,2,3)$&$-8$ & \cr 
$(4,7)$&$[1,1,0,0,0,0,0,0,0]$& $(1,1,2,4,6,8,10,12,14,9,4,7)$&$(1,0,1,2,2,2,2,2,2,2,2,3)$&$-6$ & \cr 
$(4,7)$&$[0,0,0,0,0,1,1,0,0]$& $(1,2,3,4,5,6,7,9,12,8,4,7)$&$(1,1,1,1,1,1,1,2,3,3,3,3)$&$-2$ & \cr 
$(4,7)$&$[0,0,0,0,1,0,0,1,0]$& $(1,2,3,4,5,6,8,10,12,8,4,7)$&$(1,1,1,1,1,1,2,2,2,3,3,3)$&$-4$ & \cr 
$(4,7)$&$[0,0,0,1,0,0,0,0,1]$& $(1,2,3,4,5,7,9,11,13,8,4,7)$&$(1,1,1,1,1,2,2,2,2,2,3,3)$&$-6$ & \cr 
$(4,7)$&$[0,1,0,1,0,0,1,0,0]$& $(1,1,1,2,3,5,7,9,12,8,4,7)$&$(1,0,0,1,1,2,2,2,3,3,3,3)$&$2$ & \cr 
$(4,7)$&$[0,1,1,0,0,0,0,1,0]$& $(1,1,1,2,4,6,8,10,12,8,4,7)$&$(1,0,0,1,2,2,2,2,2,3,3,3)$&$0$ & \cr 
$(4,7)$&$[0,2,0,0,0,0,0,0,1]$& $(1,1,1,3,5,7,9,11,13,8,4,7)$&$(1,0,0,2,2,2,2,2,2,2,3,3)$&$-2$ & \cr 
$(4,7)$&$[1,0,0,0,0,2,0,0,0]$& $(1,1,2,3,4,5,6,9,12,8,4,7)$&$(1,0,1,1,1,1,1,3,3,3,3,3)$&$2$ & \cr 
$(4,7)$&$[1,0,0,0,1,0,1,0,0]$& $(1,1,2,3,4,5,7,9,12,8,4,7)$&$(1,0,1,1,1,1,2,2,3,3,3,3)$&$0$ & \cr 
$(4,7)$&$[1,0,0,1,0,0,0,1,0]$& $(1,1,2,3,4,6,8,10,12,8,4,7)$&$(1,0,1,1,1,2,2,2,2,3,3,3)$&$-2$ & \cr 
$(4,7)$&$[1,0,1,0,0,0,0,0,1]$& $(1,1,2,3,5,7,9,11,13,8,4,7)$&$(1,0,1,1,2,2,2,2,2,2,3,3)$&$-4$ & \cr 
$(4,7)$&$[0,0,0,0,0,0,2,0,1]$& $(1,2,3,4,5,6,7,8,11,7,4,7)$&$(1,1,1,1,1,1,1,1,3,3,4,3)$&$2$ & \cr 
$(4,7)$&$[0,0,0,0,0,1,0,1,1]$& $(1,2,3,4,5,6,7,9,11,7,4,7)$&$(1,1,1,1,1,1,1,2,2,3,4,3)$&$0$ & \cr 
$(4,7)$&$[0,0,0,0,1,0,0,0,2]$& $(1,2,3,4,5,6,8,10,12,7,4,7)$&$(1,1,1,1,1,1,2,2,2,2,4,3)$&$-2$ & \cr 
$(4,7)$&$[0,1,1,0,0,0,0,0,2]$& $(1,1,1,2,4,6,8,10,12,7,4,7)$&$(1,0,0,1,2,2,2,2,2,2,4,3)$&$2$ & \cr 
$(4,7)$&$[1,0,0,0,1,0,0,1,1]$& $(1,1,2,3,4,5,7,9,11,7,4,7)$&$(1,0,1,1,1,1,2,2,2,3,4,3)$&$2$ & \cr 
$(4,7)$&$[1,0,0,1,0,0,0,0,2]$& $(1,1,2,3,4,6,8,10,12,7,4,7)$&$(1,0,1,1,1,2,2,2,2,2,4,3)$&$0$ & \cr 
\noalign{\hrule}
}
}\cr
Table A.2 Low level weights in the {\it IIA supergravity} of the $l_1$ representation of $E_{11}$ \cr}$$
\eject
$$\halign{\centerline{#} \cr
\vbox{\offinterlineskip
\halign{\strut \vrule \quad \hfil # \hfil\quad &\vrule \quad \hfil # \hfil\quad 
&\vrule \quad \hfil # \hfil\quad  &\vrule \quad \hfil # \hfil\quad &\vrule \quad \hfil # \hfil\quad &\vrule #
\cr
\noalign{\hrule}
Level&$A_{9}$ weights&$E_{12}$ Root&$E_{12}$ Root&Root length&\cr 
($m_{9},m_{10}$)&&($\alpha_i$ basis)&($e_i$ basis)&squared&\cr 
\noalign{\hrule}
$(0,0)$&$[1,0,0,0,0,0,0,0,0]$& $(1,0,0,0,0,0,0,0,0,0,0,0)$&$(1,-1,0,0,0,0,0,0,0,0,0,0)$&$2$ & \cr 
\noalign{\hrule}$(1,0)$&$[0,0,0,0,0,0,0,0,1]$& $(1,1,1,1,1,1,1,1,1,1,0,0)$&$(1,0,0,0,0,0,0,0,0,0,-1,0)$&$2$ & \cr 
\noalign{\hrule}$(1,1)$&$[0,0,0,0,0,0,0,0,1]$& $(1,1,1,1,1,1,1,1,1,1,1,0)$&$(1,0,0,0,0,0,0,0,0,0,0,-1)$&$2$ & \cr 
\noalign{\hrule}$(2,1)$&$[0,0,0,0,0,0,1,0,0]$& $(1,1,1,1,1,1,1,1,2,2,1,1)$&$(1,0,0,0,0,0,0,0,1,1,0,0)$&$2$ & \cr 
\noalign{\hrule}$(3,1)$&$[0,0,0,0,1,0,0,0,0]$& $(1,1,1,1,1,1,2,3,4,3,1,2)$&$(1,0,0,0,0,0,1,1,1,1,0,1)$&$2$ & \cr 
\noalign{\hrule}$(3,2)$&$[0,0,0,0,1,0,0,0,0]$& $(1,1,1,1,1,1,2,3,4,3,2,2)$&$(1,0,0,0,0,0,1,1,1,1,1,0)$&$2$ & \cr 
\noalign{\hrule}\noalign{\hrule}$(4,1)$&$[0,0,1,0,0,0,0,0,0]$& $(1,1,1,1,2,3,4,5,6,4,1,3)$&$(1,0,0,0,1,1,1,1,1,1,0,2)$&$2$ & \cr 
\noalign{\hrule}$(4,2)$&$[0,0,1,0,0,0,0,0,0]$& $(1,1,1,1,2,3,4,5,6,4,2,3)$&$(1,0,0,0,1,1,1,1,1,1,1,1)$&$0$ & \cr 
$(4,2)$&$[0,0,0,1,0,0,0,0,1]$& $(1,1,1,1,1,2,3,4,5,4,2,2)$&$(1,0,0,0,0,1,1,1,1,1,0,0)$&$2$ & \cr 
\noalign{\hrule}$(4,3)$&$[0,0,1,0,0,0,0,0,0]$& $(1,1,1,1,2,3,4,5,6,4,3,3)$&$(1,0,0,0,1,1,1,1,1,1,2,0)$&$2$ & \cr 
\noalign{\hrule}\noalign{\hrule}$(5,1)$&$[1,0,0,0,0,0,0,0,0]$& $(1,1,2,3,4,5,6,7,8,5,1,4)$&$(1,0,1,1,1,1,1,1,1,1,0,3)$&$2$ & \cr 
\noalign{\hrule}$(5,2)$&$[1,0,0,0,0,0,0,0,0]$& $(1,1,2,3,4,5,6,7,8,5,2,4)$&$(1,0,1,1,1,1,1,1,1,1,1,2)$&$-2$ & \cr 
$(5,2)$&$[0,0,1,0,0,0,0,1,0]$& $(1,1,1,1,2,3,4,5,6,5,2,3)$&$(1,0,0,0,1,1,1,1,1,2,0,1)$&$2$ & \cr 
$(5,2)$&$[0,1,0,0,0,0,0,0,1]$& $(1,1,1,2,3,4,5,6,7,5,2,3)$&$(1,0,0,1,1,1,1,1,1,1,0,1)$&$0$ & \cr 
\noalign{\hrule}$(5,3)$&$[1,0,0,0,0,0,0,0,0]$& $(1,1,2,3,4,5,6,7,8,5,3,4)$&$(1,0,1,1,1,1,1,1,1,1,2,1)$&$-2$ & \cr 
$(5,3)$&$[0,0,1,0,0,0,0,1,0]$& $(1,1,1,1,2,3,4,5,6,5,3,3)$&$(1,0,0,0,1,1,1,1,1,2,1,0)$&$2$ & \cr 
$(5,3)$&$[0,1,0,0,0,0,0,0,1]$& $(1,1,1,2,3,4,5,6,7,5,3,3)$&$(1,0,0,1,1,1,1,1,1,1,1,0)$&$0$ & \cr 
\noalign{\hrule}$(5,4)$&$[1,0,0,0,0,0,0,0,0]$& $(1,1,2,3,4,5,6,7,8,5,4,4)$&$(1,0,1,1,1,1,1,1,1,1,3,0)$&$2$ & \cr 
\noalign{\hrule}\noalign{\hrule}$(6,2)$&$[0,0,0,0,0,0,0,0,1]$& $(1,2,3,4,5,6,7,8,9,6,2,4)$&$(1,1,1,1,1,1,1,1,1,1,0,2)$&$-2$ & \cr 
$(6,2)$&$[0,1,0,0,0,0,1,0,0]$& $(1,1,1,2,3,4,5,6,8,6,2,4)$&$(1,0,0,1,1,1,1,1,2,2,0,2)$&$2$ & \cr 
$(6,2)$&$[1,0,0,0,0,0,0,1,0]$& $(1,1,2,3,4,5,6,7,8,6,2,4)$&$(1,0,1,1,1,1,1,1,1,2,0,2)$&$0$ & \cr 
$(6,2)$&$[1,0,0,0,0,0,0,0,2]$& $(1,1,2,3,4,5,6,7,8,6,2,3)$&$(1,0,1,1,1,1,1,1,1,1,-1,1)$&$2$ & \cr 
\noalign{\hrule}$(6,3)$&$[0,0,0,0,0,0,0,0,1]$& $(1,2,3,4,5,6,7,8,9,6,3,4)$&$(1,1,1,1,1,1,1,1,1,1,1,1)$&$-4$ & \cr 
$(6,3)$&$[0,0,1,0,0,1,0,0,0]$& $(1,1,1,1,2,3,4,6,8,6,3,4)$&$(1,0,0,0,1,1,1,2,2,2,1,1)$&$2$ & \cr 
$(6,3)$&$[0,1,0,0,0,0,1,0,0]$& $(1,1,1,2,3,4,5,6,8,6,3,4)$&$(1,0,0,1,1,1,1,1,2,2,1,1)$&$0$ & \cr 
$(6,3)$&$[1,0,0,0,0,0,0,1,0]$& $(1,1,2,3,4,5,6,7,8,6,3,4)$&$(1,0,1,1,1,1,1,1,1,2,1,1)$&$-2$ & \cr 
$(6,3)$&$[0,1,0,0,0,0,0,1,1]$& $(1,1,1,2,3,4,5,6,7,6,3,3)$&$(1,0,0,1,1,1,1,1,1,2,0,0)$&$2$ & \cr 
$(6,3)$&$[1,0,0,0,0,0,0,0,2]$& $(1,1,2,3,4,5,6,7,8,6,3,3)$&$(1,0,1,1,1,1,1,1,1,1,0,0)$&$0$ & \cr 
\noalign{\hrule}$(6,4)$&$[0,0,0,0,0,0,0,0,1]$& $(1,2,3,4,5,6,7,8,9,6,4,4)$&$(1,1,1,1,1,1,1,1,1,1,2,0)$&$-2$ & \cr 
$(6,4)$&$[0,1,0,0,0,0,1,0,0]$& $(1,1,1,2,3,4,5,6,8,6,4,4)$&$(1,0,0,1,1,1,1,1,2,2,2,0)$&$2$ & \cr 
$(6,4)$&$[1,0,0,0,0,0,0,1,0]$& $(1,1,2,3,4,5,6,7,8,6,4,4)$&$(1,0,1,1,1,1,1,1,1,2,2,0)$&$0$ & \cr 
$(6,4)$&$[1,0,0,0,0,0,0,0,2]$& $(1,1,2,3,4,5,6,7,8,6,4,3)$&$(1,0,1,1,1,1,1,1,1,1,1,-1)$&$2$ & \cr 
\noalign{\hrule}\noalign{\hrule}$(7,2)$&$[0,0,0,0,0,0,1,0,0]$& $(1,2,3,4,5,6,7,8,10,7,2,5)$&$(1,1,1,1,1,1,1,1,2,2,0,3)$&$0$ & \cr 
$(7,2)$&$[1,0,0,0,0,1,0,0,0]$& $(1,1,2,3,4,5,6,8,10,7,2,5)$&$(1,0,1,1,1,1,1,2,2,2,0,3)$&$2$ & \cr 
$(7,2)$&$[0,0,0,0,0,0,0,1,1]$& $(1,2,3,4,5,6,7,8,9,7,2,4)$&$(1,1,1,1,1,1,1,1,1,2,-1,2)$&$2$ & \cr 
\noalign{\hrule}$(7,3)$&$[0,0,0,0,0,0,1,0,0]$& $(1,2,3,4,5,6,7,8,10,7,3,5)$&$(1,1,1,1,1,1,1,1,2,2,1,2)$&$-4$ & \cr 
$(7,3)$&$[0,0,1,1,0,0,0,0,0]$& $(1,1,1,1,2,4,6,8,10,7,3,5)$&$(1,0,0,0,1,2,2,2,2,2,1,2)$&$2$ & \cr 
$(7,3)$&$[0,1,0,0,1,0,0,0,0]$& $(1,1,1,2,3,4,6,8,10,7,3,5)$&$(1,0,0,1,1,1,2,2,2,2,1,2)$&$0$ & \cr 
$(7,3)$&$[1,0,0,0,0,1,0,0,0]$& $(1,1,2,3,4,5,6,8,10,7,3,5)$&$(1,0,1,1,1,1,1,2,2,2,1,2)$&$-2$ & \cr 
$(7,3)$&$[0,0,0,0,0,0,0,1,1]$& $(1,2,3,4,5,6,7,8,9,7,3,4)$&$(1,1,1,1,1,1,1,1,1,2,0,1)$&$-2$ & \cr 
$(7,3)$&$[0,1,0,0,0,1,0,0,1]$& $(1,1,1,2,3,4,5,7,9,7,3,4)$&$(1,0,0,1,1,1,1,2,2,2,0,1)$&$2$ & \cr 
$(7,3)$&$[1,0,0,0,0,0,0,2,0]$& $(1,1,2,3,4,5,6,7,8,7,3,4)$&$(1,0,1,1,1,1,1,1,1,3,0,1)$&$2$ & \cr 
$(7,3)$&$[1,0,0,0,0,0,1,0,1]$& $(1,1,2,3,4,5,6,7,9,7,3,4)$&$(1,0,1,1,1,1,1,1,2,2,0,1)$&$0$ & \cr 
$(7,3)$&$[0,0,0,0,0,0,0,0,3]$& $(1,2,3,4,5,6,7,8,9,7,3,3)$&$(1,1,1,1,1,1,1,1,1,1,-1,0)$&$2$ & \cr 
\noalign{\hrule}$(7,4)$&$[0,0,0,0,0,0,1,0,0]$& $(1,2,3,4,5,6,7,8,10,7,4,5)$&$(1,1,1,1,1,1,1,1,2,2,2,1)$&$-4$ & \cr 
$(7,4)$&$[0,0,1,1,0,0,0,0,0]$& $(1,1,1,1,2,4,6,8,10,7,4,5)$&$(1,0,0,0,1,2,2,2,2,2,2,1)$&$2$ & \cr 
$(7,4)$&$[0,1,0,0,1,0,0,0,0]$& $(1,1,1,2,3,4,6,8,10,7,4,5)$&$(1,0,0,1,1,1,2,2,2,2,2,1)$&$0$ & \cr 
\noalign{\hrule}
}
}\cr
Table A.3 Low level weights in the {\it IIB supergravity} decomposition of the $l_1$ representation of $E_{11}$ (continued)\cr}$$
$$\halign{\centerline{#} \cr
\vbox{\offinterlineskip
\halign{\strut \vrule \quad \hfil # \hfil\quad &\vrule \quad \hfil # \hfil\quad 
&\vrule \quad \hfil # \hfil\quad  &\vrule \quad \hfil # \hfil\quad &\vrule \quad \hfil # \hfil\quad &\vrule #
\cr
\noalign{\hrule}
Level&$A_{9}$ weights&$E_{12}$ Root&$E_{12}$ Root&Root length&\cr 
($m_{9},m_{10}$)&&($\alpha_i$ basis)&($e_i$ basis)&squared&\cr 
\noalign{\hrule}
$(7,4)$&$[1,0,0,0,0,1,0,0,0]$& $(1,1,2,3,4,5,6,8,10,7,4,5)$&$(1,0,1,1,1,1,1,2,2,2,2,1)$&$-2$ & \cr 
$(7,4)$&$[0,0,0,0,0,0,0,1,1]$& $(1,2,3,4,5,6,7,8,9,7,4,4)$&$(1,1,1,1,1,1,1,1,1,2,1,0)$&$-2$ & \cr 
$(7,4)$&$[0,1,0,0,0,1,0,0,1]$& $(1,1,1,2,3,4,5,7,9,7,4,4)$&$(1,0,0,1,1,1,1,2,2,2,1,0)$&$2$ & \cr 
$(7,4)$&$[1,0,0,0,0,0,0,2,0]$& $(1,1,2,3,4,5,6,7,8,7,4,4)$&$(1,0,1,1,1,1,1,1,1,3,1,0)$&$2$ & \cr 
$(7,4)$&$[1,0,0,0,0,0,1,0,1]$& $(1,1,2,3,4,5,6,7,9,7,4,4)$&$(1,0,1,1,1,1,1,1,2,2,1,0)$&$0$ & \cr 
$(7,4)$&$[0,0,0,0,0,0,0,0,3]$& $(1,2,3,4,5,6,7,8,9,7,4,3)$&$(1,1,1,1,1,1,1,1,1,1,0,-1)$&$2$ & \cr 
\noalign{\hrule}$(7,5)$&$[0,0,0,0,0,0,1,0,0]$& $(1,2,3,4,5,6,7,8,10,7,5,5)$&$(1,1,1,1,1,1,1,1,2,2,3,0)$&$0$ & \cr 
$(7,5)$&$[1,0,0,0,0,1,0,0,0]$& $(1,1,2,3,4,5,6,8,10,7,5,5)$&$(1,0,1,1,1,1,1,2,2,2,3,0)$&$2$ & \cr 
$(7,5)$&$[0,0,0,0,0,0,0,1,1]$& $(1,2,3,4,5,6,7,8,9,7,5,4)$&$(1,1,1,1,1,1,1,1,1,2,2,-1)$&$2$ & \cr 
\noalign{\hrule}\noalign{\hrule}$(8,2)$&$[0,0,0,0,1,0,0,0,0]$& $(1,2,3,4,5,6,8,10,12,8,2,6)$&$(1,1,1,1,1,1,2,2,2,2,0,4)$&$2$ & \cr 
\noalign{\hrule}$(8,3)$&$[0,0,0,0,1,0,0,0,0]$& $(1,2,3,4,5,6,8,10,12,8,3,6)$&$(1,1,1,1,1,1,2,2,2,2,1,3)$&$-4$ & \cr 
$(8,3)$&$[0,1,1,0,0,0,0,0,0]$& $(1,1,1,2,4,6,8,10,12,8,3,6)$&$(1,0,0,1,2,2,2,2,2,2,1,3)$&$0$ & \cr 
$(8,3)$&$[1,0,0,1,0,0,0,0,0]$& $(1,1,2,3,4,6,8,10,12,8,3,6)$&$(1,0,1,1,1,2,2,2,2,2,1,3)$&$-2$ & \cr 
$(8,3)$&$[0,0,0,0,0,0,1,1,0]$& $(1,2,3,4,5,6,7,8,10,8,3,5)$&$(1,1,1,1,1,1,1,1,2,3,0,2)$&$0$ & \cr 
$(8,3)$&$[0,0,0,0,0,1,0,0,1]$& $(1,2,3,4,5,6,7,9,11,8,3,5)$&$(1,1,1,1,1,1,1,2,2,2,0,2)$&$-2$ & \cr 
$(8,3)$&$[0,1,0,1,0,0,0,0,1]$& $(1,1,1,2,3,5,7,9,11,8,3,5)$&$(1,0,0,1,1,2,2,2,2,2,0,2)$&$2$ & \cr 
$(8,3)$&$[1,0,0,0,0,1,0,1,0]$& $(1,1,2,3,4,5,6,8,10,8,3,5)$&$(1,0,1,1,1,1,1,2,2,3,0,2)$&$2$ & \cr 
$(8,3)$&$[1,0,0,0,1,0,0,0,1]$& $(1,1,2,3,4,5,7,9,11,8,3,5)$&$(1,0,1,1,1,1,2,2,2,2,0,2)$&$0$ & \cr 
$(8,3)$&$[0,0,0,0,0,0,1,0,2]$& $(1,2,3,4,5,6,7,8,10,8,3,4)$&$(1,1,1,1,1,1,1,1,2,2,-1,1)$&$2$ & \cr 
\noalign{\hrule}$(8,4)$&$[0,0,0,0,1,0,0,0,0]$& $(1,2,3,4,5,6,8,10,12,8,4,6)$&$(1,1,1,1,1,1,2,2,2,2,2,2)$&$-6$ & \cr 
$(8,4)$&$[0,1,1,0,0,0,0,0,0]$& $(1,1,1,2,4,6,8,10,12,8,4,6)$&$(1,0,0,1,2,2,2,2,2,2,2,2)$&$-2$ & \cr 
$(8,4)$&$[1,0,0,1,0,0,0,0,0]$& $(1,1,2,3,4,6,8,10,12,8,4,6)$&$(1,0,1,1,1,2,2,2,2,2,2,2)$&$-4$ & \cr 
$(8,4)$&$[0,0,0,0,0,0,1,1,0]$& $(1,2,3,4,5,6,7,8,10,8,4,5)$&$(1,1,1,1,1,1,1,1,2,3,1,1)$&$-2$ & \cr 
$(8,4)$&$[0,0,0,0,0,1,0,0,1]$& $(1,2,3,4,5,6,7,9,11,8,4,5)$&$(1,1,1,1,1,1,1,2,2,2,1,1)$&$-4$ & \cr 
$(8,4)$&$[0,0,2,0,0,0,0,0,1]$& $(1,1,1,1,3,5,7,9,11,8,4,5)$&$(1,0,0,0,2,2,2,2,2,2,1,1)$&$2$ & \cr 
$(8,4)$&$[0,1,0,0,1,0,0,1,0]$& $(1,1,1,2,3,4,6,8,10,8,4,5)$&$(1,0,0,1,1,1,2,2,2,3,1,1)$&$2$ & \cr 
$(8,4)$&$[0,1,0,1,0,0,0,0,1]$& $(1,1,1,2,3,5,7,9,11,8,4,5)$&$(1,0,0,1,1,2,2,2,2,2,1,1)$&$0$ & \cr 
$(8,4)$&$[1,0,0,0,0,0,2,0,0]$& $(1,1,2,3,4,5,6,7,10,8,4,5)$&$(1,0,1,1,1,1,1,1,3,3,1,1)$&$2$ & \cr 
$(8,4)$&$[1,0,0,0,0,1,0,1,0]$& $(1,1,2,3,4,5,6,8,10,8,4,5)$&$(1,0,1,1,1,1,1,2,2,3,1,1)$&$0$ & \cr 
$(8,4)$&$[1,0,0,0,1,0,0,0,1]$& $(1,1,2,3,4,5,7,9,11,8,4,5)$&$(1,0,1,1,1,1,2,2,2,2,1,1)$&$-2$ & \cr 
$(8,4)$&$[0,0,0,0,0,0,0,2,1]$& $(1,2,3,4,5,6,7,8,9,8,4,4)$&$(1,1,1,1,1,1,1,1,1,3,0,0)$&$2$ & \cr 
$(8,4)$&$[0,0,0,0,0,0,1,0,2]$& $(1,2,3,4,5,6,7,8,10,8,4,4)$&$(1,1,1,1,1,1,1,1,2,2,0,0)$&$0$ & \cr 
$(8,4)$&$[1,0,0,0,0,1,0,0,2]$& $(1,1,2,3,4,5,6,8,10,8,4,4)$&$(1,0,1,1,1,1,1,2,2,2,0,0)$&$2$ & \cr 
\noalign{\hrule}$(8,5)$&$[0,0,0,0,1,0,0,0,0]$& $(1,2,3,4,5,6,8,10,12,8,5,6)$&$(1,1,1,1,1,1,2,2,2,2,3,1)$&$-4$ & \cr 
$(8,5)$&$[0,1,1,0,0,0,0,0,0]$& $(1,1,1,2,4,6,8,10,12,8,5,6)$&$(1,0,0,1,2,2,2,2,2,2,3,1)$&$0$ & \cr 
$(8,5)$&$[1,0,0,1,0,0,0,0,0]$& $(1,1,2,3,4,6,8,10,12,8,5,6)$&$(1,0,1,1,1,2,2,2,2,2,3,1)$&$-2$ & \cr 
$(8,5)$&$[0,0,0,0,0,0,1,1,0]$& $(1,2,3,4,5,6,7,8,10,8,5,5)$&$(1,1,1,1,1,1,1,1,2,3,2,0)$&$0$ & \cr 
$(8,5)$&$[0,0,0,0,0,1,0,0,1]$& $(1,2,3,4,5,6,7,9,11,8,5,5)$&$(1,1,1,1,1,1,1,2,2,2,2,0)$&$-2$ & \cr 
$(8,5)$&$[0,1,0,1,0,0,0,0,1]$& $(1,1,1,2,3,5,7,9,11,8,5,5)$&$(1,0,0,1,1,2,2,2,2,2,2,0)$&$2$ & \cr 
$(8,5)$&$[1,0,0,0,0,1,0,1,0]$& $(1,1,2,3,4,5,6,8,10,8,5,5)$&$(1,0,1,1,1,1,1,2,2,3,2,0)$&$2$ & \cr 
$(8,5)$&$[1,0,0,0,1,0,0,0,1]$& $(1,1,2,3,4,5,7,9,11,8,5,5)$&$(1,0,1,1,1,1,2,2,2,2,2,0)$&$0$ & \cr 
$(8,5)$&$[0,0,0,0,0,0,1,0,2]$& $(1,2,3,4,5,6,7,8,10,8,5,4)$&$(1,1,1,1,1,1,1,1,2,2,1,-1)$&$2$ & \cr 
\noalign{\hrule}$(8,6)$&$[0,0,0,0,1,0,0,0,0]$& $(1,2,3,4,5,6,8,10,12,8,6,6)$&$(1,1,1,1,1,1,2,2,2,2,4,0)$&$2$ & \cr 
\noalign{\hrule}
$(9,3)$&$[0,0,1,0,0,0,0,0,0]$& $(1,2,3,4,6,8,10,12,14,9,3,7)$&$(1,1,1,1,2,2,2,2,2,2,1,4)$&$-4$ & \cr 
$(9,3)$&$[1,1,0,0,0,0,0,0,0]$& $(1,1,2,4,6,8,10,12,14,9,3,7)$&$(1,0,1,2,2,2,2,2,2,2,1,4)$&$-2$ & \cr 
$(9,3)$&$[0,0,0,0,0,1,1,0,0]$& $(1,2,3,4,5,6,7,9,12,9,3,6)$&$(1,1,1,1,1,1,1,2,3,3,0,3)$&$2$ & \cr 
$(9,3)$&$[0,0,0,0,1,0,0,1,0]$& $(1,2,3,4,5,6,8,10,12,9,3,6)$&$(1,1,1,1,1,1,2,2,2,3,0,3)$&$0$ & \cr 
\noalign{\hrule}
}
}\cr
Table A.3 Low level weights in the {\it IIB supergravity} decomposition of the $l_1$ representation of $E_{11}$ (continued) \cr}$$
$$\halign{\centerline{#} \cr
\vbox{\offinterlineskip
\halign{\strut \vrule \quad \hfil # \hfil\quad &\vrule \quad \hfil # \hfil\quad 
&\vrule \quad \hfil # \hfil\quad  &\vrule \quad \hfil # \hfil\quad &\vrule \quad \hfil # \hfil\quad &\vrule #
\cr
\noalign{\hrule}
Level&$A_{9}$ weights&$E_{12}$ Root&$E_{12}$ Root&Root length&\cr 
($m_{9},m_{10}$)&&($\alpha_i$ basis)&($e_i$ basis)&squared&\cr 
\noalign{\hrule}
$(9,3)$&$[0,0,0,1,0,0,0,0,1]$& $(1,2,3,4,5,7,9,11,13,9,3,6)$&$(1,1,1,1,1,2,2,2,2,2,0,3)$&$-2$ & \cr 
$(9,3)$&$[0,2,0,0,0,0,0,0,1]$& $(1,1,1,3,5,7,9,11,13,9,3,6)$&$(1,0,0,2,2,2,2,2,2,2,0,3)$&$2$ & \cr 
$(9,3)$&$[1,0,0,1,0,0,0,1,0]$& $(1,1,2,3,4,6,8,10,12,9,3,6)$&$(1,0,1,1,1,2,2,2,2,3,0,3)$&$2$ & \cr 
$(9,3)$&$[1,0,1,0,0,0,0,0,1]$& $(1,1,2,3,5,7,9,11,13,9,3,6)$&$(1,0,1,1,2,2,2,2,2,2,0,3)$&$0$ & \cr 
$(9,3)$&$[0,0,0,0,1,0,0,0,2]$& $(1,2,3,4,5,6,8,10,12,9,3,5)$&$(1,1,1,1,1,1,2,2,2,2,-1,2)$&$2$ & \cr 
\noalign{\hrule}$(9,4)$&$[0,0,1,0,0,0,0,0,0]$& $(1,2,3,4,6,8,10,12,14,9,4,7)$&$(1,1,1,1,2,2,2,2,2,2,2,3)$&$-8$ & \cr 
$(9,4)$&$[1,1,0,0,0,0,0,0,0]$& $(1,1,2,4,6,8,10,12,14,9,4,7)$&$(1,0,1,2,2,2,2,2,2,2,2,3)$&$-6$ & \cr 
$(9,4)$&$[0,0,0,0,0,1,1,0,0]$& $(1,2,3,4,5,6,7,9,12,9,4,6)$&$(1,1,1,1,1,1,1,2,3,3,1,2)$&$-2$ & \cr 
$(9,4)$&$[0,0,0,0,1,0,0,1,0]$& $(1,2,3,4,5,6,8,10,12,9,4,6)$&$(1,1,1,1,1,1,2,2,2,3,1,2)$&$-4$ & \cr 
$(9,4)$&$[0,0,0,1,0,0,0,0,1]$& $(1,2,3,4,5,7,9,11,13,9,4,6)$&$(1,1,1,1,1,2,2,2,2,2,1,2)$&$-6$ & \cr 
$(9,4)$&$[0,1,0,1,0,0,1,0,0]$& $(1,1,1,2,3,5,7,9,12,9,4,6)$&$(1,0,0,1,1,2,2,2,3,3,1,2)$&$2$ & \cr 
$(9,4)$&$[0,1,1,0,0,0,0,1,0]$& $(1,1,1,2,4,6,8,10,12,9,4,6)$&$(1,0,0,1,2,2,2,2,2,3,1,2)$&$0$ & \cr 
$(9,4)$&$[0,2,0,0,0,0,0,0,1]$& $(1,1,1,3,5,7,9,11,13,9,4,6)$&$(1,0,0,2,2,2,2,2,2,2,1,2)$&$-2$ & \cr 
$(9,4)$&$[1,0,0,0,0,2,0,0,0]$& $(1,1,2,3,4,5,6,9,12,9,4,6)$&$(1,0,1,1,1,1,1,3,3,3,1,2)$&$2$ & \cr 
$(9,4)$&$[1,0,0,0,1,0,1,0,0]$& $(1,1,2,3,4,5,7,9,12,9,4,6)$&$(1,0,1,1,1,1,2,2,3,3,1,2)$&$0$ & \cr 
$(9,4)$&$[1,0,0,1,0,0,0,1,0]$& $(1,1,2,3,4,6,8,10,12,9,4,6)$&$(1,0,1,1,1,2,2,2,2,3,1,2)$&$-2$ & \cr 
$(9,4)$&$[1,0,1,0,0,0,0,0,1]$& $(1,1,2,3,5,7,9,11,13,9,4,6)$&$(1,0,1,1,2,2,2,2,2,2,1,2)$&$-4$ & \cr 
$(9,4)$&$[0,0,0,0,0,0,2,0,1]$& $(1,2,3,4,5,6,7,8,11,9,4,5)$&$(1,1,1,1,1,1,1,1,3,3,0,1)$&$2$ & \cr 
$(9,4)$&$[0,0,0,0,0,1,0,1,1]$& $(1,2,3,4,5,6,7,9,11,9,4,5)$&$(1,1,1,1,1,1,1,2,2,3,0,1)$&$0$ & \cr 
$(9,4)$&$[0,0,0,0,1,0,0,0,2]$& $(1,2,3,4,5,6,8,10,12,9,4,5)$&$(1,1,1,1,1,1,2,2,2,2,0,1)$&$-2$ & \cr 
$(9,4)$&$[0,1,1,0,0,0,0,0,2]$& $(1,1,1,2,4,6,8,10,12,9,4,5)$&$(1,0,0,1,2,2,2,2,2,2,0,1)$&$2$ & \cr 
$(9,4)$&$[1,0,0,0,1,0,0,1,1]$& $(1,1,2,3,4,5,7,9,11,9,4,5)$&$(1,0,1,1,1,1,2,2,2,3,0,1)$&$2$ & \cr 
$(9,4)$&$[1,0,0,1,0,0,0,0,2]$& $(1,1,2,3,4,6,8,10,12,9,4,5)$&$(1,0,1,1,1,2,2,2,2,2,0,1)$&$0$ & \cr 
\noalign{\hrule}$(9,5)$&$[0,0,1,0,0,0,0,0,0]$& $(1,2,3,4,6,8,10,12,14,9,5,7)$&$(1,1,1,1,2,2,2,2,2,2,3,2)$&$-8$ & \cr 
$(9,5)$&$[1,1,0,0,0,0,0,0,0]$& $(1,1,2,4,6,8,10,12,14,9,5,7)$&$(1,0,1,2,2,2,2,2,2,2,3,2)$&$-6$ & \cr 
$(9,5)$&$[0,0,0,0,0,1,1,0,0]$& $(1,2,3,4,5,6,7,9,12,9,5,6)$&$(1,1,1,1,1,1,1,2,3,3,2,1)$&$-2$ & \cr 
$(9,5)$&$[0,0,0,0,1,0,0,1,0]$& $(1,2,3,4,5,6,8,10,12,9,5,6)$&$(1,1,1,1,1,1,2,2,2,3,2,1)$&$-4$ & \cr 
$(9,5)$&$[0,0,0,1,0,0,0,0,1]$& $(1,2,3,4,5,7,9,11,13,9,5,6)$&$(1,1,1,1,1,2,2,2,2,2,2,1)$&$-6$ & \cr 
$(9,5)$&$[0,1,0,1,0,0,1,0,0]$& $(1,1,1,2,3,5,7,9,12,9,5,6)$&$(1,0,0,1,1,2,2,2,3,3,2,1)$&$2$ & \cr 
$(9,5)$&$[0,1,1,0,0,0,0,1,0]$& $(1,1,1,2,4,6,8,10,12,9,5,6)$&$(1,0,0,1,2,2,2,2,2,3,2,1)$&$0$ & \cr 
$(9,5)$&$[0,2,0,0,0,0,0,0,1]$& $(1,1,1,3,5,7,9,11,13,9,5,6)$&$(1,0,0,2,2,2,2,2,2,2,2,1)$&$-2$ & \cr 
$(9,5)$&$[1,0,0,0,0,2,0,0,0]$& $(1,1,2,3,4,5,6,9,12,9,5,6)$&$(1,0,1,1,1,1,1,3,3,3,2,1)$&$2$ & \cr 
$(9,5)$&$[1,0,0,0,1,0,1,0,0]$& $(1,1,2,3,4,5,7,9,12,9,5,6)$&$(1,0,1,1,1,1,2,2,3,3,2,1)$&$0$ & \cr 
$(9,5)$&$[1,0,0,1,0,0,0,1,0]$& $(1,1,2,3,4,6,8,10,12,9,5,6)$&$(1,0,1,1,1,2,2,2,2,3,2,1)$&$-2$ & \cr 
$(9,5)$&$[1,0,1,0,0,0,0,0,1]$& $(1,1,2,3,5,7,9,11,13,9,5,6)$&$(1,0,1,1,2,2,2,2,2,2,2,1)$&$-4$ & \cr 
$(9,5)$&$[0,0,0,0,0,0,2,0,1]$& $(1,2,3,4,5,6,7,8,11,9,5,5)$&$(1,1,1,1,1,1,1,1,3,3,1,0)$&$2$ & \cr 
$(9,5)$&$[0,0,0,0,0,1,0,1,1]$& $(1,2,3,4,5,6,7,9,11,9,5,5)$&$(1,1,1,1,1,1,1,2,2,3,1,0)$&$0$ & \cr 
$(9,5)$&$[0,0,0,0,1,0,0,0,2]$& $(1,2,3,4,5,6,8,10,12,9,5,5)$&$(1,1,1,1,1,1,2,2,2,2,1,0)$&$-2$ & \cr 
$(9,5)$&$[0,1,1,0,0,0,0,0,2]$& $(1,1,1,2,4,6,8,10,12,9,5,5)$&$(1,0,0,1,2,2,2,2,2,2,1,0)$&$2$ & \cr 
$(9,5)$&$[1,0,0,0,1,0,0,1,1]$& $(1,1,2,3,4,5,7,9,11,9,5,5)$&$(1,0,1,1,1,1,2,2,2,3,1,0)$&$2$ & \cr 
$(9,5)$&$[1,0,0,1,0,0,0,0,2]$& $(1,1,2,3,4,6,8,10,12,9,5,5)$&$(1,0,1,1,1,2,2,2,2,2,1,0)$&$0$ & \cr 
\noalign{\hrule}
}
}\cr
Table A.3 Low level weights in the {\it IIB supergravity} decomposition of the $l_1$ representation of $E_{11}$ (continued)\cr}$$

\eject

\medskip
{\bf References}
\medskip

\item{[1]} P. West, {\sl $E_{11}$ and M Theory}, Class. Quant. Grav. {\bf 18 } (2001) 4443, {\tt hep-th/0104081}
\item{[2]} M. Gaberdiel, D. Olive and P. West, {\sl A class of Lorentzian Kac-Moody algebras}, Nucl. Phys. {\bf B645} (2002) 403-437, {\tt hep-th/0205068}
\item{[3]} P. West, {\sl $E_{11}$, SL(32) and Central Charges}, Phys. Lett. {\bf B 575} (2003) 333-342, {\tt hep-th/0307098}
\item{[4]} P. West, {\it  The  IIA, IIB and eleven dimensional theories and their common $E_{11}$ origin}, Nucl. Phys. {\bf B693} (2004) 76-102, {\tt hep-th/0402140} 
\item{[5]} A. Kleinschmidt and P. West, {\it  Representations of $\cal{G}^{+++}$ and the role of space-time}, JHEP  {\bf 0402} (2004) 033, {\tt hep-th/0312247} 
\item{[6]} S. Elitzur, A. Giveon, D. Kutasov, and E. Rabinovici {\it Algebraic aspects of matrix theory on $T^d$}, Nucl. Phys. {\bf B509} (1998) 122-144, {\tt hep-th/9707217}
\item{[7]} N. Obers, B. Pioline and E.  Rabinovici, {\it M-theory and U-duality on $T^d$ with gauge backgrounds}, Nucl. Phys. {\bf B525} (1998) 163-181, {\tt hep-th/9712084}
\item{[8]} N. Obers and B. Pioline, {\it U-duality and M-theory, an algebraic approach}, Corfu 1998, Quantum Aspects of Gauge Theories, Supersymmetry and Unification 421-433 {\tt hep-th/9812139}
\item{[9]} N. Obers and B. Pioline, {\it U-duality and M-theory}, Phys. Rept. {\bf 318} 113-225 (1999) {\tt hep-th/9809039}
\item{[10]} H. Lu, C. N. Pope and K. S. Stelle, {\sl Weyl Group Invariance and $p$-brane Multiplets}, Nucl. Phys. {\bf B476} (1996) 89-117 {\tt hep-th/9602140}
\item{[11]} P. West, {\sl $E_{11}$ origin of Brane charges and U-duality multiplets}, JHEP  {\bf 0408} (2004) 052, {\tt hep-th/0406150}
\item{[12]} P. West, {\sl Brane dynamics, central charges and E(11)}, JHEP  {\bf 0503} (2005) 077, {\tt hep-th/0412336}
\item{[13]} P. P. Cook and P. West, {\sl $G^{+++}$ and brane solutions}, Nucl. Phys.  {\bf B705} (2005), {\tt hep-th/0405149}
\item{[14]} F. Englert, L. Houart, A. Taormina and P. West, {\sl The Symmetry of M-Theories}, JHEP {\bf 0309} (2003) 020, {\tt hep-th/0304206}
\item{[15]} P. West, {\sl Some simple predictions from $E_{11}$ symmetry}, {\tt hep-th/0407088}
\item{[16]} P. P. Cook, {\sl Connections between Kac-Moody Algebras and M-Theory}, PhD Thesis, King's College London (2006), {\tt hep-th//0711.3498}
\item{[17]} C. Hull, {\sl U-duality and BPS Spectrum of Super Yang-Mills Theory and M-Theory}, JHEP {\bf 9810} (1998) 011, {\tt hep-th/9711179}
\item{[18]} M. Blau and M. O'Loughlin, {\sl Aspects of U-duality in Matrix Theory}, Nucl. Phys. {\bf B525} (1998) 182-214, {\tt hep-th/9712047}
\item{[19]} R. Guven, {\sl Black p-brane solutions of D=11 supergravity theory},  Phys. Lett. {\bf B276} (1992) 49-55 
\item{[20]} F. Riccioni and P. West, {\sl Dual fields and $E_{11}$},  Phys.Lett. {\bf B645} (2007) 286-292, 
{\tt hep-th/0612001}
\item{[21]} E. Lozano-Tellechea and T. Ortin, {\sl 7-Branes and Higher Kaluza-Klein Branes},  Nucl. Phys. {\bf B607} (2001) 213-236, {\tt hep-th/0012051}
\item{[22]} C. Hull, {\sl Gravitational duality, branes and charges}, Nucl. Phys. {\bf B509} (1998) 216-251, {\tt hep-th/9705162}
\item{[23]} I. C. G. Campbell and P. West, {\sl  N=2 D=10 Nonchiral Supergravity and Its Spontaneous Compactification}, Nucl. Phys. {\bf B243} (1984) 112; M. Huq, M. Namanzie, {\sl Kaluza- 
Klein supergravity in ten dimensions, Class. Quant. Grav. 2 (1985); F. Giani, M. 
Pernici, N=2 supergravity in ten dimensions}, Phys. Rev. {\bf D30} (1984), 325
\item{[24]} J. Azcarraga, J. Gauntlett, J. Izquierdo and P. Townsend, {\sl  Topological Extensions of 
the Supersymmetry Algebra for Extended Objects}, Phys. Rev. Lett. {\bf 63} no. 22 (1989)
\item{[25]} B. de Wit and H. Nicolai, {\sl Hidden symmetries, central charges and all that}, Class Quantum Gravity {\bf 18}  (2001) 3095-3112,  {\tt hep-th/0011239}
\item{[26]} M. Duff and K. Stelle, {\sl Multi-Membrane Solutions of D=11 Supergravity}, Phys. Lett. {\bf B253} (1991) 113-118 

\end